\documentclass{aa}  

\usepackage{graphicx}
\usepackage{txfonts}
\usepackage{lipsum}
\usepackage{subcaption}         
\usepackage{lscape}           
\usepackage{placeins}  
\usepackage{xcolor}

\begin{document}

\title{Transition from outside-in to inside-out at $z\sim 2$: \\
evidence from radial profiles of specific star formation rate based on JWST/HST}
\titlerunning{Transition from outside-in to inside-out at $z\sim 2$}
 
\author{Jie Song\inst{1, 2, 3}\thanks{Corresponding author: jiesong@mail.ustc.edu.cn}
\and Enci Wang\inst{1, 2}\thanks{Corresponding author: ecwang16@ustc.edu.cn}
\and Cheng Jia\inst{1, 2}
\and Cheqiu Lyu\inst{1, 2}
\and Yangyao Chen\inst{1, 2}
\and Jinyang Wang\inst{1, 2}
\and Fujia Li\inst{1, 2, 3}
\and Weiyu Ding\inst{1, 2, 3}
\and Guanwen Fang\inst{4, 5}
\and Xu Kong\inst{1, 2, 3}\thanks{Corresponding author: xkong@ustc.edu.cn}}

\institute{Department of Astronomy, University of Science and Technology of China, Hefei 230026, China
\and School of Astronomy and Space Science, University of Science and Technology of China, Hefei 230026, China
\and Institute of Deep Space Sciences, Deep Space Exploration Laboratory, Hefei 230026, China
\and School of Physics and Astronomy, Anqing Normal University, Anqing 246133, China
\and Institute of Astronomy and Astrophysics, Anqing Normal University, Anqing 246133, China}
 
\abstract
{By combining high-resolution observations from JWST and HST, we have measured the stellar masses, star formation rates (SFRs), and multi-wavelength morphologies of galaxies in the CANDELS fields. Furthermore, based on rest-frame 1 $\mu$m morphologies, we have derived spatially resolved stellar mass and SFR surface density ($\Sigma_*$ and $\Sigma_{\rm SFR}$) profiles for 46,313 galaxies with reliable structural measurements at $0<z<4$ and $\log(M_\ast /M_{\odot})>8$, and provide the corresponding catalogue.
For star-forming galaxies (SFGs), our results show excellent consistency with previous studies in terms of the star formation main sequence and the size–mass relation, demonstrating the robustness of our stellar mass and SFR measurements. 
For spatially resolved profiles, we find that at higher redshifts ($z>2.5$), the median radial profile of $\Sigma_{\rm SFR}$ is nearly parallel to but slightly steeper than that of $\Sigma_*$. This results in mildly negative gradients in the specific SFR (sSFR) profiles across all stellar mass bins considered. These findings indicate that galaxies at 
$z>2.5$ cannot grow in size via only in-situ star formation, challenging the understanding of galaxy size evolution beyond the cosmic noon. In contrast, at 
$z<2.0$, the sSFR profiles transition to exhibit more and more positive gradients at lower redshifts, consistent with an inside-out growth scenario where star formation preferentially expands the galactic outskirts.}

\maketitle
\nolinenumbers

\section{Introduction} \label{sec:1}
Understanding how galaxies assemble their stellar mass and grow in size over cosmic time remains a central question in galaxy formation and evolution. Over the past several decades, extensive efforts have been devoted to characterizing the growth process of galaxies, revealing that their structures form hierarchically, involving processes such as baryonic cooling, gas accretion, star formation, galaxy mergers, and a wide variety of feedback mechanisms (e.g., \citealt{2000MNRAS.319..168C, Lilly-13, 2014Natur.516...68G, 2014A&ARv..22...71S, 2017ASSL..430..145K, 2018ARA&A..56..435W, Wang-19, 2022MNRAS.517L..92E, 2022MNRAS.515.1430D, Wang-21, Wang-22a, Wang-22b, 2024ARNPS..74..173P, 2024OJAp....7E.121E, Chen-25, Jia-25, Lyu-25}).

As one of the most important aspects of galaxy growth, numerous studies have found that the majority of star formation activity occurs along the so-called Star-Forming Main Sequence (SFMS), which represents a tight correlation between stellar mass and SFR (e.g., \citealt{2007ApJ...670..156D, 2014ApJS..214...15S, 2014ApJ...795..104W, 2017ApJ...847...76S, 2023MNRAS.519.1526P, 2024A&A...691A.164K, 2024ApJ...977..133C, 2025ApJ...981..161R, He-25}). This relation exhibits a typical scatter of only about 0.3 dex, and the scatter remains nearly constant across different redshift epochs (e.g., \citealt{2012ApJ...754L..29W, 2014ApJS..214...15S, 2024A&A...691A.164K}), which may suggest that galaxies grow in mass over cosmic time in a state of self-regulated semiequilibrium (e.g., \citealt{2010ApJ...713..686D, 2010MNRAS.407.2091G, Wang-19, 2020MNRAS.497..698T}). Additionally, many studies have demonstrated that galaxy sizes increase with stellar mass, forming the well-known size–mass relation (e.g., \citealt{2014ApJ...788...28V, 2020ApJ...905..170M, 2021MNRAS.506..928N, 2024ApJ...960...53V}). By combining constraints from both the SFMS and the size–mass relation, some previous works have reconstructed the characteristic size–growth trajectories of individual galaxies (e.g., \citealt{2015ApJ...813...23V, 2024ApJ...977..165J}).

However, since star formation is not uniformly distributed within galaxies, studies based solely on integrated physical properties are insufficient for capturing the full picture of galaxy growth. A comprehensive understanding of galaxy growth and quenching mechanisms requires not only knowledge of galaxies’ integrated properties, but also their spatially resolved information (e.g., \citealt{2017ApJ...840...47B, 2017ApJ...834...81J, 2018MNRAS.479.5083A, Wang-18, 2020ARA&A..58...99S, 2022MNRAS.510.3622B, 2023ApJ...945..117A}). Spatial distributions of SFR and stellar mass provide direct insight into where growth occurs within galaxies (e.g., \citealt{2012ApJ...747L..28N, 2015Sci...348..314T, 2016ApJ...828...27N, 2018MNRAS.479.5083A, 2020MNRAS.497..698T, 2023ApJ...945..117A}). In low-redshift universe, spatially resolved studies have suggested that galaxies can assemble their mass in different patterns (e.g., \citealt{2013ApJ...764L...1P, Wang-18}). A minor population of galaxies exhibit rapid stellar mass assembly within their central regions, characterized by elevated central sSFR and relatively younger central stellar populations. This phenomenological profile corresponds to an outside-in growth mode. Conversely, most star-forming galaxies display preferential mass accumulation in their outskirts, featuring higher sSFR at large radii and comparatively older central populations, a signature consistent with the inside-out growth scenario.

Since galaxy growth is inherently continuous process, the increase in stellar mass (potentially along the SFMS) is accompanied by size growth, as well as corresponding evolution in stellar mass surface-density and SFR surface-density profiles (e.g., \citealt{2009Natur.457..451D, 2015Sci...348..314T, 2016ApJ...828...27N, 2017MNRAS.465..722F, 2020ARA&A..58..661F, 2024OJAp....7E.113J, 2025ApJ...980..168D}). Therefore, obtaining a more complete understanding of galaxy evolution requires combining both the integrated and spatially resolved physical properties (e.g., \citealt{2024ApJ...975..252H}). For example, by analyzing the spatial profiles of stellar mass and SFR, one may predict the future size of a galaxy after a given period of star formation. 

Nevertheless, studying the spatially resolved properties of galaxies at intermediate to high redshifts critically depends on either high-quality IFS or high-resolution, multi-wavelength imaging. In the case of IFS, the available sample sizes are typically small and often subject to selection biases (e.g., \citealt{2015Sci...348..314T}). For imaging-based approaches, most prior studies have relied predominantly on Hubble Space Telescope (HST) observations (e.g., \citealt{2012ApJ...753..114W, 2018MNRAS.479.5083A}). However, the limited wavelength coverage of HST restricts observations to the rest-frame ultraviolet and optical at $z\sim 3$, thereby hindering accurate measurements of stellar masses and other key physical properties (e.g., \citealt{2010ApJ...709..644I, 2023ApJ...958...82S, 2025ApJ...978L..42C}). The successful launch of the James Webb Space Telescope (JWST) marks a transformative advancement, enabling more detailed investigations of resolved galaxy properties. With its deep, high-resolution imaging in the near-infrared, JWST facilitates significantly more precise measurements of galaxies at intermediate to high redshifts, thereby allowing robust studies of their resolved physical structures and star formation histories (e.g., \citealt{2023ApJ...958...82S, 2025ApJ...978L..42C, 2025arXiv250314591M, 2025arXiv250405244H}).

Some recent studies have begun to investigate the spatially resolved physical properties of galaxies by leveraging combined observations from HST and the JWST (e.g., \citealt{2023ApJ...948..126G, 2023ApJ...945..117A, 2024A&A...686A..63G}). However, these efforts have generally been constrained by the limited availability of deep JWST observations. Over the past three observing cycles, JWST has significantly expanded its coverage by conducting deep, multi-band imaging across a wide range of extragalactic fields (e.g., \citealt{2025arXiv250104085F, 2023arXiv230602465E, 2021jwst.prop.1837D}). 

In this work, we compile JWST multi-wavelength imaging from the well-established Cosmic Assembly Near-infrared Deep Extragalactic Legacy Survey (CANDELS; \citealt{2011ApJS..197...35G, 2011ApJS..197...36K}), including observations in EGS \citep{2007ApJ...660L...1D}, GOODS-S and GOODS-N \citep{2004ApJ...600L..93G}, COSMOS \citep{2007ApJS..172....1S}, and UDS \citep{2007MNRAS.379.1599L}, along with the existing HST observations in the same fields. 
Adopting a self-consistent analysis framework, we derive both integrated and spatially resolved physical properties (including stellar mass and SFR) of galaxies, and measure morphological parameters (e.g., half-light radius along the semimajor axis ($R_{\rm e}$), S\'{e}rsic index ($n$), axis ratio ($q$), and some nonparametric parameters) across each photometric bands\footnote{The catalog will be made publicly available at \url{https://github.com/jsong-astro/JWST-CANDELS} upon the publication of this paper. A detailed description of the catalog contents can be found in the README file provided on the associated webpage.}\citep{2024ApJ...977..165J}. As the first paper in a series, this work provides a comprehensive introduction to the methodology employed for deriving the integrated and spatially resolved physical properties of galaxies and compare our results with previous studies. By examining the SFMS and the size–mass relation within our galaxy sample, we find good consistency with previous studies, thereby validating the robustness of our analysis methodology. Furthermore, the investigation of spatially resolved profiles of key physical properties provides compelling evidence for the transition from outside-in to 
inside-out growth mode driven-by in-situ star formation at redshift of 2. Based on the measurements established in this study, our forthcoming work will involve carefully constructing galaxy samples that trace progenitor–descendant connections. These samples will be used to constrain models of galaxy growth and to study the physical mechanisms that may lead to the quenching of star formation.

This paper is organized as follows. In Section \ref{sec:2}, we outline the data acquisition and sample selection criteria. Section \ref{sec:3} provides a detailed description of the methodology employed to measure both the integrated and spatially resolved physical properties of galaxies. The results are presented in Section \ref{sec:4}. Section \ref{sec:5} discusses the possible physical mechanisms. Finally, a summary and conclusions are provided in Section \ref{sec:6}. Throughout this paper, we adopt a flat $\Lambda$CDM cosmology with $\rm H_0= 70km\ s^{-1}\ Mpc^{-1}$, $\Omega_m = 0.3$, and $\Omega_{\Lambda}=0.7$, and a \cite{2003PASP..115..763C} initial mass function.

\section{Data set} \label{sec:2}
\subsection{Multiband image}\label{sec:2.1}
In this study, we employ JWST observations from the CANDELS fields, incorporating data from the CEERS \citep{2025arXiv250104085F}, JADES \citep{2023arXiv230602465E}, and PRIMER \citep{2021jwst.prop.1837D} surveys. The CEERS survey covers an area of $\rm 94.6\ arcmin^2$ within the CANDELS/EGS field including JWST/NIRCam observations in the F115W, F150W, F200W, F277W, F356W, and F444W filters. The JADES survey spans
$\rm 83.0\ arcmin^2$ in the CANDELS/GOODS-N field (hereafter referred to as JADES-GDN) and $\rm 84.5\ arcmin^2$ in the CANDELS/GOODS-S field (JADES-GDS hereafter). In addition to the same filters as those in the CEERS field, JADES also includes JWST/NIRCam observations in the F090W filter. The PRIMER program covers $\rm 141.8\ arcmin^2$ in the CANDELS/COSMOS field (PRIMER-COSMOS hereafter) and $\rm 251.2\ arcmin^2$ in the CANDELS/UDS field (PRIMER-UDS hereafter), which utilizes the same broad-band filters as the JADES survey. In addition, these fields also include some medium-band observations. However, because the medium-band data are relatively shallow, we do not incorporate them into our analysis. As demonstrated in Appendix \ref{sec:7.2}, excluding these bands does not have a significant impact on our results.

All of these fields have also been observed by HST in multiple bands, including F435W, F606W, F814W, F125W, F140W, and F160W. However, given the proximity of the central wavelength of HST/F125W to that of JWST/F115W, and the similarity between the central wavelengths of HST/F140W and F160W with that of JWST/F150W, we include only the F435W, F606W, and F814W filters from HST in our subsequent analysis.

Both HST and JWST images have been carefully processed by the Cosmic Dawn Center \citep{2023ApJ...947...20V}, and the corresponding data are publicly available through the DJA website\footnote{\url{https://dawn-cph.github.io/dja/index.html}}. The pixel scale of the final mosaic image is $0\farcs04$/pixel. In this study, we utilize the most recent data releases available at the time of analysis: v7.0 for the PRIMER-COSMOS field, v7.2 for the CEERS, JADES-GDS, and PRIMER-UDS fields, and v7.3 for the JADES-GDN field.

\subsection{Photometric catalogue}
In addition to the carefully processed multi-wavelength imaging data, \citet{2023ApJ...947...20V} also provided photometric catalogs for all available JWST and HST filters in these fields. Source detection was performed on combined images of all NIRCam long-wavelength filters (typically F277W+F356W+F444W), followed by source extraction using SEP \citep{2016JOSS....1...58B}. Photometry was measured within circular apertures of diameter $0\farcs5$, with aperture corrections applied to estimate the ``total'' flux within elliptical Kron apertures. Based on these total flux measurements, photometric redshifts ($z_{\rm phot}$) were derived using EAZY-PY \citep{2008ApJ...686.1503B}. Additionally, spectroscopic redshift data were compiled for these fields, providing spectroscopic redshifts ($z_{\rm spec}$) for a subset of sources.

With the high quality of the HST and JWST data, \citet{2023ApJ...947...20V} demonstrated that the photometric redshifts are highly reliable. By comparing $z_{\rm phot}$ with $z_{\rm spec}$ for the spectroscopic sample, they found normalized median absolute deviations ($\sigma_{\rm NMAD}$) of 0.026, 0.019, 0.016, 0.022, and 0.023 for the CEERS, JADES-GDS, JADES-GDN, PRIMER-COSMOS, and PRIMER-UDS fields, respectively, where $\sigma_{\rm NMAD}$ is defined as the normalized median absolute deviation of $\Delta z / (1 + z_{\rm spec})$. Both the photometric and redshift catalogs are also publicly available on the DJA website. For further details regarding the photometric measurements and redshift estimation, we refer the reader to \citet{2023ApJ...947...20V}. For this work, we adopt the best redshifts reported by \citet{2023ApJ...947...20V}, using $z_{\rm spec}$ when available and $z_{\rm phot}$ otherwise.

\begin{figure}[htb!]
\centering
\includegraphics[width=0.45\textwidth, height=0.45\textwidth]{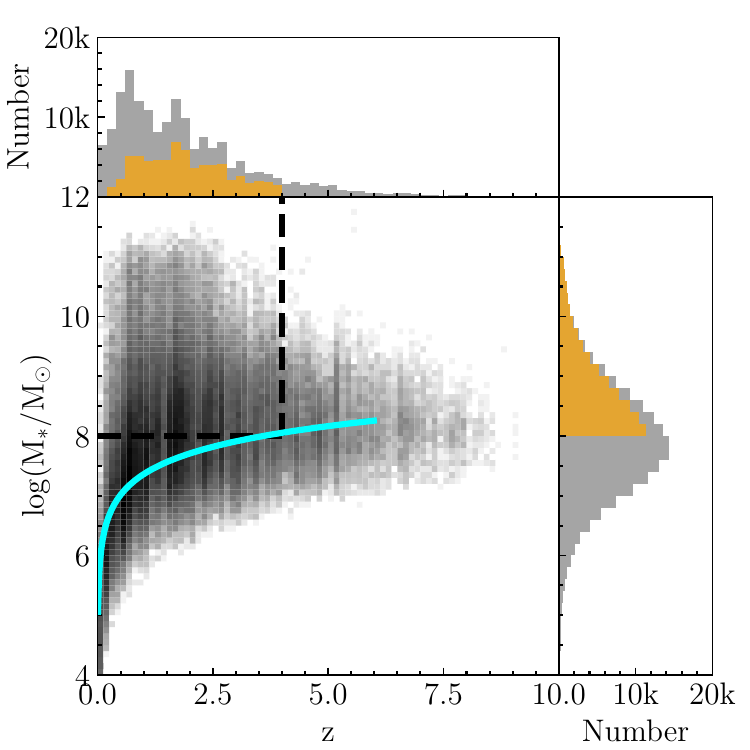}
\caption{The redshift and stellar mass distribution of our total good sample. The region enclosed by the black dashed lines indicates the selected sample used in this study, defined by $0<z<4$ and $\log(M_{\ast}/M_{\odot}) > 8$.  The cyan solid curve represents the 90\% stellar mass completeness limit corresponding to a magnitude limit of $\rm F444W_{lim}$ = 28 mag. The gray (yellow) histograms in the top and right panels show the redshift and stellar mass distributions of the good sample (selected sample), respectively.}
\label{fig1}
\end{figure}

\subsection{Sample selection} \label{sec:2.3}
Numerous studies have demonstrated that point sources occupy a well-defined sequence in size–magnitude space (e.g., \citealt{2014ApJS..214...24S, 2022ApJS..258...11W}). Utilizing the half-light radius versus magnitude diagram provided in the photometric catalog, we first exclude point sources from our parent sample. To further remove potential spurious detections and sources with unreliable redshift estimates, we apply the following selection criteria to ensure a robust galaxy sample: (1) signal-to-noise ratio (S/N) greater than 10 in the detection image ($\rm S/N_{\rm det} > 10$); (2) reduced chi-squared value satisfying $\chi^2 / N_{\rm filt} \leq 8$ with at least six filters available for photometric redshift estimation ($N_{\rm filt} \geq 6$), where $N_{\rm filt}$ is the number of filters used in the fit. As we aim to estimate the physical properties of galaxies in Section \ref{sec:3.1}, we further limit our sample to relatively bright sources by applying the following additional criteria: (3) F444W magnitude brighter than 28.5 mag; (4) S/N greater than 3 in all six JWST broad bands available across all fields (F115W, F150W, F200W, F277W, F356W, and F444W), ensuring reliable spectral energy distribution (SED) fitting. Applying these criteria yields a final sample of 206,790 galaxies.

\section{Method} \label{sec:3}
\subsection{The estimation of physical properties} \label{sec:3.1}
We derive the integrated physical properties of the galaxies in our sample using CIGALE program \citep{2019A&A...622A.103B}, based on 9-band photometry for the CEERS field and 10-band photometry for the remaining fields. The fitting setup of CIGALE used in this work closely follows that of \citet{2024ApJ...963L..49S}. We first correct the photometric catalog for Milky Way extinction using the extinction curve of \citet{1999PASP..111...63F} and the extinction map from \citet{2011ApJ...737..103S}. The SED fitting assumes a delayed-$\tau$ star formation history, \citet{2003MNRAS.344.1000B} stellar population with solar metallicity, the dust attenuation law of \citet{2000ApJ...533..682C}, and the nebular emission models of \citet{2011MNRAS.415.2920I}. Due to the lack of mid-infrared observations, we do not include a dust emission component in our modeling.

CIGALE provides multiple estimates of SFR, including the average SFR over the past 100 Myr ($\rm SFR_{\rm 100Myrs}$), the past 10 Myr ($\rm SFR_{\rm 10Myrs}$), and the instantaneous SFR ($\rm SFR_{\rm instant}$). Unless otherwise stated, we adopt $\rm SFR_{\rm 100Myrs}$ as the representative SFR in our analysis. Although, as discussed in Appendix~\ref{sec:7.2}, the absence of F090W data in the CEERS field has a negligible impact on our estimates of stellar mass and SFR. To alleviate potential concerns regarding differences in band coverage across different survey fields. we exclude the CEERS sample for further analysis. In fact, we have examined the results after incorporating the CEERS data and find that our main conclusions remain essentially unchanged.

Figure~\ref{fig1} shows the redshift and stellar mass distributions of the good sample defined in Section \ref{sec:2.3}, excluding sources from the CEERS field. The gray histograms along the top and right axes represent the redshift and stellar mass distributions, respectively, for the full good sample. Previous studies have emphasized the importance of MIRI data for accurately constraining the physical properties of galaxies at $z > 4$ (e.g., \citealt{2023ApJ...949L..18P, 2023ApJ...958...82S}). However, \citet{2025ApJ...978L..42C} had also demonstrated that robust estimates of galaxy stellar mass can still be achieved up to $z \sim 10$ using only HST and JWST/NIRCam data. In Appendix~\ref{sec:7.1}, we validate the reliability of our methodology in recovering galaxy physical properties at $z < 4$ using mock observations. To adopt a conservative approach, we limit our subsequent analysis to galaxies at $z < 4$, where the rest-frame $\sim1\mu m$ broadband imaging is available. The physical properties of galaxies at higher redshifts will be investigated in future work.

Given the varying depths of observations across different fields, we assess the mass completeness of our sample by estimating the 90\% completeness limit using the method described in \citet{2010A&A...523A..13P}. Based on a source injection analysis, \citet{2024A&A...691A.240M} showed that the detection rate reaches 100\% for objects brighter than 28 mag in the combined F356W+F444W detection images in these fields. To ensure sample completeness, we therefore adopt a conservative magnitude limit of $\rm F444W_{lim} = 28\ mag$ across all surveys included in this study. In each redshift bin of width $\Delta z = 0.25$, we identify the faintest 20\% of galaxies and use them to calculate the stellar mass completeness limit. Using the typical mass-to-light ratio for each galaxy, the stellar mass limit $M_{\rm lim}$
at a given redshift slice represents the mass a galaxy would have if its apparent magnitude equaled the magnitude limit. Specifically, the stellar mass limit at a particular redshift is derived as: $\log (M_{\rm lim}) = \log (M_{\ast}) + 0.4 (\rm F444W - F444W_{\rm lim})$. Then $M_{\rm comp}$ is defined as the upper envelope of the $M_{\rm lim}$ distribution below which 90\% of the $M_{\rm lim}$ values are located at a given redshift. We estimate the stellar mass completeness across the redshift range $0 < z < 6$, and empirically parametrize it as a function of redshift: $M_{\rm comp}(z) = 7.36 + 0.5 \ln(z)$. Restricting the analysis to $z < 4$ does not significantly alter this result. The resulting completeness limit is shown as the cyan line in Figure~\ref{fig1}. To ensure a mass-complete sample, we only include galaxies with $\log(M_{\ast}/M_{\odot}) > 8$ in our subsequent analysis, which is indicated by the black dashed line in Figure~\ref{fig1}. The orange histograms on the top and right panels of the figure show the redshift and stellar mass distributions of our selected sample with $\log(M_{\ast}/M_{\odot}) > 8$ and $z<4$, which includes 67,986 galaxies in total.

\begin{figure}[htb!]
\centering
\includegraphics[width=0.45\textwidth]{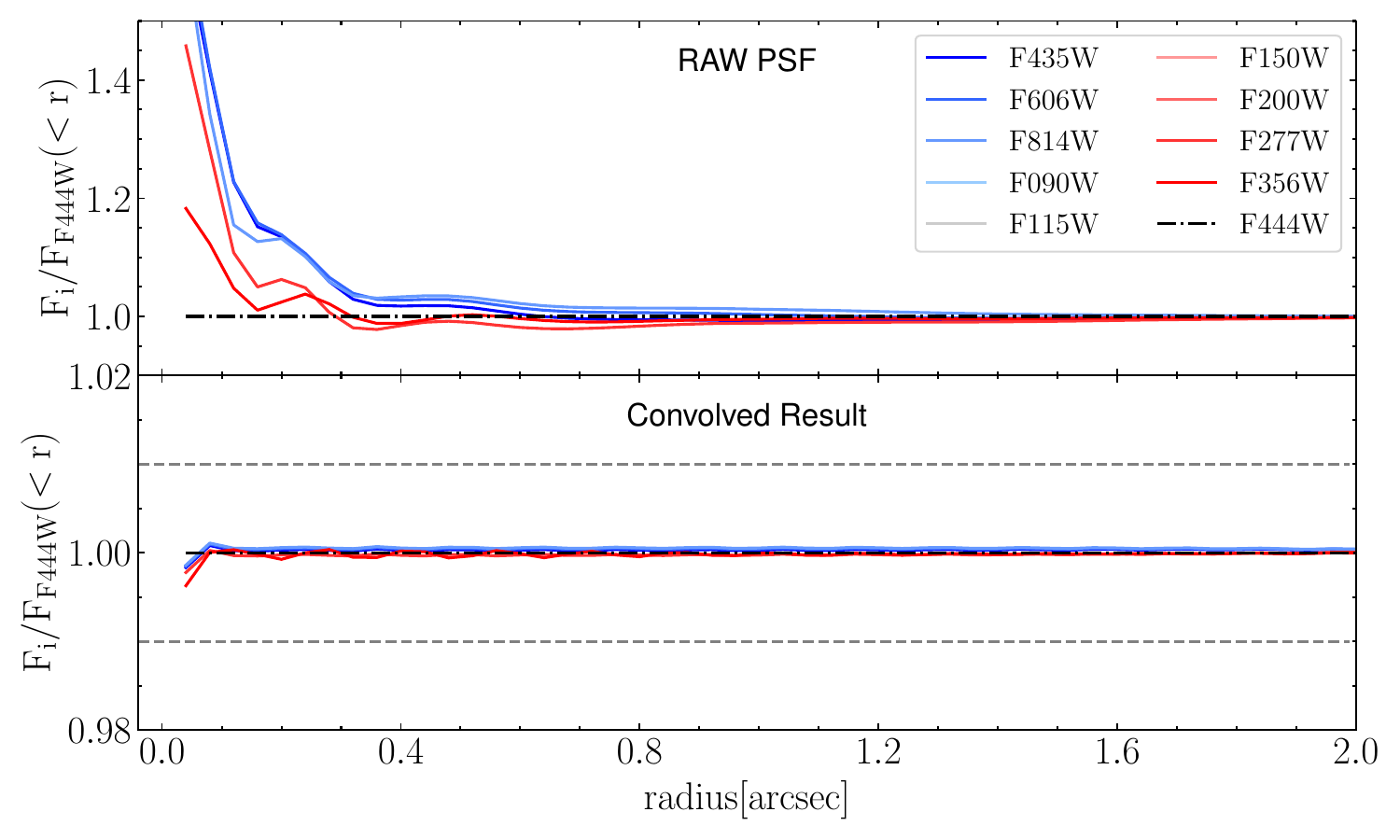}
\caption{The fraction of enclosed light as a function of radius for each filter relative to F444W in the JADES-GDS field. The upper and lower panels present the results before and after PSF matching, respectively.}
\label{fig2}
\end{figure}

\subsection{PSF match}\label{sec:3.2}
Since our goal is to perform spatially resolved SED fitting, it is crucial to homogenize the spatial resolution across all filters. To achieve this, we select 10–20 unsaturated point sources brighter than 24 mag in the F444W band in each field by identifying the point source sequence in the size–magnitude diagram. Using these sources, we construct empirical PSFs for each band with the PHOTUTILS package \citep{larry_bradley_2024_13989456}. Then, by optimizing the performance metrics defined in \citet{2011PASP..123.1218A}, the PSF matching kernels are generated to match the PSF of shorter-wavelength images in each field to that of the F444W band, which has the largest full width at half maximum (FWHM is about $0\farcs14$). All images are then convolved with the corresponding kernels to ensure uniform spatial resolution across all bands for each field.

To visually assess the effectiveness of our PSF matching procedure, Figure~\ref{fig2} shows the fraction of enclosed light as a function of radius for each filter relative to F444W band in the JADES-GDS field. The upper and lower panels show the results before and after PSF matching, respectively. As shown, after PSF matching, the encircled energy profiles of PSFs in different bands exhibit remarkable similarity, with deviations mostly within 0.01. Similar consistency is observed across all other fields, confirming the robustness and reliability of our PSF homogenization method.

\subsection{The estimation of galaxy morphology} \label{sec:3.3}
To derive the spatially resolved physical property profiles of galaxies, we first estimate their morphologies to inform the construction of measurement apertures. We fit a single S\'{e}rsic model using GALFIT \citep{2002AJ....124..266P, 2010AJ....139.2097P} to obtain key structural parameters of galaxies such as 
$R_{\rm e}$, $n$, $q$, and position angle (PA) for each available band \citep{2024ApJ...977..165J}.

Following the approach of \citet{2024ApJ...977..165J}, we estimate the morphology at rest-frame $1 \ \mu$m by linearly interpolating the S\'{e}rsic parameters from the two observed bands that are nearest in wavelength. If only one suitable band is available, we adopt its fitted parameters directly. In the following analysis, unless stated otherwise, galaxy morphology all refers to the morphology at rest-frame $1 \ \mu$m. Galaxies for which S\'{e}rsic modeling fails in both bands near rest-frame $1 \ \mu$m are excluded. In addition, \cite{2022ApJ...934L..35C} had demonstrated that single-S\'{e}rsic fitting with GALFIT can recover $R_{\rm e}$ with an accuracy better than 20\% for galaxies whose $R_{\rm e}$ exceeds one-third of the PSF FWHM, while galaxies below this threshold are considered unresolved. Therefore, we also exclude sources with $R_{\rm e} < 1$ pixel (approximately one-third of the PSF FWHM) from our analysis.
We have also measured non-parametric morphological parameters using the statmorph\_csst package \citep{2023ApJ...954..113Y}, applied to the F444W-band images. To exclude the possible merger systems, we follow the recommendations of \citet{2003AJ....126.1183C} and \citet{2014ARA&A..52..291C} and remove galaxies with asymmetry values greater than 0.35. Finally, the morphology of 46,313 galaxies are well estimated in total.

\begin{figure*}[htb!]
\centering
\includegraphics[width=0.99\textwidth, height=0.29\textwidth]{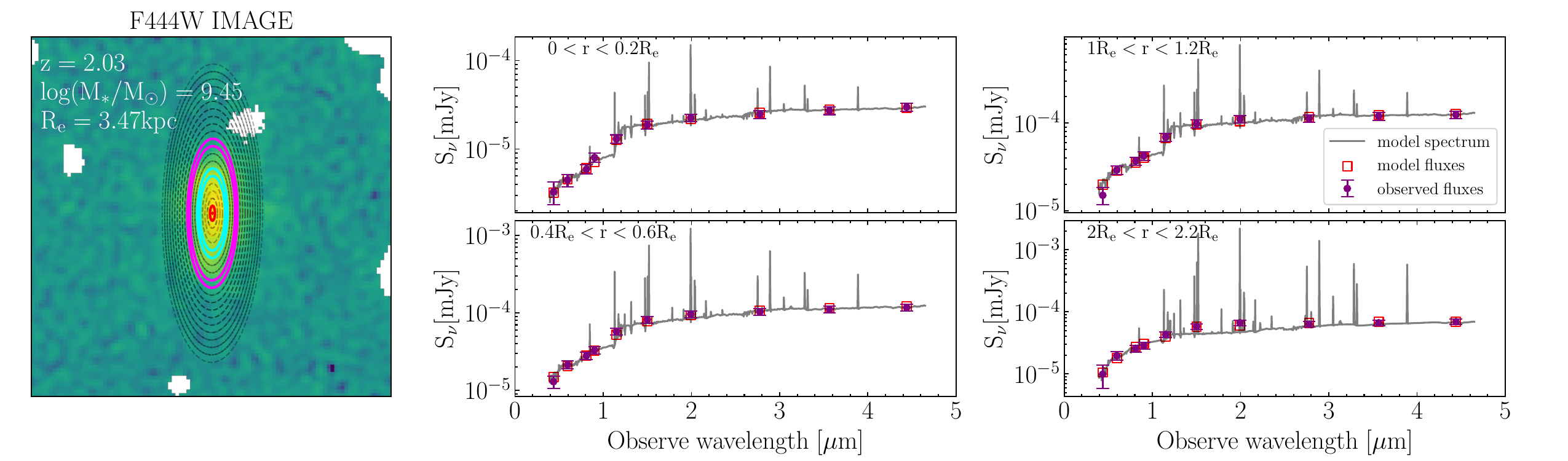}
\caption{ Example of our spatially resolved SED fitting for a randomly selected galaxy. The left panel shows the F444W-band image, with black dashed lines marking elliptical annuli spaced at intervals of $0.2 R_{\rm e}$. Four representative radial regions are highlighted in red, orange, cyan, and magenta, corresponding to $0<r<0.2R_{\rm e}$, $0.4R_{\rm e}<r<0.6R_{\rm e}$, $1R_{\rm e}<r<1.2R_{\rm e}$, and $2 R_{\rm e}<r<2.2R_{\rm e}$, respectively. The middle and right columns display the SED fitting results for these four radial bins.}
\label{fig3}
\end{figure*}

\subsection{The estimation of galaxy physical property profiles} \label{sec:3.4}
Figure~\ref{fig3} illustrates the procedure for deriving spatially resolved physical property profiles of our galaxy sample. For each galaxy, we extract a $300 \times 300$ pixel cutout from the PSF-matched images and mask all other sources within the region. Using the galaxy center and the morphological parameters $R_{\rm e}$, $q$, and $\rm PA$ derived at rest-frame 1 $\mu$m from GALFIT, we construct a series of concentric elliptical annuli with a radial step size of $0.2 R_{\rm e}$, extending out to $5 R_{\rm e}$. Fluxes and associated uncertainties in each band are measured within these annuli using the PHOTUTILS package. Although this radial binning strategy may mix young and old stellar populations, our previous work demonstrated that in SED fitting, as long as observations at rest-frame wavelengths longer than 1$\mu m$ are included, the impact of the outshining effect is substantially reduced \citep{2023ApJ...958...82S}. Furthermore, to ensure consistency with our scientific objective of characterizing radial profiles of galaxy physical properties, we do not adopt other more granular binning strategies.

We then perform spatially resolved SED fitting in each elliptical annulus using CIGALE, adopting the same configurations described in Section~\ref{sec:3.1}. To ensure the reliability of the derived physical properties, we retain only those annuli where the S/N exceeds 3 in all six JWST broad bands, consistent with the selection criteria outlined in Section~\ref{sec:2.3}.

Although some studies have shown that profiles derived from PSF-matched images can be affected by PSF smearing (e.g., \citealt{2013ApJ...763...73S, 2019ApJ...877..103S, 2020ApJ...905..170M, 2022ApJ...937L..33S, 2023ApJ...945..155M}) and have suggested the deconvolved images are considered more appropriate, the reliability of such deconvolution techniques has not been thoroughly assessed. Therefore, we still adopt the PSF-matched images in this work. In Appendix~\ref{sec:7.3}, we also present results obtained from deconvolved images and find that they are nearly identical to those derived from the PSF-matched data.

The left panel of Figure~\ref{fig3} shows an example galaxy in the F444W band, with black dashed lines denoting elliptical annuli spaced at intervals of $0.2 R_{\rm e}$. The red, orange, cyan, and magenta annuli highlight four representative radial regions: $0<r<0.2R_{\rm e}$, $0.4R_{\rm e}<r<0.6R_{\rm e}$, $1R_{\rm e}<r<1.2R_{\rm e}$, and $2 R_{\rm e}<r<2.2R_{\rm e}$, respectively. The middle and right panels present the corresponding SED fitting results for these four annuli. As illustrated, the SED fitting achieves high-quality results in annuli with sufficient S/N.

\section{Result}\label{sec:4}
In the previous section, we have applied a self-consistent method to measure the physical properties, morphologies, and spatially resolved profiles of galaxies observed with JWST in the CANDELS fields. To prepare for the subsequent studies in this series, we use the measurements obtained in this work to investigate the SFMS, the size–mass relation, and the evolution of galaxy physical property profiles.

\begin{figure*}[htb!]
\centering
\includegraphics[width=1.0\textwidth, height=0.48\textwidth]{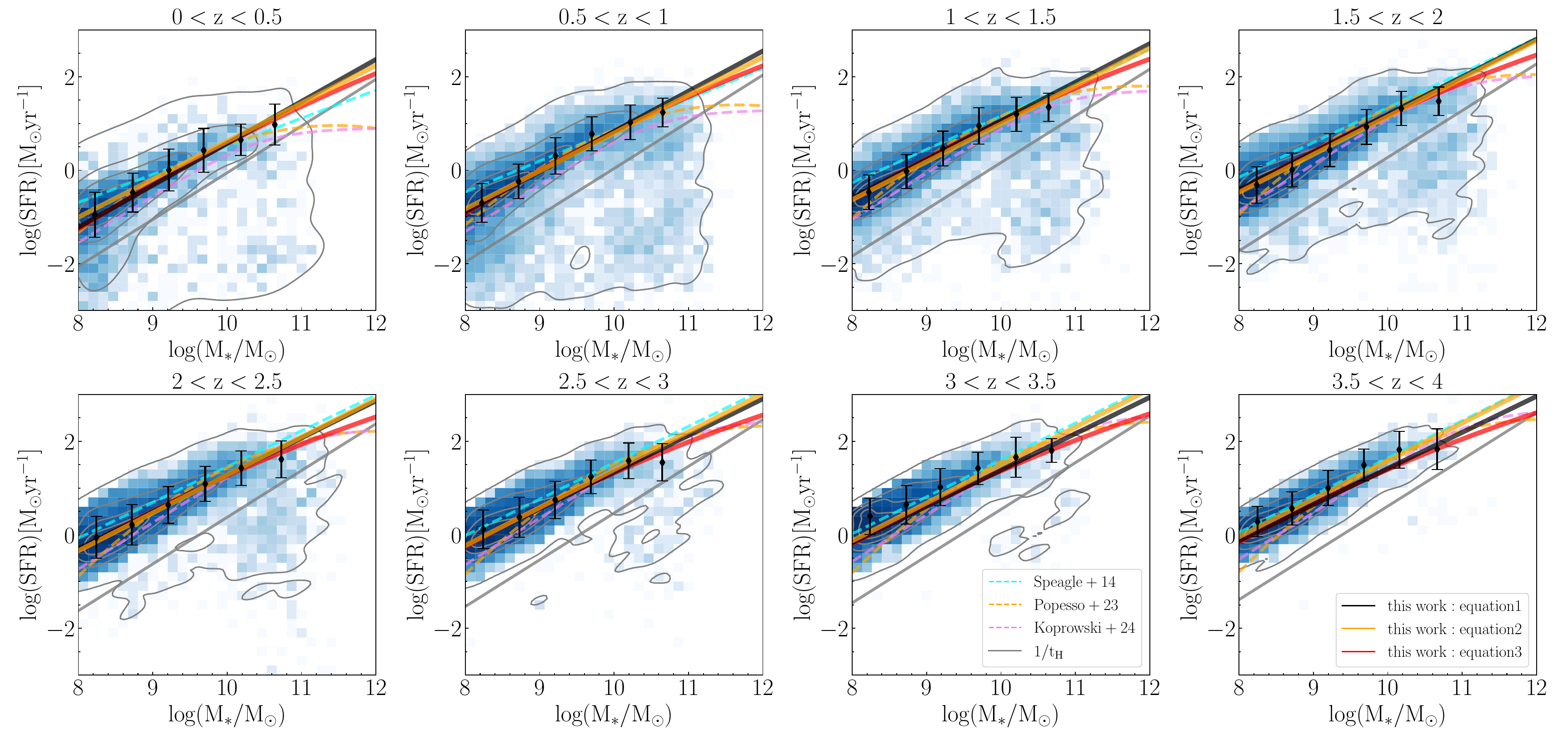}
\caption{Distribution of galaxy SFR as a function of stellar mass across different redshift bins, with each panel corresponding to different redshift interval. The gray contours enclose 25\%, 50\%, 75\%, and 99\% of the galaxy population, respectively. The gray solid line in each panel indicates the sSFR threshold of $1/t_{\rm H}$, where $t_{\rm H}$ is the Hubble time at the median redshift of the bin; galaxies above this threshold are classified as SFGs. Black points with error bars represent the median SFR and the corresponding scatter for SFGs in different stellar mass bins. The black solid line shows the best-fit SFMS assuming the  formulation in Equation (\ref{eq:1}), while the orange and red solid line corresponds to the best-fit results  derived from Equations (\ref{eq:2}) and (\ref{eq:3}), respectively. For comparison, we include results from previous studies: the cyan dashed line denotes the SFMS from \cite{2014ApJS..214...15S}, while the orange and violet dashed lines represent results from \cite{2023MNRAS.519.1526P} and \cite{2024A&A...691A.164K}, respectively. Overall, our derived SFMS is consistent with these previous studies, demonstrating the reliability of our measurements.}
\label{fig4}
\end{figure*} 

\begin{figure*}[htb!]
\centering
\includegraphics[width=1.0\textwidth, height=0.26\textwidth]{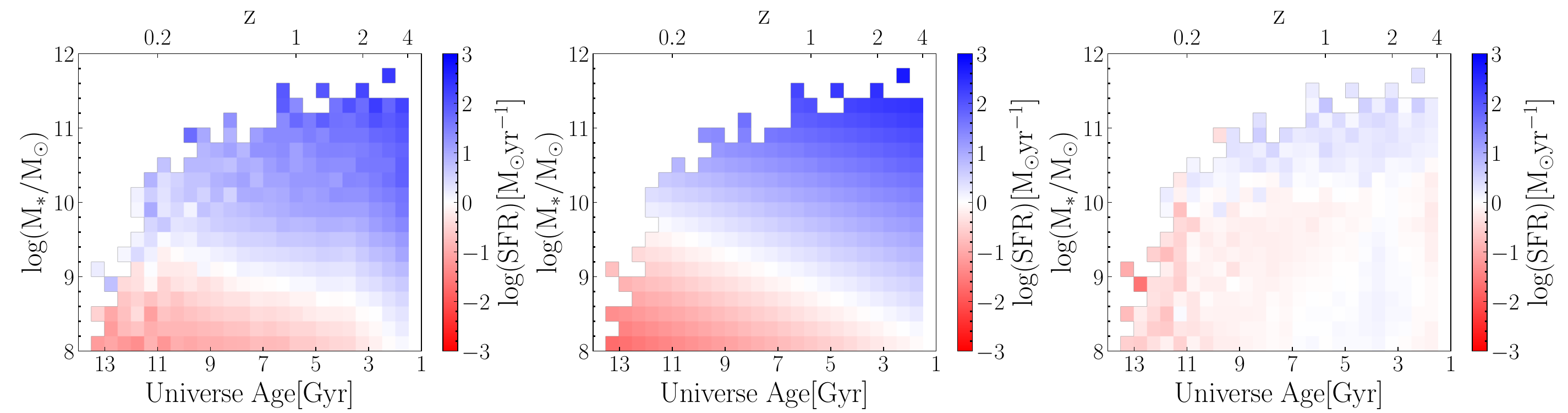}
\caption{Left panel: The distribution of SFR for SFGs mapped onto the stellar mass versus universe age plane, where the color scale indicates the median SFR within each square bin. Middle panel: The best-fit SFMS derived using a linear relation. Right panel: The residuals between the observed SFRs and the best-fit linear model. The small residuals demonstrate that a linear relation provides a good representation of the SFMS for our sample across the explored stellar mass and cosmic time ranges.}
\label{fig5}
\end{figure*}

\subsection{Star formation main sequence}\label{sec:4.1}

Over the past decade, the SFMS, including its slope, normalization, scatter, and evolution over cosmic time, has been extensively studied (e.g., \citealt{2014ApJS..214...15S, 2022ApJ...936..165L, 2023MNRAS.519.1526P}). These investigations have extended the characterization of the SFMS up to the epoch of cosmic reionization (e.g., \citealt{2024ApJ...977..133C, 2025ApJ...979..193C}). Overall, the SFMS is found to follow a tight, nearly linear relation between stellar mass and SFR. However, several studies have reported a deviation from linearity at the high-mass end (e.g., $M_{\ast} > 10^{11}M_{\odot}$), commonly referred to as the ``bending'' of the SFMS (e.g., \citealt{2015A&A...575A..74S, 2022A&A...661L...7D, 2023MNRAS.519.1526P}). This trend has also been confirmed recently by early observations from Euclid \citep{2025arXiv250315314E}.

In this study, we examine the SFMS of galaxies at $z < 4$ using the mass-complete sample defined in Section~\ref{sec:3.1}. The results are shown in Figure~\ref{fig4} with each panel presenting the results for different redshift bins. In each redshift bin, the blue shaded regions represent the distribution of galaxies in the stellar mass–SFR plane, while the gray contours enclose 25\%, 50\%, 75\%, and 99\% of the galaxy population, respectively. To characterize the SFMS, it is essential to separate 
SFGs from QGs. \cite{2024ApJ...961..163B} demonstrated, through both simulations and observations, that a threshold of $\mathrm{sSFR} < 1/t_{\rm H}$ (where $t_{\rm H}$ is the Hubble time at a given redshift) reliably identifies quenched galaxies. Following this criterion, we classify galaxies with $\mathrm{sSFR} > 1/t_{\rm H}$ as star-forming and include only these galaxies in our SFMS analysis. The adopted quenching threshold is shown as a gray solid line in Figure~\ref{fig4}.

Based on a compilation of data from 25 studies, \cite{2014ApJS..214...15S} explored the SFMS up to $z \sim 6$ and found that for galaxies with stellar masses above $10^{10} M_{\odot}$, the SFMS can be well represented by a linear relation. In this study, we adopt the same functional form to describe the SFMS of our galaxy sample:

\begin{equation}
    \log({\rm SFR})=(\alpha_1 + \beta_1 t) \log(M_\ast) - (\alpha_2+\beta_2 \times t)
\label{eq:1}
\end{equation}

where $t$ represents the age of the Universe at galaxy's redshift. To model the SFMS in a way that accounts for both stellar mass and Universe age, we perform a two-dimensional fit using the Markov Chain Monte Carlo (MCMC) method, implemented via the emcee package \citep{2013PASP..125..306F}. To minimize the impact of outliers on the fitting process, we bin the galaxies by both stellar mass and Universe age, as illustrated in Figure \ref{fig5}. The median SFR is calculated within each bin, which is then used for the fitting. In computing the likelihood function, we include the uncertainties in the median SFR for each bin, expressed as $\sigma({\rm SFR})/\sqrt{N}$, where $\sigma({\rm SFR})$ is the scatter in SFR within the bin and $N$ is the number of galaxies in the bin. The best-fitting parameters are: $\alpha_1 = 0.745 \pm 0.005$, $\beta_1 = 0.0167 \pm 0.001$, $\alpha_2 = 5.845 \pm 0.042$, and $\beta_2 = 0.278 \pm 0.009$.

In Figure \ref{fig5}, the left panel shows the map of galaxy SFRs in the stellar mass–Universe age plane. The middle panel displays the best-fit SFMS surface obtained from our two-dimensional fitting. The right panel illustrates the residuals between the observed median SFRs and the model predictions. As shown in the figure, this linear form of the SFMS provides a good description of our sample across the full range of stellar mass and cosmic time. The corresponding results are also shown in Figure \ref{fig4}: the black solid lines represent our fitting results, while the cyan dashed lines indicate relations reported by \cite{2014ApJS..214...15S}. Additionally, the black points with error bars represent the median SFR and the corresponding scatter for SFGs in different stellar mass bins. Overall, our results show good agreement with those of \cite{2014ApJS..214...15S}, except in the redshift range of $0<z<0.5$. This discrepancy may be attributed to the different stellar mass ranges considered in the analyses, as well as the relatively small sample size in this redshift bin.

Moreover, several studies have reported that within the redshift range $0<z<3$, the sSFR scales approximately as $(1+z)^{2.5 \sim 3.5}$ (e.g., \citealt{2010MNRAS.405.2279O, 2015A&A...579A...2I, 2015A&A...581A..54T}). Motivated by this trend, some studies have also described the SFMS using the following functional form (e.g., \citealt{2018A&A...619A..27B}):
\begin{equation}
\rm log(SFR) = \alpha log(M_{\ast}/M_{\odot}) + \beta log(1 + z) + \gamma
\label{eq:2}
\end{equation}
We also apply this functional form to our sample and performed a two-dimensional fitting using the \texttt{emcee} package. The results show that this parametrization also provides a good description of the SFMS in our dataset. The best-fit parameters are: $\alpha = 0.812\pm 0.002$, $\beta= 1.811\pm 0.009$, and $\gamma=-7.762\pm 0.018$. The corresponding result is shown as the orange solid line in Figure \ref{fig5}. As illustrated in the figure, this parameterization yields a remarkably good agreement with the results derived using Equation (\ref{eq:1}).

Many previous studies have demonstrated that the SFMS deviates from a linear relation at the high-mass end (e.g., \citealt{2015A&A...575A..74S, 2022A&A...661L...7D, 2023MNRAS.519.1526P, 2024A&A...691A.164K}). By compiling data from several studies, \cite{2023MNRAS.519.1526P} investigated the SFMS for galaxies within the redshift range $0<z<6$ and stellar masses between $10^{8.5}M_{\odot}$ and $10^{10.5} M_{\odot}$. They found that SFR can be well described by a polynomial function of $\log(M_{\ast})$. Although the bending of the SFMS at the high-mass end is not prominent in our study due to the limited number of massive galaxies in our sample, we still performed a fit to our data using a polynomial function of $\log(M_{\ast})$ to capture any potential non-linearity. In this work, we adopt the same model as used in \cite{2023MNRAS.519.1526P}: 
\begin{equation}
  \log({\rm SFR}) = (a_1 t + b_1)\log(M_{\ast}) + b_2 \log^2(M_{\ast}) + (b_0 + a_0 t)
\label{eq:3}
\end{equation}
where $t$ also represents the Universe age. Following the same procedure as described earlier, the best-fitting parameters are: $a_0 = -0.295\pm 0.009$, $a_1=0.019\pm 0.001$, $b_0=-9.298\pm 0.197$, $b_1=1.514\pm 0.143$, and $b_2=-0.043\pm0.002$. The corresponding best-fit result is shown in Figure \ref{fig4} as a red solid line. For comparison, we also represent the results from \cite{2023MNRAS.519.1526P} and \cite{2024A&A...691A.164K}, shown as orange and violet dashed lines, respectively. 

Compared to the linear relation, this polynomial relation exhibits a slight bending at the high-mass end, which can also be seen directly from the sample distribution. When compared with the results from \cite{2023MNRAS.519.1526P}, our results show good consistency with theirs in the intermediate-mass range ($9 < \log(M_{\ast}/M_{\odot}) < 10$). When considering the high-mass end, our result also show good consistency with theirs at $z > 2$. However, at lower redshifts, \cite{2023MNRAS.519.1526P} revealed a more pronounced bending phenomenon. Furthermore, at the low-mass end, our results also show some deviations when compared to those of \cite{2023MNRAS.519.1526P}. Considering that our sample lacks massive galaxies, while previous studies are deficient in low-mass galaxies, these differences can be reasonably understood. 

\begin{figure*}[htb!]
\centering
\includegraphics[width=1.0\textwidth, height=0.48\textwidth]{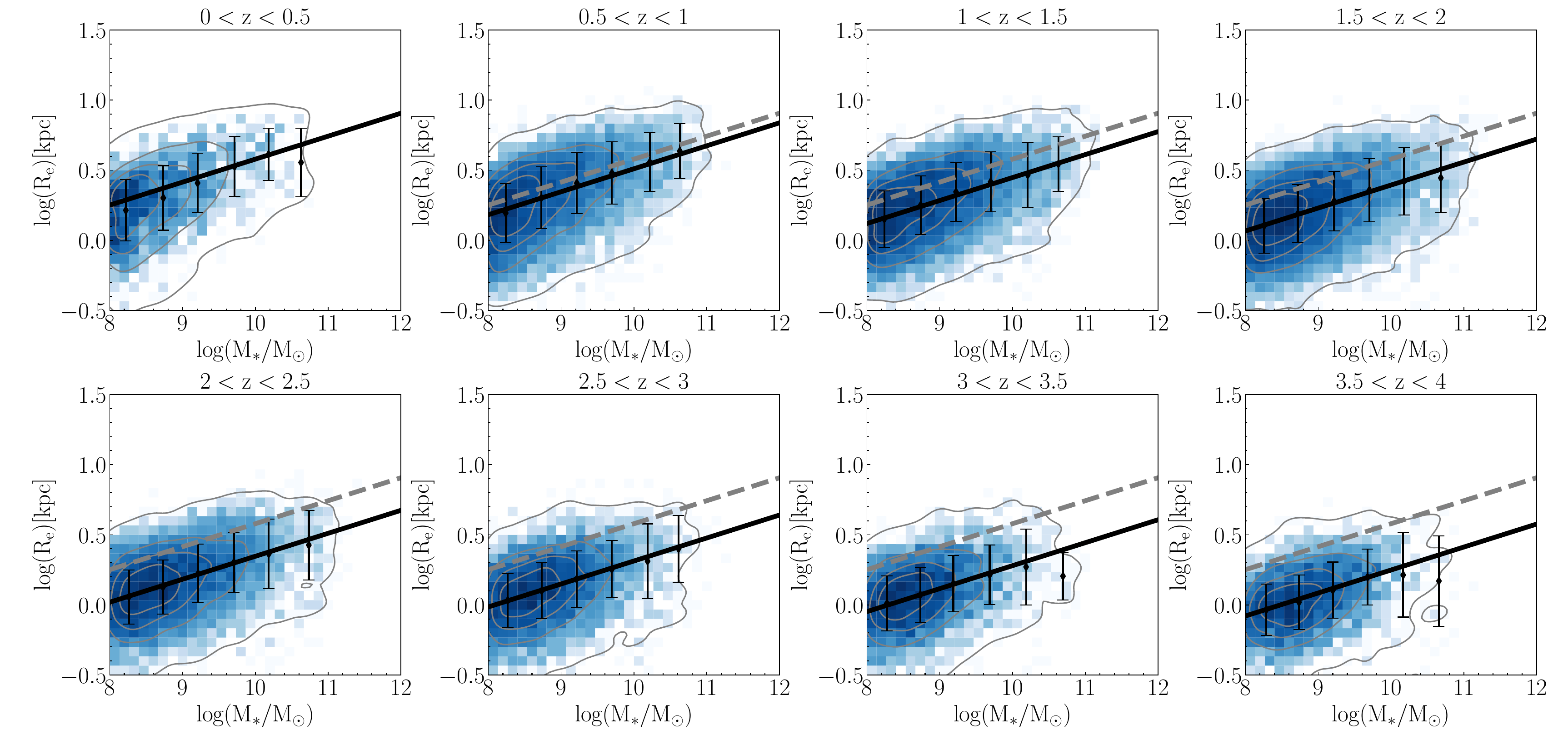}
\caption{Distribution of galaxy sizes as a function of stellar mass for SFGs across different redshift bins. In each panel, the gray contours enclose 25\%, 50\%, 75\%, and 99\% of the sample, respectively. The black points with error bars denote the median effective radius and the corresponding standard deviation within stellar mass bins. The black solid line represents the best-fit size–mass relation derived from our analysis. For comparison, we show the results for the $0 < z < 0.5$ bin as gray dashed lines in the remaining panels.}
\label{fig6}
\end{figure*} 

\begin{figure*}[htb!]
\centering
\includegraphics[width=1.0\textwidth, height=0.26\textwidth]{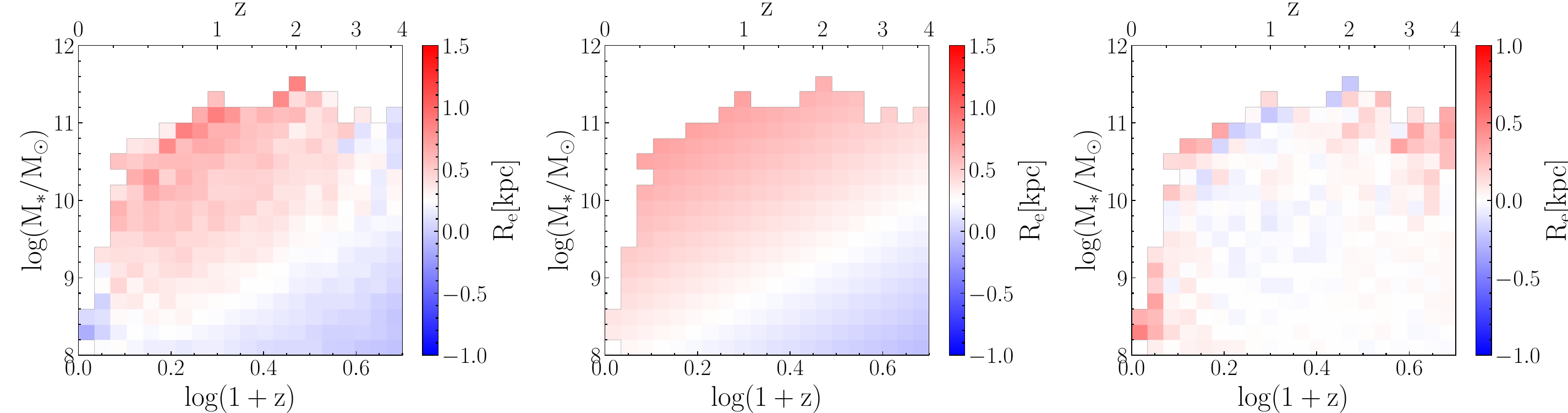}
\caption{Left panel: The distribution of size for SFGs mapped onto the stellar mass versus redshift plane, where the color scale indicates the median size within each square bin. Middle panel: The best-fit models of the size mass relation. Right panel: The residuals between the observed size mass relation and the best-fit model.}
\label{fig7}
\end{figure*} 

\begin{figure}[htb!]
\centering
\includegraphics[width=0.45\textwidth]{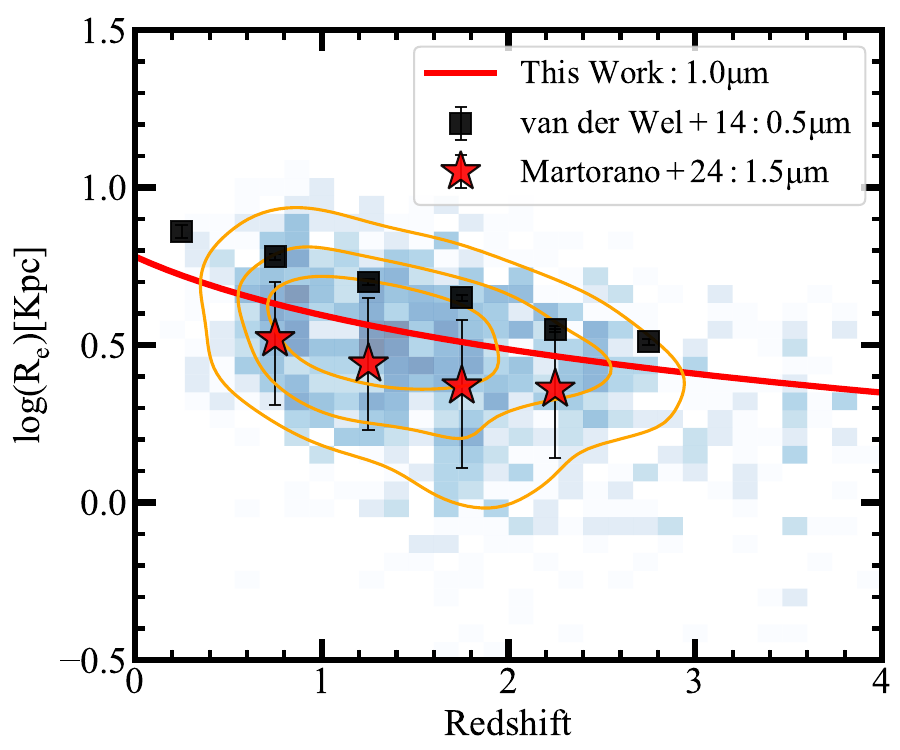}
\caption{The blue two-dimensional histogram shows the distribution of our galaxies with stellar mass $\sim 5\times 10^{10}M_{\odot}$, while the orange contours enclose 25\%, 50\%, and 75\% of the sample. The red solid line indicates the best-fitting relation derived in Section~\ref{sec:4.2}. To facilitate comparison with previous studies, the results from van der \cite{2014ApJ...788...28V}, derived from rest-frame 0.5 $\mu$m sizes, are shown as black squares, while those from \cite{2024ApJ...972..134M}, derived from rest-frame 1.5 $\mu$m sizes, are shown as red stars.}
\label{fig8}
\end{figure}

\subsection{Size-mass relation}\label{sec:4.2}
Understanding the size growth of galaxies is essential for uncovering their evolutionary pathways, as galaxy size evolution reflects the underlying processes of mass assembly. In this work, we revisit this topic using the latest JWST observations. While many previous investigations have primarily focused on rest-frame optical sizes, it is now well studied that rest-frame optical morphologies are influenced by spatial variations in stellar populations, dust attenuation, and metallicity gradients. In contrast, rest-frame near-infrared morphologies provide a more direct probe of the underlying stellar mass distribution (e.g., \citealt{2019ApJ...877..103S, 2020ApJ...905..170M, 2022ApJ...937L..33S, 2023ApJ...945..155M, 2024ApJ...960...53V, 2024ApJ...977..165J}). Therefore, in this study, we focus on the morphology of galaxies at rest-frame 1$\mu$m to better trace their stellar mass structure.

In Figure \ref{fig6}, we present the size–mass relations of our SFGs sample across various redshift intervals. The blue shaded regions represent the overall distribution of galaxy sizes, while the gray contours enclose 25\%, 50\%, 75\%, and 99\% of the sample, respectively. The black points denote the median $R_e$ in different stellar mass bins, with error bars indicating the standard deviation within each bin. Following the approach of \cite{2024ApJ...977..165J}, we model the dependence of galaxy size on both stellar mass and redshift using the following functional form:
\begin{equation}
    \rm log(R_e/kpc) = \alpha log(M_{\ast}/M_{\odot}) + \beta log(1 + z) + k
\end{equation}
where $\alpha$, $\beta$, and $k$ are free parameters. We also perform a two-dimensional fitting using the MCMC method, adopting the same methodology as described in Section \ref{sec:4.1}. Similar to Figure \ref{fig5}, Figure \ref{fig7} presents the mapping of the median galaxy size across the stellar mass versus redshift plane. Given that at $z>3.5$, only the F444W band can be used to approximate galaxy morphology at rest-frame 1$\mu$m, we restrict our analysis of the size–mass relation to galaxies at $z<3.5$, where rest-frame 1$\mu$m morphologies are more reliably probed. The best-fit parameters are $\alpha = 0.164 \pm 0.002$, $\beta = -0.618 \pm 0.012$, and $k = -0.974 \pm 0.017$. As shown in Figure \ref{fig7}, The small residuals indicates that our methodology provides a robust characterization of how galaxy size varies with stellar mass and redshift.

The size–mass relation of galaxies has also been extensively investigated in previous studies. Using data from the CANDELS fields, \cite{2014ApJ...788...28V} reported a relation of $r_e \propto M_{\ast}^{0.2}$ for SFGs with stellar masses above $10^{9} M_{\odot}$ in the redshift range $0 < z < 3$, which is slightly steeper than the slope found in this work. Similar slightly steeper trends have also been observed in other studies (e.g., \citealt{2019ApJ...880...57M, 2021MNRAS.506..928N}). However, more recently, based on JWST observations in the CEERS field, \cite{2024ApJ...962..176W} reported $\alpha$ values ranging from 0.15 to 0.19 over $0<z<4$, which is in good agreement with our results. Although these studies typically rely on rest-frame optical morphologies—which are known to be systematically larger than those measured at rest-frame 1$\mu$m (e.g., \citealt{2022ApJ...937L..33S, 2024ApJ...960...53V})—they nonetheless lend support to the robustness of our measurements. In our previous work, using rest-frame 1$\mu$m morphologies from the JADES field, \cite{2024ApJ...977..165J} obtained a slightly higher value of $\alpha = 0.19$. This discrepancy may arise from differences in the stellar mass estimation methods.

The size evolution of SFGs offers key insights into the processes governing their growth and assembly. In previous studies, the evolution of galaxy sizes has been parameterized as $r_e \propto (1+z)^{\beta}$, with reported values of $\beta$ typically ranging from -1.3 to -0.5 (e.g., \citealt{2014ApJ...788...28V, 2015ApJS..219...15S, 2023ApJ...950..130S, 2024ApJ...962..176W, 2024MNRAS.527.6110O, 2024MNRAS.533.3724V, 2025arXiv250407185Y}). For instance, using HST data from the CANDELS fields, \cite{2014ApJ...788...28V} found $\beta \approx -0.75$ for late-type galaxies. More recently, JWST-based analyses of CEERS data yielded $\beta = -0.63$ by \cite{2024ApJ...962..176W}, and $\beta = -0.71$ by \cite{2024MNRAS.527.6110O}. Our own best-fit value of $\beta = -0.618 \pm 0.012$ is in good agreement with these findings, which may suggests that galaxy morphologies in the near-infrared may exhibit evolutionary trends similar to those observed in the optical bands. To more clearly illustrate the redshift evolution of galaxy size derived from our sample, we take galaxies with stellar mass $\sim  5 \times 10^{10}M_{\odot}$ as a representative example to illustrate the evolutionary trend of our sample, which is shown in Figure \ref{fig8}. The blue two-dimensional histogram shows the distribution of our galaxies, while the orange contours enclose 25\%, 50\%, and 75\% of the sample. The red solid line indicates the best-fitting relation. To facilitate comparison with previous studies, we also show the corresponding results from \cite{2014ApJ...788...28V}, based on galaxy sizes measured at rest-frame 0.5 $\mu$m, and from \cite{2024ApJ...972..134M}, derived from rest-frame 1.5 $\mu$m size measurements. It can be seen that they exhibit similar evolutionary trends, albeit with some systematic offsets. However, this is reasonable given that numerous studies have shown that galaxy sizes measured at longer wavelengths are systematically smaller than those derived at shorter wavelengths (e.g., \citealt{2024ApJ...977..165J, 2026ApJ...999L...6M}). A more comprehensive exploration of the differences between rest-frame optical and near-infrared morphologies—including their redshift and stellar mass dependencies—will be presented in a forthcoming study.

\begin{figure*}[htb!]
\centering
\includegraphics[width=1.0\textwidth]{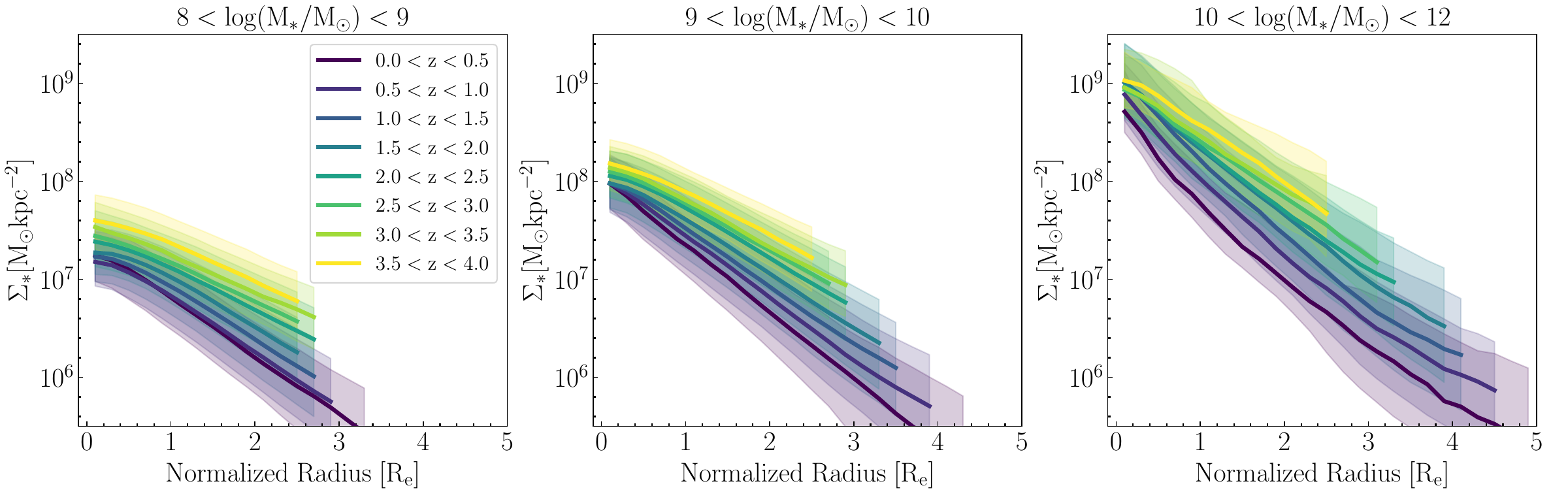}\\[1em]
\includegraphics[width=1.0\textwidth]{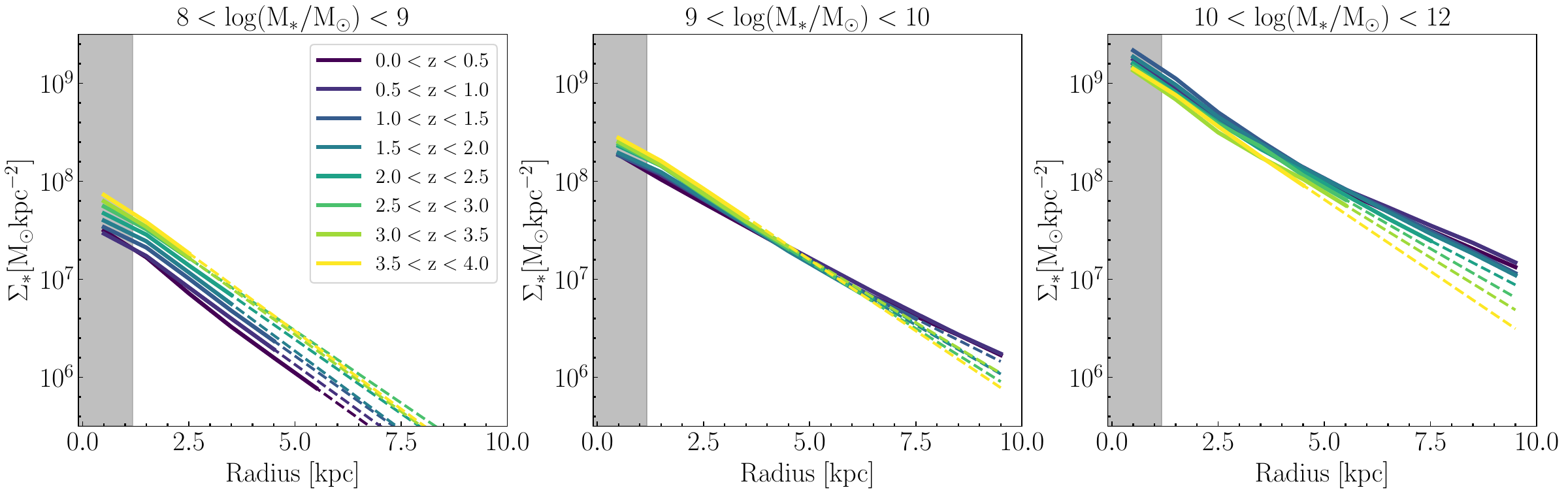}
\caption{Radial profiles of stellar mass surface density for SFGs, categorized into three stellar mass intervals and eight redshift intervals. For each bin, we present the median $\Sigma_{\ast}$ as a function of radius, normalized by the $R_{\rm e}$ (top row) and in physical units of kpc (bottom row). To ensure the reliability of the profiles, we limit the radial range to regions where more than 50\% of galaxies in the bin satisfy the adopted signal-to-noise criterion. The shaded regions in the top panels denote the 25th–75th percentile distribution of $\Sigma_{\ast}$ at each radius. Due to the close of the profiles in physical units across different redshifts, we omit the corresponding distribution in the bottom panels for clarity. We also show the extent of the PSF FWHM using a shaded gray region in the bottom panel. For results in physical units, we also performed a linear extrapolation of the outer profiles, which is shown as a dashed line, to infer the stellar mass surface density in the outskirts of galaxies.}
\label{fig9}
\end{figure*} 

\subsection{Stellar mass profile} \label{sec:4.3}
To investigate the dependence of galaxy physical property profiles on redshift and stellar mass, we divide our sample into three stellar mass bins ($10^8 \sim 10^9\ M_{\odot}$, $10^9\sim 10^{10}\ M_{\odot}$, and $10^{10}\sim 10^{12}\ M_{\odot}$) and eight redshift bins (with an interval of $\Delta z = 0.5$). For each galaxy, we adopt the same S/N criterion described in Section \ref{sec:2.3} to ensure the reliability of measurements: only annuli with S/N exceeding 3 in all six JWST bands are retained. We define the quantity max\_high\_snr\_radius as the largest radius within which this criterion is satisfied. For each redshift–stellar mass bin, we derive the median physical property profile by computing the median surface density of the relevant quantity across all galaxies at each radius. If the radius under consideration exceeds a galaxy’s max\_high\_snr\_radius, we linearly extrapolate that galaxy’s profile to estimate its physical property at that radius. To maintain robustness, we restrict our analysis to the radial range where more than 50\% of galaxies have a max\_high\_snr\_radius exceeding this limit.

Figure \ref{fig9} presents the $\Sigma_{\ast}$ profiles of our sample. In each panel, different colors represent results from different redshift bins. The first and second rows display the profiles as a function of radius normalized by $R_{\rm e}$ and in physical units (kpc), respectively. The shaded regions in the first row represent the 25th–75th percentile distribution of $\Sigma_{\ast}$ at different radius. In addition, we have also estimated the standard error of $\Sigma_{\ast}$ at each radius to quantify the uncertainties in our radial profiles, which are typically smaller than 0.05 dex. Since the $\Sigma_{\ast}$ profiles in kpc units are close across different redshifts, we have opted not to include the corresponding shaded regions for better visualization. Since many previous work has demonstrated that radial profiles within radii smaller than the PSF FWHM are strongly impacted by PSF smearing (e.g., \citealt{2018ApJ...860...60L, 2024ApJ...975..252H}), we also show the extent of the PSF FWHM using a shaded gray region in the bottom panel. For results in physical units, we also performed a linear extrapolation of the outer profiles, which is shown as a dashed line, to infer the stellar mass surface density in the outskirts of galaxies. 

As shown in Figure \ref{fig9}, the $\Sigma_{\ast}$ profiles exhibit clear negative gradients across all redshift and stellar mass bins, with $\Sigma_{\ast}$ decreasing as a function of radius. This results has also been reported in many previous studies (e.g., \citealt{2016ApJ...828...27N, 2018MNRAS.479.5083A, 2020ApJ...905..170M, 2021MNRAS.508..219N, 2023ApJ...945..117A, 2023ApJ...945..155M, 2024ApJ...960...53V, 2024ApJ...975..252H, 2025ApJ...980..168D}). At fixed redshift, more massive systems consistently exhibit higher $\Sigma_{\ast}$, even in their central regions. This trend is consistent with many previous findings (e.g., \citealt{2023ApJ...945..117A, 2024ApJ...975..252H, 2025ApJ...980..168D}). Such a pattern implies that, during the course of star formation, galaxies progressively build up their central stellar mass. These results provide important constraints on galaxy growth models, such as that proposed by \cite{2016ApJ...828...27N}, which suggests that galaxies form compact cores at early epochs and subsequently grow their outer regions over time. However, it is important to emphasize that our current analysis focuses solely on the variation of $\Sigma_{\ast}$ as a function of redshift and stellar mass, rather than directly tracing individual evolutionary pathways. In future work, we aim to construct more carefully selected samples that may allow for the identification of progenitor–descendant connections, thereby enabling a more direct investigation of galaxy growth mechanisms.

When considering galaxies at fixed stellar mass, galaxies at higher redshifts tend to have higher $\Sigma_{\ast}$ values, which is consistent with the expectation that galaxies at earlier cosmic times are generally more compact. Similar trends have been reported in the literature (e.g., \citealt{2017ApJ...840...47B, 2017ApJ...834...81J}). Using HST observations, \cite{2017ApJ...840...47B} found that galaxies with $M_{\ast} \sim 10^9 M_{\odot}$ show a decline of approximately 0.3 dex in central stellar mass surface density within 1 kpc from $z \sim 3$ to $z \sim 0.5$, in agreement with our results. However, the underlying physical mechanisms responsible for the increased compactness of high-redshift galaxies remain uncertain.

\begin{figure*}[htb!]
\centering
\includegraphics[width=1.0\textwidth]{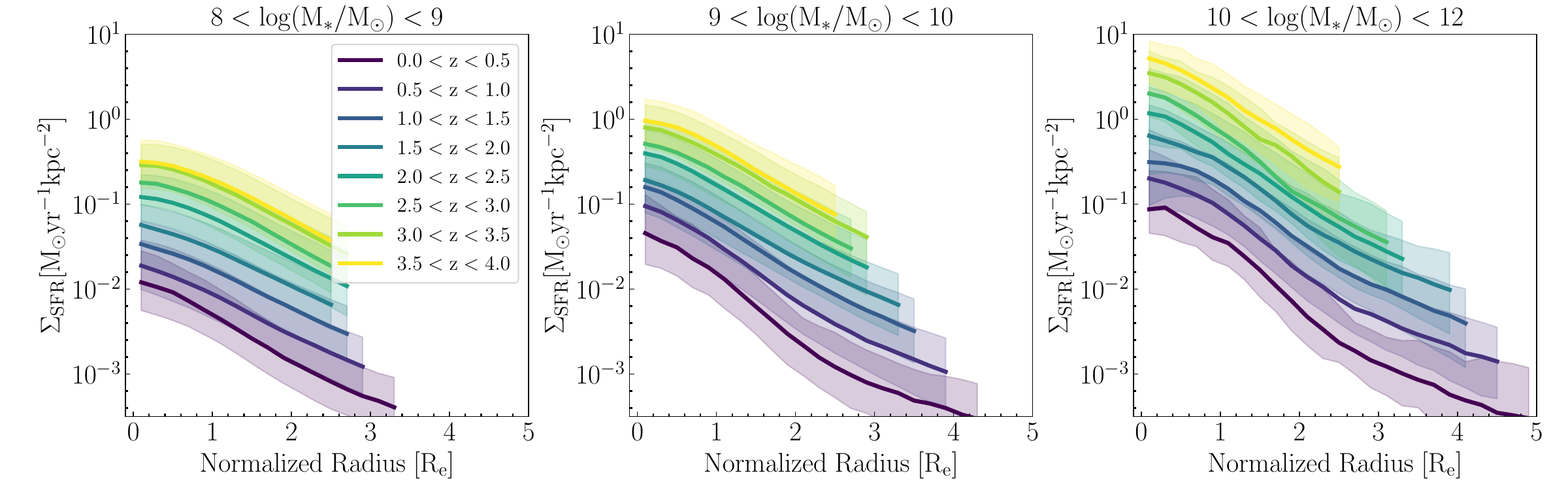}\\[1em]
\includegraphics[width=1.0\textwidth]{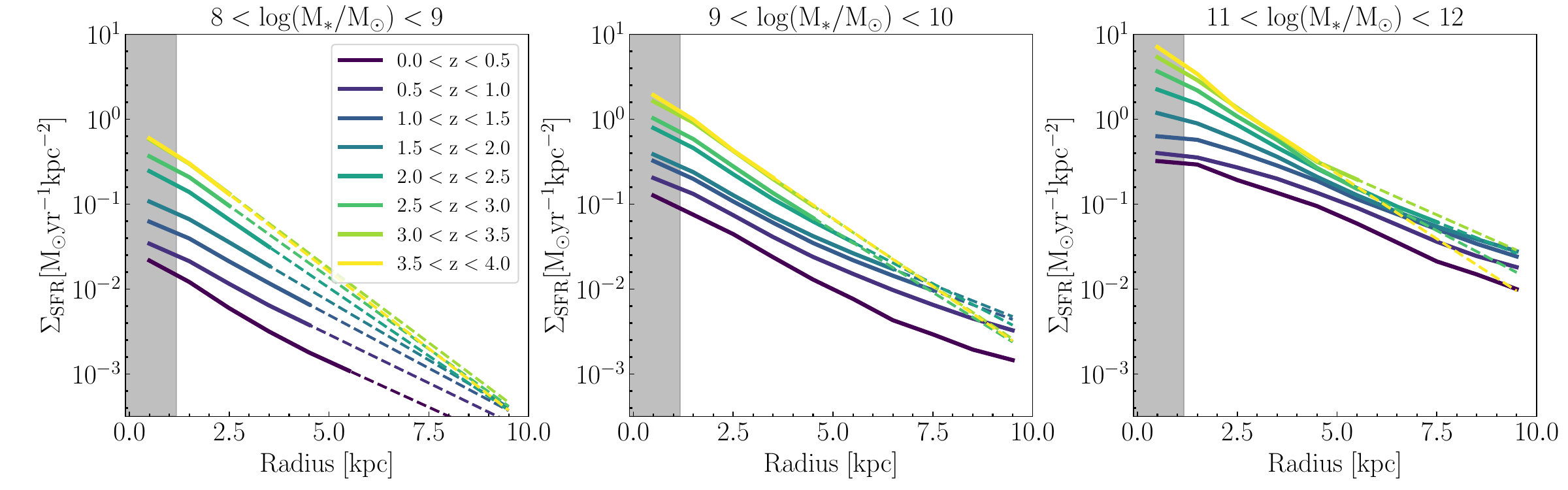}
\caption{Similar to Figure \ref{fig9} but for radial profiles of $\Sigma_{\rm SFR}$.}
\label{fig10}
\end{figure*} 

\begin{figure*}[htb!]
\centering
\includegraphics[width=1.0\textwidth]{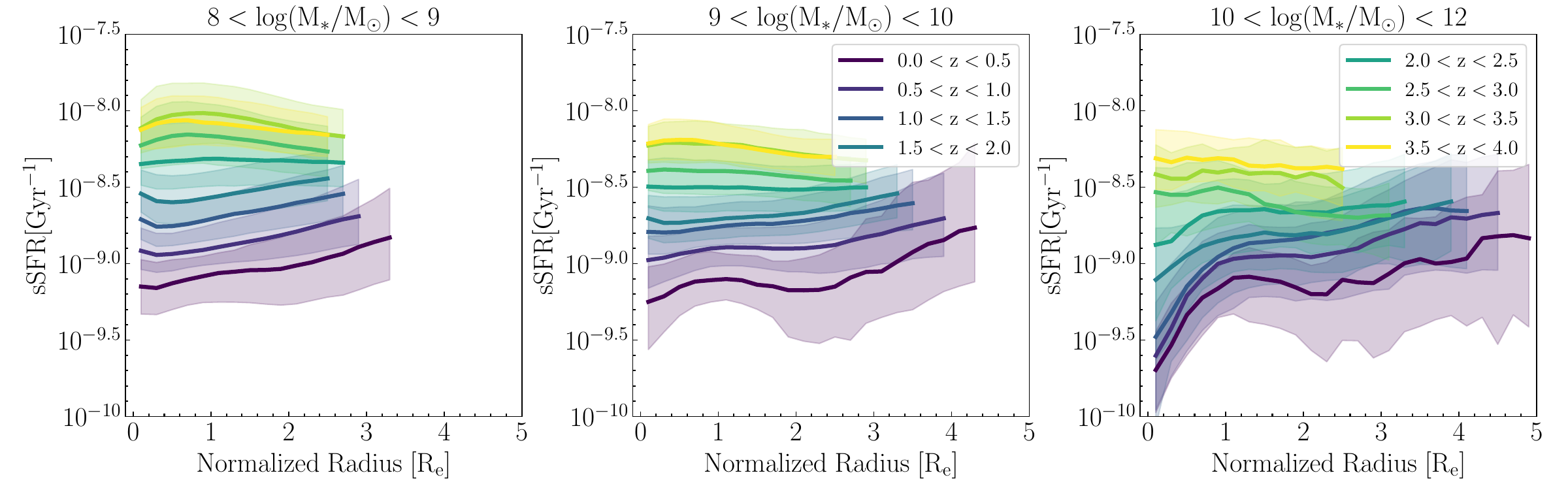}\\[1em]
\includegraphics[width=1.0\textwidth]{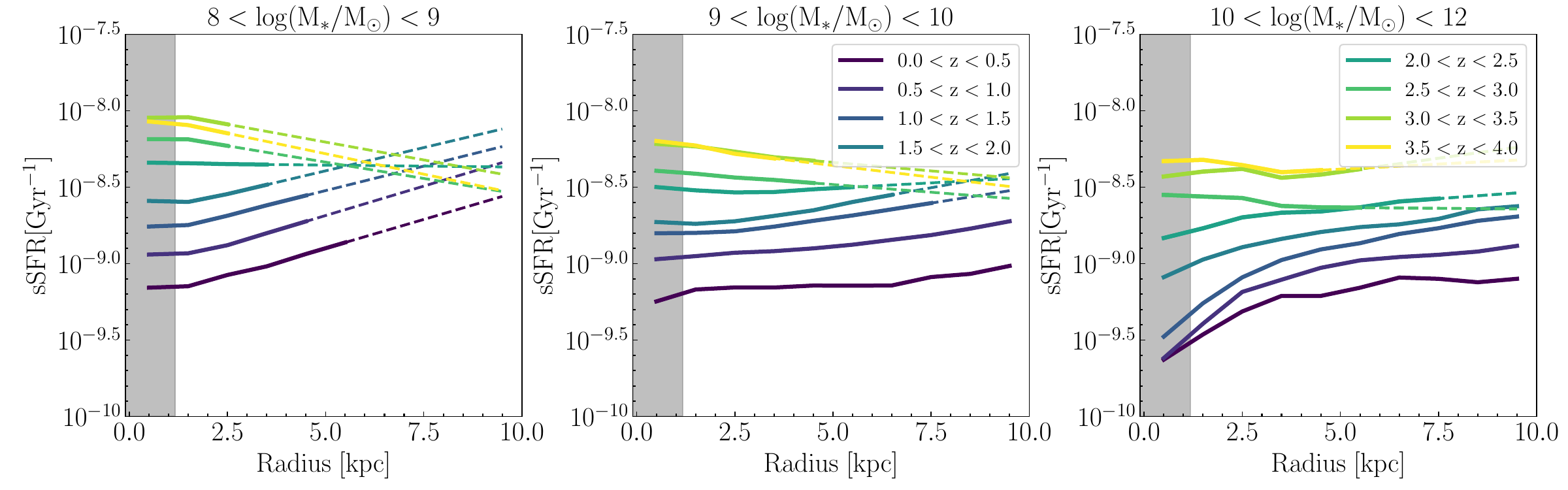}
\caption{Similar to Figure \ref{fig9} but for radial profiles of ${\rm sSFR}$.}
\label{fig11}
\end{figure*} 

\begin{table*}
\centering
\caption{sSFR gradient at different stellar mass and redshift bins}
\fontsize{7pt}{11pt}\selectfont
\label{tab:1}
\begin{tabular}{c|c|c|c|c|c|c|c|c}
\hline\hline
\multicolumn{9}{c}{sSFR gradient between $0.5R_e$ and $2.5 R_e$ in unit of dex/$R_e$}\\
\hline
& $0<z<0.5$ & $0.5<z<1$ & $1<z<1.5$ & $1.5 < z < 2$ & $2<z<2.5$ & $2.5<z<3$ & $3<z<3.5$ & $3.5<z<4$ \\
\hline
$8<\log(M_{\ast}/M_{\odot})<9$ & $0.083\pm0.010$ & $0.101\pm0.003$ & $0.096\pm0.003$ & 
    $0.078\pm 0.002$ & $-0.003\pm0.004$ & $-0.051\pm0.005$ & $-0.064\pm 0.006$ & $-0.046\pm0.006$ \\
$9<\log(M_{\ast}/M_{\odot})<10$ & $0.001\pm0.002$ & $0.029\pm0.006$ & $0.050\pm0.004$ & 
    $0.055\pm 0.004$ & $-0.003\pm0.004$ & $-0.034\pm0.005$ & $-0.049\pm0.006$ & $-0.035\pm 0.005$\\
$10<\log(M_{\ast}/M_{\odot})<12$ & $0.115\pm 0.005$ & $0.144\pm0.017$ & $0.184\pm0.015$ &
    $0.080\pm0.015$ & $0.058\pm0.019$ & $-0.071\pm0.025$ & $-0.030\pm 0.034$ & $-0.037\pm 0.038$ \\
\hline
\multicolumn{9}{c}{sSFR gradient between $0.5$ Kpc and $2.5$ Kpc in unit of dex/kpc}\\
\hline
& $0<z<0.5$ & $0.5<z<1$ & $1<z<1.5$ & $1.5 < z < 2$ & $2<z<2.5$ & $2.5<z<3$ & $3<z<3.5$ & $3.5<z<4$ \\
\hline
$8<\log(M_{\ast}/M_{\odot})<9$ & $0.041\pm0.008$ & $0.031\pm0.003$ & $0.035\pm0.003$ & 
    $0.023\pm 0.002$ & $-0.004\pm 0.004$ & $-0.022\pm0.005$ & $-0.021\pm 0.005$ & $-0.039\pm 0.006$ \\
$9<\log(M_{\ast}/M_{\odot})<10$ & $0.045\pm0.019$ & $0.021\pm 0.006$ & $0.006\pm0.004$ & 
    $0.002\pm 0.003$& $-0.017\pm 0.004$ & $-0.022\pm0.004$ & $-0.025\pm0.005$ & $-0.041\pm0.005$ \\
$10<\log(M_{\ast}/M_{\odot})<12$ & $0.157\pm0.045$ & $0.216\pm0.016$ & $0.194\pm0.017$ &
    $0.098\pm0.018$ & $0.067\pm 0.020$ & $-0.010\pm0.031$ & $0.025\pm0.033$ & $-0.013\pm0.029$ \\
\hline
\multicolumn{9}{c}{sSFR gradient derived from deconvolved images between $0.5$ Kpc and $2.5$ Kpc in unit of dex/kpc}\\
\hline
$8<\log(M_{\ast}/M_{\odot})<9$ & $0.064\pm0.009$ & $0.057\pm0.004$ & $0.065\pm0.004$ & 
    $0.061\pm 0.004$ & $-0.034\pm 0.005$ & $-0.063\pm0.006$ & $-0.064\pm 0.007$ & $-0.046\pm 0.009$ \\
$9<\log(M_{\ast}/M_{\odot})<10$ & $0.082\pm0.020$ & $0.044\pm 0.007$ & $0.018\pm0.006$ & 
    $-0.012\pm 0.005$& $-0.056\pm 0.005$ & $-0.072\pm0.006$ & $-0.059\pm0.009$ & $-0.090\pm0.008$ \\
$10<\log(M_{\ast}/M_{\odot})<12$ & $0.338\pm0.050$ & $0.304\pm0.019$ & $0.311\pm0.023$ &
    $0.186\pm0.027$ & $0.137\pm 0.028$ & $0.011\pm0.031$ & $0.060\pm0.046$ & $-0.013\pm0.060$ \\
\hline
\end{tabular}
\end{table*}

\subsection{SFR and sSFR profile evolution} \label{sec:4.4}
The stellar mass of a galaxy reflects the cumulative result of its past star formation activities. In this section, we investigate the recent star formation activity of galaxies by examining their $\Sigma_{\rm SFR}$ and sSFR (defined as $\Sigma_{\rm SFR}/\Sigma_{*}$) profiles, shown in Figure \ref{fig10} and Figure \ref{fig11}. As shown in Figure \ref{fig10}, the $\Sigma_{\rm SFR}$ profiles exhibit negative gradients across all redshift and stellar mass bins, with a flattening trend observed in the central regions. Many studies have highlighted the existence of a resolved SFMS (rSFMS)—a tight correlation between the SFR surface density and the stellar mass surface density (e.g., \citealt{2019ApJ...884L..33L, 2022MNRAS.510.3622B}). The negative $\Sigma_{\rm SFR}$ profiles observed here can be naturally explained by this rSFMS, in conjunction with the negative gradient of the stellar mass profile. 

When comparing the $\Sigma_{\rm SFR}$ profiles of galaxies at different redshifts, we observe that higher-redshift galaxies exhibit elevated $\Sigma_{\rm SFR}$ across all radii. This trend likely reflects the higher gas fractions or increased cold gas accretion rates that are characteristic of galaxies at earlier cosmic times. Regarding the dependence on stellar mass, galaxies with higher stellar masses consistently show larger $\Sigma_{\rm SFR}$ at all radii compared to their lower-mass counterparts. This is expected, as more massive galaxies have higher $\Sigma_{\ast}$ at all radii, in line with the existence of the rSFMS. Several studies suggest that the integrated SFMS arises from the underlying rSFMS, indicating that the relationship between $\Sigma_{\ast}$ and $\Sigma_{\rm SFR}$ is more fundamental. In future work, we will further investigate the connection between the resolved and integrated SFMS using the spatially resolved physical properties derived in this study.

As for the ${\rm sSFR}$ profiles, we find distinct behavior across different redshift ranges. As shown in Figure \ref{fig11}, for galaxies at $z < 2$, we observe a positive ${\rm sSFR}$ gradient, with this trend becoming more pronounced in more massive galaxies. This trend has been observed in many previous studies (e.g., \citealt{2016ApJ...828...27N, 2019A&A...626A..61M, 2024ApJ...963L..49S, 2024ApJ...975..252H}). Given that galaxies across all redshift and stellar mass bins exhibit negative $\Sigma_{\rm SFR}$ gradients, the observed positive ${\rm sSFR}$ gradients reflect an inside-out growth scenario. In contrast, for galaxies at $z > 2.5$, there is a negative ${\rm sSFR}$ gradient, although it's less pronounced for more massive galaxies. To quantitatively characterize the sSFR gradient, we have measured the sSFR gradient between $0.5 R_e$ and $2.5 Re$. We choose this radial range to ensure that the separation is larger than the PSF FWHM, thereby minimizing the impact of PSF smearing. At the same time, this range allows us to estimate sSFR gradients consistently across different mass bins. Similar to $\Sigma_\ast$, we also estimate the standard error at each radius to characterize the uncertainties in the sSFR profile, and subsequently determine the corresponding errors in the inferred gradients. Additionally, using physical units, we also evaluate the sSFR gradients bwtween 0.5 kpc and 2.5 kpc. The results are summarized at Table \ref{tab:1}. As indicated in the table, galaxies with $\log(M_\ast/M_\odot)<10$ display a significant shift from positive sSFR gradients at $z<2$ to negative gradients at $z>2.5$. Massive galaxies show a qualitatively similar trend, but the associated uncertainties are substantially larger because the number of high-redshift massive galaxies is limited. Additionally, since high-redshift star formation is more stochastic, we have also examined the case of adopting a shorter timescale (10 Myr) for the SFR tracer and found that the results remain unchanged.

While positive sSFR gradients at $z<2$ have been reported in numerous studies, however, results at higher redshifts, particularly for low-mass galaxies, have not been systematically investigated prior to JWST. Using data from VELA simulation \citep{2014MNRAS.442.1545C, 2015MNRAS.450.2327Z}, \cite{2016MNRAS.458..242T} had demonstrated that galaxies begin to exhibit positive ${\rm sSFR}$ gradients at redshifts below $z\sim 3$. By combining HST and JWST data, \cite{2023ApJ...945..117A} studied the sSFR profiles of 219 SFGs and found that the sSFR radial profiles of the majority of galaxies at $z>2.5$ are broadly flat. In a recent study based on JWST slitless data, \cite{2024A&A...690A..64M} reported that galaxies at $4.8<z<6.5$ exhibit roughly positive $\rm EW(H\alpha)$, which may indicate a positive sSFR gradient. However, in another study using data from the JADES field, \cite{2024A&A...692A.184T} found that galaxies at $z>4$ exhibit negative $\rm EW(H{\beta})$ profiles, which imply a negative sSFR gradient at this redshift range. Meanwhile, using a sample of 669 galaxies at $4<z<8$ in the CEERS field, \cite{2025ApJ...993..225J} found that high-redshift galaxies predominantly exhibit positive color gradients, which also indicates centrally concentrated star formation. These findings are consistent with the results presented in our work. However, the physical mechanisms driving this transition and the dependence on stellar mass remain unclear and warrant further investigation. 

Moreover, since sSFR measures the ratio of recently formed stellar mass to the total stellar mass, a positive ${\rm sSFR}$ gradient indicates that the outskirts of galaxies are growing more rapidly in stellar mass, potentially expanding their overall size. On the other hand, a negative ${\rm sSFR}$ gradient would indicate more efficient stellar mass build-up in the central regions, which may lead to a more compact morphology. The transition from a negative to a positive sSFR gradient may lead to different size growth status. At lower redshifts, star formation contributes significantly to galaxy size growth, particularly for high-mass systems. In contrast, at higher redshifts, the role of in-situ star formation in driving size growth appears to be less prominent or even result in a smaller effective radius. If this trend continues to even higher redshifts, it could lead to a negative size–mass relation, which has already been observed in both observation (e.g., \citealt{2024MNRAS.527.6110O}) and some cosmological simulations (e.g., \citealt{2018MNRAS.474.3976G, 2023ApJ...946...71C}). A more thorough and comprehensive investigation may be needed for understanding the results at higher redshifts.

\section{Discussion} \label{sec:5}
Our analysis of the spatially resolved sSFR profiles reveals a pivotal transition in the mode of galaxy assembly around $z\sim 2$. Specifically, the shift from negative sSFR gradients at high redshifts ($z>2.5$) to positive gradients at lower redshifts ($z<2$) suggests a fundamental change in how star-forming galaxies acquire gas and build their stellar mass. The observed evolution may reflect the interplay between the angular momentum of accreting gas, the dynamical stability of galactic disks, and feedback processes that regulate star formation.

The prevalence of negative sSFR gradients observed in our sample at $z>2.5$ points to a phase where mass assembly is centrally concentrated. In the high-redshift Universe, the accretion of cold gas streams is characterized by relatively low specific angular momentum (e.g., \citealt{2011MNRAS.418.2493P, 2012MNRAS.423.3616D}). When combined with violent disk instabilities prevalent in these highly gas-rich and turbulent disks, gas clumps lose angular momentum rapidly and migrate inward toward the galaxy center (e.g., \citealt{2015MNRAS.450.2327Z, 2020MNRAS.493.4126D}). This efficient funneling of gas triggers intense nuclear starbursts, naturally resulting in the negative sSFR gradients we observe, where the specific star formation rate is significantly elevated in the core compared to the outskirts. As new stellar mass is preferentially added to the central region, the galaxy’s mass distribution becomes highly concentrated. 

At $z < 2$, the emergence of positive sSFR gradients signals a transition to the classical "inside-out" growth mode. Several physical processes can conspire to drive this shift. First, as the central bulge assembled becomes sufficiently massive, it deepens the central potential and stabilizes the inner disk against further violent inflows (e.g., \citealt{2009ApJ...707..250M}). Simultaneously, the nature of gas accretion evolves with cosmic time. Cosmological models dictate that subsequent gas accretion onto the dark matter halo carries significantly higher specific angular momentum due to the expansion of the Universe and the growth of the halo virial radius (e.g., \citealt{2011MNRAS.418.2493P, 2017ASSL..430..249S}). This high-angular-momentum gas cannot penetrate the deep central potential well; instead, it settles into an extended disk structure in the galaxy's outskirts \citep{2014ApJ...784...26O}. Consequently, star formation becomes most active in these outer regions, resulting in the positive sSFR gradients observed in our lower-redshift sample. This marks the onset of secular disk growth, where the continuous addition of mass at large radii leads to the gradual build-up of a stellar envelope and a steady increase in the galaxy's effective size.

\section{Summary}\label{sec:6}
By combining high-resolution imaging data from JWST and HST, we have measured both the integrated and spatially resolved physical properties of galaxies with $\log(M_{\ast}/M_{\odot})>8$ over the redshift range $0<z<4$ in the CANDELS fields. This work represents the first in our series of studies. In this paper, through investigating the relationships between galaxy integrated and spatially resolved physical properties, we have drawn the following conclusions:

(1) For galaxies at $z<4$,  combining HST and JWST data allows for a reliable recovery of stellar mass and SFRs. Additionally, within the range $\log(M_{\ast}/M_{\odot})>8$, the star-forming main sequence of galaxies exhibits a roughly linear relationship.

(2) Regarding the morphologies of galaxies in the rest-frame $1 \mu m$ band, the size-mass relation and size evolution of galaxies show trends similar to those in optical bands, following the relations $R_e \propto M_{\ast}^{0.17}(1+z)^{-0.62}$.

(3) By comparing the physical property profiles across different redshift and stellar mass bins, we find that, across all redshift and stellar mass ranges, galaxies consistently exhibit negative gradients in both stellar mass surface density and SFR surface density. Furthermore, at higher redshifts and stellar masses, galaxies tend to show higher stellar mass and SFR surface densities.

(4) For the sSFR profiles, we observe a distinct trend: at $z>2.5$, galaxies display a mild negative gradient in sSFR profiles, whereas at lower redshifts, they exhibit a positive gradient. These findings suggest that galaxies undergo a transition in their stellar assembly mode, driven by in-situ star formation, shifting from an outside-in to an inside-out growth pattern at $z\sim2$, marking a pivotal shift in their structural evolution. This finding deepens the mystery of how galaxies grow in size beyond the cosmic noon, demanding more research to resolve the puzzle.

\section*{Data availability}
The data used in this work are available in electronic form at the CDS via http://cdsweb.u-strasbg.fr/cgi-bin/qcat?J/A+A/.

\begin{acknowledgements}
This work is supported by the National Science Foundation of China (NSFC, Grant No. 12233008, No. 12473008), the National Key R\&D Program of China (2023YFA1608100), the Strategic Priority Research Program of the Chinese Academy of Sciences (Grant No. XDB0550200), the Cyrus Chun Ying Tang Foundations, and the 111 Project for ``Observational and Theoretical Research on Dark Matter and Dark Energy'' (B23042), and the Start-up Fund of the University of Science and Technology of China (No. KY2030000200).

The data products presented herein were retrieved from the Dawn JWST Archive (DJA). DJA is an initiative of the Cosmic Dawn Center (DAWN), which is funded by the Danish National Research Foundation under grant DNRF140. This work is based on observations made with the NASA/ESA/CSA James Webb Space Telescope. The data were obtained from the Mikulski Archive for Space Telescopes at the Space Telescope Science Institute, which is operated by the Association of Universities for Research in Astronomy, Inc., under NASA contract NAS 5-03127 for JWST.
The data described here can be obtained from \url{https://dx.doi.org/10.17909/z7p0-8481}, \url{https://dx.doi.org/10.17909/8tdj-8n28}, and \url{https://dx.doi.org/10.17909/T94S3X}.
\end{acknowledgements}

\bibliographystyle{aa}
\bibliography{ref}

@ARTICLE{2025arXiv250104085F,
       author = {{Finkelstein}, Steven L. and {Bagley}, Micaela B. and {Arrabal Haro}, Pablo and {Dickinson}, Mark and {Ferguson}, Henry C. and {Kartaltepe}, Jeyhan S. and {Kocevski}, Dale D. and {Koekemoer}, Anton M. and {Lotz}, Jennifer M. and {Papovich}, Casey and {P{\'e}rez-Gonz{\'a}lez}, Pablo G. and {Pirzkal}, Nor and {Somerville}, Rachel S. and {Trump}, Jonathan R. and {Yang}, Guang and {Yung}, L.~Y. Aaron and {Fontana}, Adriano and {Grazian}, Andrea and {Grogin}, Norman A. and {Kewley}, Lisa J. and {Kirkpatrick}, Allison and {Larson}, Rebecca L. and {Pentericci}, Laura and {Ravindranath}, Swara and {Wilkins}, Stephen M. and {Almaini}, Omar and {Amor{\'\i}n}, Ricardo O. and {Barro}, Guillermo and {Bhatawdekar}, Rachana and {Bisigello}, Laura and {Brooks}, Madisyn and {Buat}, V{\'e}ronique and {Buitrago}, Fernando and {Burgarella}, Denis and {Calabr{\`o}}, Antonello and {Castellano}, Marco and {Cheng}, Yingjie and {Cleri}, Nikko J. and {Cole}, Justin W. and {Cooper}, M.~C. and {Cooper}, Olivia R. and {Costantin}, Luca and {Cox}, Isa G. and {Croton}, Darren and {Daddi}, Emanuele and {Davis}, Kelcey and {Dekel}, Avishai and {Elbaz}, David and {Fern{\'a}ndez}, Vital and {Fujimoto}, Seiji and {Gandolfi}, Giovanni and {Gardner}, Jonathan P. and {Gawiser}, Eric and {Giavalisco}, Mauro and {G{\'o}mez-Guijarro}, Carlos and {Guo}, Yuchen and {Gupta}, Ansh R. and {Hathi}, Nimish P. and {Harish}, Santosh and {Henry}, Aur{\'e}lien and {Hirschmann}, Michaela and {Hu}, Weida and {Hutchison}, Taylor A. and {Iyer}, Kartheik G. and {Jaskot}, Anne E. and {Jha}, Saurabh W. and {Jung}, Intae and {Kassin}, Susan A. and {Kokorev}, Vasily and {Kurczynski}, Peter and {Leung}, Gene C.~K. and {Llerena}, Mario and {Long}, Arianna S. and {Lucas}, Ray A. and {Lu}, Shiying and {McGrath}, Elizabeth J. and {McIntosh}, Daniel H. and {Merlin}, Emiliano and {Mobasher}, Bahram and {Morales}, Alexa M. and {Napolitano}, Lorenzo and {Pacucci}, Fabio and {Pandya}, Viraj and {Rafelski}, Marc and {Rodighiero}, Giulia and {Rose}, Caitlin and {Santini}, Paola and {Seill{\'e}}, Lise-Marie and {Simons}, Raymond C. and {Shen}, Lu and {Straughn}, Amber N. and {Tacchella}, Sandro and {Taylor}, Anthony J. and {Vanderhoof}, Brittany N. and {Vega-Ferrero}, Jes{\'u}s and {Weiner}, Benjamin J. and {Willmer}, Christopher N.~A. and {Zhu}, Peixin and {Bell}, Eric F. and {Wuyts}, Stijn and {Holwerda}, Benne W. and {Wang}, Xin and {Wang}, Weichen and {Zavala}, Jorge A. and {CEERS Collaboration}},
        title = "{The Cosmic Evolution Early Release Science Survey (CEERS)}",
      journal = {\apjl},
         year = 2025,
        month = apr,
       volume = {983},
       number = {1},
          eid = {L4},
        pages = {L4},
          doi = {10.3847/2041-8213/adbbd3},
archivePrefix = {arXiv},
       eprint = {2501.04085},
 primaryClass = {astro-ph.GA},
       adsurl = {https://ui.adsabs.harvard.edu/abs/2025ApJ...983L...4F}
}

@ARTICLE{Lyu-25,
       author = {{Lyu}, Cheqiu and {Wang}, Enci and {Zhang}, Hongxin and {Peng}, Yingjie and {Wang}, Xin and {Li}, Haixin and {Ma}, Chengyu and {Yu}, Haoran and {Chen}, Zeyu and {Jia}, Cheng and {Kong}, Xu},
        title = "{Dominant Role of Coplanar Inflows in Driving Disk Evolution Revealed by Gas-phase Metallicity Gradients}",
      journal = {\apjl},
         year = 2025,
        month = mar,
       volume = {981},
       number = {1},
          eid = {L6},
        pages = {L6},
          doi = {10.3847/2041-8213/adb4ed},
archivePrefix = {arXiv},
       eprint = {2502.12409},
 primaryClass = {astro-ph.GA},
       adsurl = {https://ui.adsabs.harvard.edu/abs/2025ApJ...981L...6L}
}

@ARTICLE{He-25,
       author = {{He}, Zhicheng and {Wang}, Enci and {Ho}, Luis C. and {Wang}, Huiyuan and {Shi}, Yong and {Kong}, Xu and {Wang}, Tinggui},
        title = "{Symmetry in Fundamental Parameters of Galaxies on the Star-forming Main Sequence}",
      journal = {\mnras},
         year = 2026,
        month = mar,
          doi = {10.1093/mnras/stag443},
 primaryClass = {astro-ph.GA},
       adsurl = {https://ui.adsabs.harvard.edu/abs/2026MNRAS.tmp..402H}}

@ARTICLE{Jia-25,
       author = {{Jia}, Cheng and {Wang}, Enci and {Lyu}, Cheqiu and {Ma}, Chengyu and {Song}, Jie and {Chen}, Yangyao and {Wang}, Kai and {Yu}, Haoran and {Chen}, Zeyu and {Wang}, Jinyang and {Wang}, Yifan and {Kong}, Xu},
        title = "{Potential-driven Metal Cycling: JADES Census of Gas-phase Metallicity for Galaxies at 1 < z < 7}",
      journal = {\apjl},
         year = 2025,
        month = jun,
       volume = {986},
       number = {2},
          eid = {L24},
        pages = {L24},
          doi = {10.3847/2041-8213/addfd9},
archivePrefix = {arXiv},
       eprint = {2504.18820},
 primaryClass = {astro-ph.GA},
       adsurl = {https://ui.adsabs.harvard.edu/abs/2025ApJ...986L..24J}
}

@ARTICLE{Chen-25,
       author = {{Chen}, Zeyu and {Wang}, Enci and {Zou}, Hu and {Zou}, Siwei and {Gao}, Yang and {Wang}, Huiyuan and {Yu}, Haoran and {Jia}, Cheng and {Li}, Haixin and {Ma}, Chengyu and {Yao}, Yao and {Ding}, Weiyu and {Zhu}, Runyu},
        title = "{The Circumgalactic Medium Traced by Mg II Absorption with DESI: Dependence on Galaxy Stellar Mass, Star Formation Rate, and Azimuthal Angle}",
      journal = {\apj},
         year = 2025,
        month = mar,
       volume = {981},
       number = {1},
          eid = {81},
        pages = {81},
          doi = {10.3847/1538-4357/ada942},
archivePrefix = {arXiv},
       eprint = {2411.08485},
 primaryClass = {astro-ph.GA},
       adsurl = {https://ui.adsabs.harvard.edu/abs/2025ApJ...981...81C}
}

@ARTICLE{Lilly-13,
       author = {{Lilly}, Simon J. and {Carollo}, C. Marcella and {Pipino}, Antonio and {Renzini}, Alvio and {Peng}, Yingjie},
        title = "{Gas Regulation of Galaxies: The Evolution of the Cosmic Specific Star Formation Rate, the Metallicity-Mass-Star-formation Rate Relation, and the Stellar Content of Halos}",
      journal = {\apj},
         year = 2013,
        month = aug,
       volume = {772},
       number = {2},
          eid = {119},
        pages = {119},
          doi = {10.1088/0004-637X/772/2/119},
archivePrefix = {arXiv},
       eprint = {1303.5059},
 primaryClass = {astro-ph.CO},
       adsurl = {https://ui.adsabs.harvard.edu/abs/2013ApJ...772..119L}
}

@ARTICLE{Wang-18,
       author = {{Wang}, Enci and {Li}, Cheng and {Xiao}, Ting and {Lin}, Lin and {Bershady}, Matthew and {Law}, David R. and {Merrifield}, Michael and {Sanchez}, Sebastian F. and {Riffel}, Rogemar A. and {Riffel}, Rogerio and {Yan}, Renbin},
        title = "{SDSS-IV MaNGA: Star Formation Cessation in Low-redshift Galaxies. I. Dependence on Stellar Mass and Structural Properties}",
      journal = {\apj},
         year = 2018,
        month = apr,
       volume = {856},
       number = {2},
          eid = {137},
        pages = {137},
          doi = {10.3847/1538-4357/aab263},
archivePrefix = {arXiv},
       eprint = {1710.07569},
 primaryClass = {astro-ph.GA},
       adsurl = {https://ui.adsabs.harvard.edu/abs/2018ApJ...856..137W}
}

@ARTICLE{Wang-19,
       author = {{Wang}, Enci and {Lilly}, Simon J. and {Pezzulli}, Gabriele and {Matthee}, Jorryt},
        title = "{On the Elevation and Suppression of Star Formation within Galaxies}",
      journal = {\apj},
         year = 2019,
        month = jun,
       volume = {877},
       number = {2},
          eid = {132},
        pages = {132},
          doi = {10.3847/1538-4357/ab1c5b},
archivePrefix = {arXiv},
       eprint = {1901.10276},
 primaryClass = {astro-ph.GA},
       adsurl = {https://ui.adsabs.harvard.edu/abs/2019ApJ...877..132W}
}

@ARTICLE{Wang-22b,
       author = {{Wang}, Enci and {Lilly}, Simon J.},
        title = "{Gas-phase Metallicity Profiles of Star-forming Galaxies in the Modified Accretion Disk Framework}",
      journal = {\apj},
         year = 2022,
        month = apr,
       volume = {929},
       number = {1},
          eid = {95},
        pages = {95},
          doi = {10.3847/1538-4357/ac5e31},
archivePrefix = {arXiv},
       eprint = {2201.04151},
 primaryClass = {astro-ph.GA},
       adsurl = {https://ui.adsabs.harvard.edu/abs/2022ApJ...929...95W}
}

@ARTICLE{Wang-22a,
       author = {{Wang}, Enci and {Lilly}, Simon J.},
        title = "{The Origin of Exponential Star-forming Disks}",
      journal = {\apj},
         year = 2022,
        month = mar,
       volume = {927},
       number = {2},
          eid = {217},
        pages = {217},
          doi = {10.3847/1538-4357/ac49ed},
archivePrefix = {arXiv},
       eprint = {2201.04148},
 primaryClass = {astro-ph.GA},
       adsurl = {https://ui.adsabs.harvard.edu/abs/2022ApJ...927..217W}
}

@ARTICLE{Wang-21,
       author = {{Wang}, Enci and {Lilly}, Simon J.},
        title = "{Gas-phase Metallicity as a Diagnostic of the Drivers of Star Formation on Different Spatial Scales}",
      journal = {\apj},
         year = 2021,
        month = apr,
       volume = {910},
       number = {2},
          eid = {137},
        pages = {137},
          doi = {10.3847/1538-4357/abe413},
archivePrefix = {arXiv},
       eprint = {2009.01935},
 primaryClass = {astro-ph.GA},
       adsurl = {https://ui.adsabs.harvard.edu/abs/2021ApJ...910..137W}
}

@ARTICLE{2023arXiv230602465E,
author = {{Eisenstein}, Daniel J. and {Willott}, Chris and {Alberts}, Stacey and {Arribas}, Santiago and {Bonaventura}, Nina and {Bunker}, Andrew J. and {Cameron}, Alex J. and {Carniani}, Stefano and {Charlot}, Stephane and {Curtis-Lake}, Emma and {D'Eugenio}, Francesco and {Ferruit}, Pierre and {Giardino}, Giovanna and {Hainline}, Kevin and {Hausen}, Ryan and {Jakobsen}, Peter and {Johnson}, Benjamin D. and {Maiolino}, Roberto and {Rauscher}, Bernard J. and {Rieke}, Marcia and {Rieke}, George and {Rix}, Hans-Walter and {Robertson}, Brant and {Stark}, Daniel P. and {Tacchella}, Sandro and {Williams}, Christina C. and {Willmer}, Christopher N.~A. and {Baker}, William M. and {Baum}, Stefi and {Bhatawdekar}, Rachana and {Boyett}, Kristan and {Chen}, Zuyi and {Chevallard}, Jacopo and {Circosta}, Chiara and {Curti}, Mirko and {Danhaive}, A. Lola and {DeCoursey}, Christa and {Endsley}, Ryan and {de Graaff}, Anna and {Dressler}, Alan and {Egami}, Eiichi and {Helton}, Jakob M. and {Hviding}, Raphael E. and {Ji}, Zhiyuan and {Jones}, Gareth C. and {Kumari}, Nimisha and {L{\"u}tzgendorf}, Nora and {Laseter}, Isaac and {Looser}, Tobias J. and {Lyu}, Jianwei and {Maseda}, Michael V. and {Nelson}, Erica and {Parlanti}, Eleonora and {Perna}, Michele and {Pusk{\'a}s}, D{\'a}vid and {Rawle}, Tim and {Rodr{\'\i}guez Del Pino}, Bruno and {Rujopakarn}, Wiphu and {Sandles}, Lester and {Saxena}, Aayush and {Scholtz}, Jan and {Sharpe}, Katherine and {Shivaei}, Irene and {Silcock}, Maddie S. and {Simmonds}, Charlotte and {Skarbinski}, Maya and {Smit}, Renske and {Stone}, Meredith and {Suess}, Katherine A. and {Sun}, Fengwu and {Tang}, Mengtao and {Topping}, Michael W. and {{\"U}bler}, Hannah and {Villanueva}, Natalia C. and {Wallace}, Imaan E.~B. and {Whitler}, Lily and {Witstok}, Joris and {Woodrum}, Charity},
        title = "{Overview of the JWST Advanced Deep Extragalactic Survey (JADES)}",
      journal = {\apjs},
         year = 2026,
        month = mar,
       volume = {283},
       number = {1},
          eid = {6},
        pages = {6},
          doi = {10.3847/1538-4365/ae3163},
archivePrefix = {arXiv},
       eprint = {2306.02465},
 primaryClass = {astro-ph.GA},
       adsurl = {https://ui.adsabs.harvard.edu/abs/2026ApJS..283....6E}
       }

@MISC{2021jwst.prop.1837D,
       author = {{Dunlop}, James S. and {Abraham}, Roberto G. and {Ashby}, Matthew L.~N. and {Bagley}, Micaela and {Best}, Philip N. and {Bongiorno}, Angela and {Bouwens}, Rychard and {Bowler}, Rebecca A.~A. and {Brammer}, Gabriel and {Bremer}, Malcolm and {Calabro'}, Antonello and {Carnall}, Adam and {Castellano}, Marco and {Cirasuolo}, Michele and {Conselice}, Christopher and {Cullen}, Fergus and {Dave}, Romeel and {Dayal}, Pratika and {Dekel}, Avishai and {Dickinson}, Mark and {Duncan}, Kenneth James and {Elbaz}, David and {Ellis}, Richard S. and {Ferguson}, Harry C. and {Ferrara}, Andrea and {Finkelstein}, Steven L. and {Fontana}, Adriano and {Furlanetto}, Steven and {Fynbo}, Johan P.~U. and {Gallerani}, Simona and {Gardner}, Jonathan P. and {Giavalisco}, Mauro and {Grazian}, Andrea and {Grogin}, Norman and {Harikane}, Yuichi and {Hopkins}, Philip F. and {Ilbert}, Olivier and {Illingworth}, Garth D. and {Juneau}, Stephanie and {Jung}, Intae and {Kartaltepe}, Jeyhan and {Kassin}, Susan and {Kauffmann}, Olivier Benjamin and {Khochfar}, Sadegh and {Kirkpatrick}, Allison and {Kocevski}, Dale D. and {Koekemoer}, Anton M. and {Labbe}, Ivo and {Laporte}, Nicolas and {Larson}, Rebecca L. and {Lucas}, Ray A. and {Magee}, Daniel K. and {Mason}, Charlotte and {McCracken}, Henry Joy and {McLeod}, Derek and {McLure}, Ross and {Merlin}, Emiliano and {Mesinger}, Andrei and {Milvang-Jensen}, Bo and {Newman}, Jeffrey Allen and {Oesch}, Pascal and {Ouchi}, Masami and {Pacifici}, Camilla and {Papovich}, Casey and {Peacock}, John and {Peeples}, Molly and {Pentericci}, Laura and {Perez-Gonzalez}, Pablo G. and {Pirzkal}, Norbert and {Pope}, Alexandra and {Pye}, John P. and {Reddy}, Naveen A. and {Robertson}, Brant and {Salvato}, Mara and {Santini}, Paola and {Schaerer}, Daniel and {Shapley}, Alice E. and {Simons}, Raymond and {Smit}, Renske and {Smith}, Britton D. and {Snyder}, Greg and {Somerville}, Rachel S. and {Stanway}, Elizabeth R. and {Stefanon}, Mauro and {Tasca}, Lidia and {Tikkanen}, Tuomo and {Tresse}, Laurence and {Trump}, Jonathan R. and {Whitaker}, Katherine E. and {Wilkins}, Stephen Matthew and {Wright}, Gillian and {Wyithe}, J. Stuart B. and {van Dokkum}, Pieter and {van der Werf}, Paul},
        title = "{PRIMER: Public Release IMaging for Extragalactic Research}",
 howpublished = {JWST Proposal. Cycle 1, ID. \#1837},
         year = 2021,
        month = mar,
        pages = {1837},
       adsurl = {https://ui.adsabs.harvard.edu/abs/2021jwst.prop.1837D}
}

@ARTICLE{2023ApJ...947...20V,
       author = {{Valentino}, Francesco and {Brammer}, Gabriel and {Gould}, Katriona M.~L. and {Kokorev}, Vasily and {Fujimoto}, Seiji and {Jespersen}, Christian Kragh and {Vijayan}, Aswin P. and {Weaver}, John R. and {Ito}, Kei and {Tanaka}, Masayuki and {Ilbert}, Olivier and {Magdis}, Georgios E. and {Whitaker}, Katherine E. and {Faisst}, Andreas L. and {Gallazzi}, Anna and {Gillman}, Steven and {Gim{\'e}nez-Arteaga}, Clara and {G{\'o}mez-Guijarro}, Carlos and {Kubo}, Mariko and {Heintz}, Kasper E. and {Hirschmann}, Michaela and {Oesch}, Pascal and {Onodera}, Masato and {Rizzo}, Francesca and {Lee}, Minju and {Strait}, Victoria and {Toft}, Sune},
        title = "{An Atlas of Color-selected Quiescent Galaxies at z > 3 in Public JWST Fields}",
      journal = {\apj},
         year = 2023,
        month = apr,
       volume = {947},
       number = {1},
          eid = {20},
        pages = {20},
          doi = {10.3847/1538-4357/acbefa},
archivePrefix = {arXiv},
       eprint = {2302.10936},
 primaryClass = {astro-ph.GA},
       adsurl = {https://ui.adsabs.harvard.edu/abs/2023ApJ...947...20V}
}

@ARTICLE{2014ApJS..214...24S,
       author = {{Skelton}, Rosalind E. and {Whitaker}, Katherine E. and {Momcheva}, Ivelina G. and {Brammer}, Gabriel B. and {van Dokkum}, Pieter G. and {Labb{\'e}}, Ivo and {Franx}, Marijn and {van der Wel}, Arjen and {Bezanson}, Rachel and {Da Cunha}, Elisabete and {Fumagalli}, Mattia and {F{\"o}rster Schreiber}, Natascha and {Kriek}, Mariska and {Leja}, Joel and {Lundgren}, Britt F. and {Magee}, Daniel and {Marchesini}, Danilo and {Maseda}, Michael V. and {Nelson}, Erica J. and {Oesch}, Pascal and {Pacifici}, Camilla and {Patel}, Shannon G. and {Price}, Sedona and {Rix}, Hans-Walter and {Tal}, Tomer and {Wake}, David A. and {Wuyts}, Stijn},
        title = "{3D-HST WFC3-selected Photometric Catalogs in the Five CANDELS/3D-HST Fields: Photometry, Photometric Redshifts, and Stellar Masses}",
      journal = {\apjs},
         year = 2014,
        month = oct,
       volume = {214},
       number = {2},
          eid = {24},
        pages = {24},
          doi = {10.1088/0067-0049/214/2/24},
archivePrefix = {arXiv},
       eprint = {1403.3689},
 primaryClass = {astro-ph.GA},
       adsurl = {https://ui.adsabs.harvard.edu/abs/2014ApJS..214...24S}
}

@ARTICLE{2022ApJS..258...11W,
       author = {{Weaver}, J.~R. and {Kauffmann}, O.~B. and {Ilbert}, O. and {McCracken}, H.~J. and {Moneti}, A. and {Toft}, S. and {Brammer}, G. and {Shuntov}, M. and {Davidzon}, I. and {Hsieh}, B.~C. and {Laigle}, C. and {Anastasiou}, A. and {Jespersen}, C.~K. and {Vinther}, J. and {Capak}, P. and {Casey}, C.~M. and {McPartland}, C.~J.~R. and {Milvang-Jensen}, B. and {Mobasher}, B. and {Sanders}, D.~B. and {Zalesky}, L. and {Arnouts}, S. and {Aussel}, H. and {Dunlop}, J.~S. and {Faisst}, A. and {Franx}, M. and {Furtak}, L.~J. and {Fynbo}, J.~P.~U. and {Gould}, K.~M.~L. and {Greve}, T.~R. and {Gwyn}, S. and {Kartaltepe}, J.~S. and {Kashino}, D. and {Koekemoer}, A.~M. and {Kokorev}, V. and {Le F{\`e}vre}, O. and {Lilly}, S. and {Masters}, D. and {Magdis}, G. and {Mehta}, V. and {Peng}, Y. and {Riechers}, D.~A. and {Salvato}, M. and {Sawicki}, M. and {Scarlata}, C. and {Scoville}, N. and {Shirley}, R. and {Silverman}, J.~D. and {Sneppen}, A. and {Smolc̆i{\'c}}, V. and {Steinhardt}, C. and {Stern}, D. and {Tanaka}, M. and {Taniguchi}, Y. and {Teplitz}, H.~I. and {Vaccari}, M. and {Wang}, W. -H. and {Zamorani}, G.},
        title = "{COSMOS2020: A Panchromatic View of the Universe to z{\ensuremath{\sim}}10 from Two Complementary Catalogs}",
      journal = {\apjs},
         year = 2022,
        month = jan,
       volume = {258},
       number = {1},
          eid = {11},
        pages = {11},
          doi = {10.3847/1538-4365/ac3078},
archivePrefix = {arXiv},
       eprint = {2110.13923},
 primaryClass = {astro-ph.GA},
       adsurl = {https://ui.adsabs.harvard.edu/abs/2022ApJS..258...11W}
}

@ARTICLE{2019A&A...622A.103B,
       author = {{Boquien}, M. and {Burgarella}, D. and {Roehlly}, Y. and {Buat}, V. and {Ciesla}, L. and {Corre}, D. and {Inoue}, A.~K. and {Salas}, H.},
        title = "{CIGALE: a python Code Investigating GALaxy Emission}",
      journal = {\aap},
         year = 2019,
        month = feb,
       volume = {622},
          eid = {A103},
        pages = {A103},
          doi = {10.1051/0004-6361/201834156},
archivePrefix = {arXiv},
       eprint = {1811.03094},
 primaryClass = {astro-ph.GA},
       adsurl = {https://ui.adsabs.harvard.edu/abs/2019A&A...622A.103B}
}

@ARTICLE{2024ApJ...963L..49S,
       author = {{Shen}, Lu and {Papovich}, Casey and {Matharu}, Jasleen and {Pirzkal}, Nor and {Hu}, Weida and {Backhaus}, Bren E. and {Bagley}, Micaela B. and {Cheng}, Yingjie and {Cleri}, Nikko J. and {Finkelstein}, Steven L. and {Huertas-Company}, Marc and {Giavalisco}, Mauro and {Grogin}, Norman A. and {Jung}, Intae and {Kartaltepe}, Jeyhan S. and {Koekemoer}, Anton M. and {Lotz}, Jennifer M. and {Maseda}, Michael V. and {P{\'e}rez-Gonz{\'a}lez}, Pablo G. and {Rothberg}, Barry and {Simons}, Raymond C. and {Tacchella}, Sandro and {Williams}, Christina C. and {Yung}, L.~Y. Aaron},
        title = "{NGDEEP Epoch 1: Spatially Resolved H{\ensuremath{\alpha}} Observations of Disk and Bulge Growth in Star-forming Galaxies at z {\ensuremath{\sim}} 0.6{\textendash}2.2 from JWST NIRISS Slitless Spectroscopy}",
      journal = {\apjl},
         year = 2024,
        month = mar,
       volume = {963},
       number = {2},
          eid = {L49},
        pages = {L49},
          doi = {10.3847/2041-8213/ad28bd},
archivePrefix = {arXiv},
       eprint = {2310.13745},
 primaryClass = {astro-ph.GA},
       adsurl = {https://ui.adsabs.harvard.edu/abs/2024ApJ...963L..49S}
}

@ARTICLE{1999PASP..111...63F,
       author = {{Fitzpatrick}, Edward L.},
        title = "{Correcting for the Effects of Interstellar Extinction}",
      journal = {\pasp},
         year = 1999,
        month = jan,
       volume = {111},
       number = {755},
        pages = {63-75},
          doi = {10.1086/316293},
archivePrefix = {arXiv},
       eprint = {astro-ph/9809387},
 primaryClass = {astro-ph},
       adsurl = {https://ui.adsabs.harvard.edu/abs/1999PASP..111...63F}
}

@ARTICLE{2011ApJ...737..103S,
       author = {{Schlafly}, Edward F. and {Finkbeiner}, Douglas P.},
        title = "{Measuring Reddening with Sloan Digital Sky Survey Stellar Spectra and Recalibrating SFD}",
      journal = {\apj},
         year = 2011,
        month = aug,
       volume = {737},
       number = {2},
          eid = {103},
        pages = {103},
          doi = {10.1088/0004-637X/737/2/103},
archivePrefix = {arXiv},
       eprint = {1012.4804},
 primaryClass = {astro-ph.GA},
       adsurl = {https://ui.adsabs.harvard.edu/abs/2011ApJ...737..103S}
}

@ARTICLE{2003MNRAS.344.1000B,
       author = {{Bruzual}, G. and {Charlot}, S.},
        title = "{Stellar population synthesis at the resolution of 2003}",
      journal = {\mnras},
         year = 2003,
        month = oct,
       volume = {344},
       number = {4},
        pages = {1000-1028},
          doi = {10.1046/j.1365-8711.2003.06897.x},
archivePrefix = {arXiv},
       eprint = {astro-ph/0309134},
 primaryClass = {astro-ph},
       adsurl = {https://ui.adsabs.harvard.edu/abs/2003MNRAS.344.1000B}
}

@ARTICLE{2000ApJ...533..682C,
       author = {{Calzetti}, Daniela and {Armus}, Lee and {Bohlin}, Ralph C. and {Kinney}, Anne L. and {Koornneef}, Jan and {Storchi-Bergmann}, Thaisa},
        title = "{The Dust Content and Opacity of Actively Star-forming Galaxies}",
      journal = {\apj},
         year = 2000,
        month = apr,
       volume = {533},
       number = {2},
        pages = {682-695},
          doi = {10.1086/308692},
archivePrefix = {arXiv},
       eprint = {astro-ph/9911459},
 primaryClass = {astro-ph},
       adsurl = {https://ui.adsabs.harvard.edu/abs/2000ApJ...533..682C}
}

@ARTICLE{2011MNRAS.415.2920I,
       author = {{Inoue}, Akio K.},
        title = "{Rest-frame ultraviolet-to-optical spectral characteristics of extremely metal-poor and metal-free galaxies}",
      journal = {\mnras},
         year = 2011,
        month = aug,
       volume = {415},
       number = {3},
        pages = {2920-2931},
          doi = {10.1111/j.1365-2966.2011.18906.x},
archivePrefix = {arXiv},
       eprint = {1102.5150},
 primaryClass = {astro-ph.CO},
       adsurl = {https://ui.adsabs.harvard.edu/abs/2011MNRAS.415.2920I}
}

@ARTICLE{2025ApJ...978L..42C,
       author = {{Cochrane}, R.~K. and {Katz}, H. and {Begley}, R. and {Hayward}, C.~C. and {Best}, P.~N.},
        title = "{High-z Stellar Masses Can Be Recovered Robustly with JWST Photometry}",
      journal = {\apjl},
         year = 2025,
        month = jan,
       volume = {978},
       number = {2},
          eid = {L42},
        pages = {L42},
          doi = {10.3847/2041-8213/ad9a4d},
archivePrefix = {arXiv},
       eprint = {2412.02622},
 primaryClass = {astro-ph.GA},
       adsurl = {https://ui.adsabs.harvard.edu/abs/2025ApJ...978L..42C}
}

@ARTICLE{2023ApJ...949L..18P,
       author = {{Papovich}, Casey and {Cole}, Justin W. and {Yang}, Guang and {Finkelstein}, Steven L. and {Barro}, Guillermo and {Buat}, V{\'e}ronique and {Burgarella}, Denis and {P{\'e}rez-Gonz{\'a}lez}, Pablo G. and {Santini}, Paola and {Seill{\'e}}, Lise-Marie and {Shen}, Lu and {Arrabal Haro}, Pablo and {Bagley}, Micaela B. and {Bell}, Eric F. and {Bisigello}, Laura and {Calabr{\`o}}, Antonello and {Casey}, Caitlin M. and {Castellano}, Marco and {Chworowsky}, Katherine and {Cleri}, Nikko J. and {Costantin}, Luca and {Cooper}, M.~C. and {Dickinson}, Mark and {Ferguson}, Henry C. and {Fontana}, Adriano and {Giavalisco}, Mauro and {Grazian}, Andrea and {Grogin}, Norman A. and {Hathi}, Nimish P. and {Holwerda}, Benne W. and {Hutchison}, Taylor A. and {Kartaltepe}, Jeyhan S. and {Kewley}, Lisa J. and {Kirkpatrick}, Allison and {Kocevski}, Dale D. and {Koekemoer}, Anton M. and {Larson}, Rebecca L. and {Long}, Arianna S. and {Lucas}, Ray A. and {Pentericci}, Laura and {Pirzkal}, Nor and {Ravindranath}, Swara and {Somerville}, Rachel S. and {Trump}, Jonathan R. and {Urbano Stawinski}, Stephanie M. and {Weiner}, Benjamin J. and {Wilkins}, Stephen M. and {Yung}, L.~Y. Aaron and {Zavala}, Jorge A.},
        title = "{CEERS Key Paper. V. Galaxies at 4 < z < 9 Are Bluer than They Appear{\textendash}Characterizing Galaxy Stellar Populations from Rest-frame {\ensuremath{\sim}}1 {\ensuremath{\mu}}m Imaging}",
      journal = {\apjl},
         year = 2023,
        month = jun,
       volume = {949},
       number = {2},
          eid = {L18},
        pages = {L18},
          doi = {10.3847/2041-8213/acc948},
archivePrefix = {arXiv},
       eprint = {2301.00027},
 primaryClass = {astro-ph.GA},
       adsurl = {https://ui.adsabs.harvard.edu/abs/2023ApJ...949L..18P}
}

@ARTICLE{2023ApJ...958...82S,
       author = {{Song}, Jie and {Fang}, GuanWen and {Lin}, Zesen and {Gu}, Yizhou and {Kong}, Xu},
        title = "{Solution to the Conflict between the Estimations of Resolved and Unresolved Galaxy Stellar Mass from the Perspective of JWST}",
      journal = {\apj},
         year = 2023,
        month = nov,
       volume = {958},
       number = {1},
          eid = {82},
        pages = {82},
          doi = {10.3847/1538-4357/ad0365},
archivePrefix = {arXiv},
       eprint = {2310.12228},
 primaryClass = {astro-ph.GA},
       adsurl = {https://ui.adsabs.harvard.edu/abs/2023ApJ...958...82S}
}

@ARTICLE{2010A&A...523A..13P,
       author = {{Pozzetti}, L. and {Bolzonella}, M. and {Zucca}, E. and {Zamorani}, G. and {Lilly}, S. and {Renzini}, A. and {Moresco}, M. and {Mignoli}, M. and {Cassata}, P. and {Tasca}, L. and {Lamareille}, F. and {Maier}, C. and {Meneux}, B. and {Halliday}, C. and {Oesch}, P. and {Vergani}, D. and {Caputi}, K. and {Kova{\v{c}}}, K. and {Cimatti}, A. and {Cucciati}, O. and {Iovino}, A. and {Peng}, Y. and {Carollo}, M. and {Contini}, T. and {Kneib}, J. -P. and {Le F{\'e}vre}, O. and {Mainieri}, V. and {Scodeggio}, M. and {Bardelli}, S. and {Bongiorno}, A. and {Coppa}, G. and {de la Torre}, S. and {de Ravel}, L. and {Franzetti}, P. and {Garilli}, B. and {Kampczyk}, P. and {Knobel}, C. and {Le Borgne}, J. -F. and {Le Brun}, V. and {Pell{\`o}}, R. and {Perez Montero}, E. and {Ricciardelli}, E. and {Silverman}, J.~D. and {Tanaka}, M. and {Tresse}, L. and {Abbas}, U. and {Bottini}, D. and {Cappi}, A. and {Guzzo}, L. and {Koekemoer}, A.~M. and {Leauthaud}, A. and {Maccagni}, D. and {Marinoni}, C. and {McCracken}, H.~J. and {Memeo}, P. and {Porciani}, C. and {Scaramella}, R. and {Scarlata}, C. and {Scoville}, N.},
        title = "{zCOSMOS - 10k-bright spectroscopic sample. The bimodality in the galaxy stellar mass function: exploring its evolution with redshift}",
      journal = {\aap},
         year = 2010,
        month = nov,
       volume = {523},
          eid = {A13},
        pages = {A13},
          doi = {10.1051/0004-6361/200913020},
archivePrefix = {arXiv},
       eprint = {0907.5416},
 primaryClass = {astro-ph.CO},
       adsurl = {https://ui.adsabs.harvard.edu/abs/2010A&A...523A..13P}
}

@ARTICLE{2024A&A...691A.240M,
       author = {{Merlin}, E. and {Santini}, P. and {Paris}, D. and {Castellano}, M. and {Fontana}, A. and {Treu}, T. and {Finkelstein}, S.~L. and {Dunlop}, J.~S. and {Arrabal Haro}, P. and {Bagley}, M. and {Boyett}, K. and {Calabr{\`o}}, A. and {Correnti}, M. and {Davis}, K. and {Dickinson}, M. and {Donnan}, C.~T. and {Ferguson}, H.~C. and {Fortuni}, F. and {Giavalisco}, M. and {Glazebrook}, K. and {Grazian}, A. and {Grogin}, N.~A. and {Hathi}, N. and {Hirschmann}, M. and {Kartaltepe}, J.~S. and {Kewley}, L.~J. and {Kirkpatrick}, A. and {Kocevski}, D.~D. and {Koekemoer}, A.~M. and {Leung}, G. and {Lotz}, J.~M. and {Lucas}, R.~A. and {Magee}, D.~K. and {Marchesini}, D. and {Mascia}, S. and {McLeod}, D.~J. and {McLure}, R.~J. and {Nanayakkara}, T. and {Napolitano}, L. and {Nonino}, M. and {Papovich}, C. and {Pentericci}, L. and {P{\'e}rez-Gonz{\'a}lez}, P.~G. and {Pirzkal}, N. and {Ravindranath}, S. and {Roberts-Borsani}, G. and {Somerville}, R.~S. and {Trenti}, M. and {Trump}, J.~R. and {Vulcani}, B. and {Wang}, X. and {Watson}, P.~J. and {Wilkins}, S.~M. and {Yang}, G. and {Yung}, L.~Y.~A.},
        title = "{ASTRODEEP-JWST: NIRCam-HST multi-band photometry and redshifts for half a million sources in six extragalactic deep fields}",
      journal = {\aap},
         year = 2024,
        month = nov,
       volume = {691},
          eid = {A240},
        pages = {A240},
          doi = {10.1051/0004-6361/202451409},
archivePrefix = {arXiv},
       eprint = {2409.00169},
 primaryClass = {astro-ph.GA},
       adsurl = {https://ui.adsabs.harvard.edu/abs/2024A&A...691A.240M}
}

@software{larry_bradley_2024_13989456,
  author       = {Larry Bradley and
                  Brigitta Sip{\H o}cz and
                  Thomas Robitaille and
                  Erik Tollerud and
                  Z\`e Vin{\'{\i}}cius and
                  Christoph Deil and
                  Kyle Barbary and
                  Tom J Wilson and
                  Ivo Busko and
                  Axel Donath and
                  Hans Moritz G{\"u}nther and
                  Mihai Cara and
                  P. L. Lim and
                  Sebastian Me{\ss}linger and
                  Simon Conseil and
                  Zach Burnett and
                  Azalee Bostroem and
                  Michael Droettboom and
                  E. M. Bray and
                  Lars Andersen Bratholm and
                  Adam Ginsburg and
                  William Jamieson and
                  Geert Barentsen and
                  Matt Craig and
                  Brett M. Morris and
                  Marshall Perrin and
                  Shivangee Rathi and
                  Sergio Pascual and
                  Iskren Y. Georgiev},
  title        = {astropy/photutils: 2.0.2},
  month        = oct,
  year         = 2024,
  publisher    = {Zenodo},
  version      = {2.0.2},
  doi          = {10.5281/zenodo.13989456},
  url          = {https://doi.org/10.5281/zenodo.13989456},
}

@ARTICLE{2011PASP..123.1218A,
       author = {{Aniano}, G. and {Draine}, B.~T. and {Gordon}, K.~D. and {Sandstrom}, K.},
        title = "{Common-Resolution Convolution Kernels for Space- and Ground-Based Telescopes}",
      journal = {\pasp},
         year = 2011,
        month = oct,
       volume = {123},
       number = {908},
        pages = {1218},
          doi = {10.1086/662219},
archivePrefix = {arXiv},
       eprint = {1106.5065},
 primaryClass = {astro-ph.IM},
       adsurl = {https://ui.adsabs.harvard.edu/abs/2011PASP..123.1218A}
}

@ARTICLE{2024ApJ...977..165J,
       author = {{Jia}, Cheng and {Wang}, Enci and {Wang}, Huiyuan and {Li}, Hui and {Yao}, Yao and {Song}, Jie and {Zhang}, Hongxin and {Rong}, Yu and {Chen}, Yangyao and {Yu}, Haoran and {Chen}, Zeyu and {Li}, Haixin and {Ma}, Chengyu and {Kong}, Xu},
        title = "{Size Growth on Short Timescales of Star-forming Galaxies: Insights from Size Variation with Rest-frame Wavelength with JADES}",
      journal = {\apj},
         year = 2024,
        month = dec,
       volume = {977},
       number = {2},
          eid = {165},
        pages = {165},
          doi = {10.3847/1538-4357/ad919a},
archivePrefix = {arXiv},
       eprint = {2411.07458},
 primaryClass = {astro-ph.GA},
       adsurl = {https://ui.adsabs.harvard.edu/abs/2024ApJ...977..165J}
}

@ARTICLE{2002AJ....124..266P,
       author = {{Peng}, Chien Y. and {Ho}, Luis C. and {Impey}, Chris D. and {Rix}, Hans-Walter},
        title = "{Detailed Structural Decomposition of Galaxy Images}",
      journal = {\aj},
         year = 2002,
        month = jul,
       volume = {124},
       number = {1},
        pages = {266-293},
          doi = {10.1086/340952},
archivePrefix = {arXiv},
       eprint = {astro-ph/0204182},
 primaryClass = {astro-ph},
       adsurl = {https://ui.adsabs.harvard.edu/abs/2002AJ....124..266P}
}

@ARTICLE{2010AJ....139.2097P,
       author = {{Peng}, Chien Y. and {Ho}, Luis C. and {Impey}, Chris D. and {Rix}, Hans-Walter},
        title = "{Detailed Decomposition of Galaxy Images. II. Beyond Axisymmetric Models}",
      journal = {\aj},
         year = 2010,
        month = jun,
       volume = {139},
       number = {6},
        pages = {2097-2129},
          doi = {10.1088/0004-6256/139/6/2097},
archivePrefix = {arXiv},
       eprint = {0912.0731},
 primaryClass = {astro-ph.CO},
       adsurl = {https://ui.adsabs.harvard.edu/abs/2010AJ....139.2097P}
}

@ARTICLE{2023ApJ...954..113Y,
       author = {{Yao}, Yao and {Song}, Jie and {Kong}, Xu and {Fang}, Guanwen and {Zhang}, Hong-Xin and {Chen}, Xinkai},
        title = "{Evolution of Nonparametric Morphology of Galaxies in the JWST CEERS Field at z ≃ 0.8-3.0}",
      journal = {\apj},
         year = 2023,
        month = sep,
       volume = {954},
       number = {2},
          eid = {113},
        pages = {113},
          doi = {10.3847/1538-4357/ace7b5},
archivePrefix = {arXiv},
       eprint = {2307.13975},
 primaryClass = {astro-ph.GA},
       adsurl = {https://ui.adsabs.harvard.edu/abs/2023ApJ...954..113Y}
}

@ARTICLE{2003AJ....126.1183C,
       author = {{Conselice}, Christopher J. and {Bershady}, Matthew A. and {Dickinson}, Mark and {Papovich}, Casey},
        title = "{A Direct Measurement of Major Galaxy Mergers at z<\raisebox{-0.5ex}\textasciitilde3}",
      journal = {\aj},
         year = 2003,
        month = sep,
       volume = {126},
       number = {3},
        pages = {1183-1207},
          doi = {10.1086/377318},
archivePrefix = {arXiv},
       eprint = {astro-ph/0306106},
 primaryClass = {astro-ph},
       adsurl = {https://ui.adsabs.harvard.edu/abs/2003AJ....126.1183C}
}

@ARTICLE{2014ARA&A..52..291C,
       author = {{Conselice}, Christopher J.},
        title = "{The Evolution of Galaxy Structure Over Cosmic Time}",
      journal = {\araa},
         year = 2014,
        month = aug,
       volume = {52},
        pages = {291-337},
          doi = {10.1146/annurev-astro-081913-040037},
archivePrefix = {arXiv},
       eprint = {1403.2783},
 primaryClass = {astro-ph.GA},
       adsurl = {https://ui.adsabs.harvard.edu/abs/2014ARA&A..52..291C}
}

@ARTICLE{2014ApJS..214...15S,
       author = {{Speagle}, J.~S. and {Steinhardt}, C.~L. and {Capak}, P.~L. and {Silverman}, J.~D.},
        title = "{A Highly Consistent Framework for the Evolution of the Star-Forming ``Main Sequence'' from z \raisebox{-0.5ex}\textasciitilde 0-6}",
      journal = {\apjs},
         year = 2014,
        month = oct,
       volume = {214},
       number = {2},
          eid = {15},
        pages = {15},
          doi = {10.1088/0067-0049/214/2/15},
archivePrefix = {arXiv},
       eprint = {1405.2041},
 primaryClass = {astro-ph.GA},
       adsurl = {https://ui.adsabs.harvard.edu/abs/2014ApJS..214...15S}
}

@ARTICLE{2023MNRAS.519.1526P,
       author = {{Popesso}, P. and {Concas}, A. and {Cresci}, G. and {Belli}, S. and {Rodighiero}, G. and {Inami}, H. and {Dickinson}, M. and {Ilbert}, O. and {Pannella}, M. and {Elbaz}, D.},
        title = "{The main sequence of star-forming galaxies across cosmic times}",
      journal = {\mnras},
         year = 2023,
        month = feb,
       volume = {519},
       number = {1},
        pages = {1526-1544},
          doi = {10.1093/mnras/stac3214},
archivePrefix = {arXiv},
       eprint = {2203.10487},
 primaryClass = {astro-ph.GA},
       adsurl = {https://ui.adsabs.harvard.edu/abs/2023MNRAS.519.1526P}
}

@ARTICLE{2024ApJ...977..133C,
       author = {{Clarke}, Leonardo and {Shapley}, Alice E. and {Sanders}, Ryan L. and {Topping}, Michael W. and {Brammer}, Gabriel B. and {Bento}, Trinity and {Reddy}, Naveen A. and {Kehoe}, Emily},
        title = "{The Star-forming Main Sequence in JADES and CEERS at z > 1.4: Investigating the Burstiness of Star Formation}",
      journal = {\apj},
         year = 2024,
        month = dec,
       volume = {977},
       number = {1},
          eid = {133},
        pages = {133},
          doi = {10.3847/1538-4357/ad8ba4},
archivePrefix = {arXiv},
       eprint = {2406.05178},
 primaryClass = {astro-ph.GA},
       adsurl = {https://ui.adsabs.harvard.edu/abs/2024ApJ...977..133C}
}

@ARTICLE{2025ApJ...979..193C,
       author = {{Cole}, Justin W. and {Papovich}, Casey and {Finkelstein}, Steven L. and {Bagley}, Micaela B. and {Dickinson}, Mark and {Iyer}, Kartheik G. and {Yung}, L.~Y. Aaron and {Ciesla}, Laure and {Amor{\'\i}n}, Ricardo O. and {Arrabal Haro}, Pablo and {Bhatawdekar}, Rachana and {Calabr{\`o}}, Antonello and {Cleri}, Nikko J. and {de la Vega}, Alexander and {Dekel}, Avishai and {Endsley}, Ryan and {Gawiser}, Eric and {Giavalisco}, Mauro and {Hathi}, Nimish P. and {Hirschmann}, Michaela and {Holwerda}, Benne W. and {Kartaltepe}, Jeyhan S. and {Koekemoer}, Anton M. and {Lucas}, Ray A. and {Mascia}, Sara and {Mobasher}, Bahram and {P{\'e}rez-Gonz{\'a}lez}, Pablo G. and {Rodighiero}, Giulia and {Ronayne}, Kaila and {Tacchella}, Sandro and {Weiner}, Benjamin J. and {Wilkins}, Stephen M.},
        title = "{CEERS: Increasing Scatter along the Star-forming Main Sequence Indicates Early Galaxies Form in Bursts}",
      journal = {\apj},
         year = 2025,
        month = feb,
       volume = {979},
       number = {2},
          eid = {193},
        pages = {193},
          doi = {10.3847/1538-4357/ad9a6a},
archivePrefix = {arXiv},
       eprint = {2312.10152},
 primaryClass = {astro-ph.GA},
       adsurl = {https://ui.adsabs.harvard.edu/abs/2025ApJ...979..193C}
}

@ARTICLE{2015A&A...575A..74S,
       author = {{Schreiber}, C. and {Pannella}, M. and {Elbaz}, D. and {B{\'e}thermin}, M. and {Inami}, H. and {Dickinson}, M. and {Magnelli}, B. and {Wang}, T. and {Aussel}, H. and {Daddi}, E. and {Juneau}, S. and {Shu}, X. and {Sargent}, M.~T. and {Buat}, V. and {Faber}, S.~M. and {Ferguson}, H.~C. and {Giavalisco}, M. and {Koekemoer}, A.~M. and {Magdis}, G. and {Morrison}, G.~E. and {Papovich}, C. and {Santini}, P. and {Scott}, D.},
        title = "{The Herschel view of the dominant mode of galaxy growth from z = 4 to the present day}",
      journal = {\aap},
         year = 2015,
        month = mar,
       volume = {575},
          eid = {A74},
        pages = {A74},
          doi = {10.1051/0004-6361/201425017},
archivePrefix = {arXiv},
       eprint = {1409.5433},
 primaryClass = {astro-ph.GA},
       adsurl = {https://ui.adsabs.harvard.edu/abs/2015A&A...575A..74S}
}

@ARTICLE{2022A&A...661L...7D,
       author = {{Daddi}, E. and {Delvecchio}, I. and {Dimauro}, P. and {Magnelli}, B. and {Gomez-Guijarro}, C. and {Coogan}, R. and {Elbaz}, D. and {Kalita}, B.~S. and {Le Bail}, A. and {Rich}, R.~M. and {Tan}, Q.},
        title = "{The bending of the star-forming main sequence traces the cold- to hot-accretion transition mass over 0 < z < 4}",
      journal = {\aap},
         year = 2022,
        month = may,
       volume = {661},
          eid = {L7},
        pages = {L7},
          doi = {10.1051/0004-6361/202243574},
archivePrefix = {arXiv},
       eprint = {2203.10880},
 primaryClass = {astro-ph.CO},
       adsurl = {https://ui.adsabs.harvard.edu/abs/2022A&A...661L...7D}
}

@ARTICLE{2025arXiv250315314E,
       author = {{Euclid Collaboration} and {Enia}, A. and {Pozzetti}, L. and {Bolzonella}, M. and {Bisigello}, L. and {Hartley}, W.~G. and {Saulder}, C. and {Daddi}, E. and {Siudek}, M. and {Zamorani}, G. and {Cassata}, P. and {Gentile}, F. and {Wang}, L. and {Rodighiero}, G. and {Allevato}, V. and {Corcho-Caballero}, P. and {Dom{\'\i}nguez S{\'a}nchez}, H. and {Tortora}, C. and {Baes}, M. and {Abdurro'uf} and {Nersesian}, A. and {Spinoglio}, L. and {Schaye}, J. and {Ascasibar}, Y. and {Scott}, D. and {Duran-Camacho}, E. and {Quai}, S. and {Talia}, M. and {Mao}, Z. and {Aghanim}, N. and {Altieri}, B. and {Amara}, A. and {Andreon}, S. and {Auricchio}, N. and {Aussel}, H. and {Baccigalupi}, C. and {Baldi}, M. and {Balestra}, A. and {Bardelli}, S. and {Basset}, A. and {Battaglia}, P. and {Bender}, R. and {Biviano}, A. and {Bonchi}, A. and {Branchini}, E. and {Brescia}, M. and {Brinchmann}, J. and {Camera}, S. and {Ca{\~n}as-Herrera}, G. and {Capobianco}, V. and {Carbone}, C. and {Carretero}, J. and {Casas}, S. and {Castander}, F.~J. and {Castellano}, M. and {Castignani}, G. and {Cavuoti}, S. and {Chambers}, K.~C. and {Cimatti}, A. and {Colodro-Conde}, C. and {Congedo}, G. and {Conselice}, C.~J. and {Conversi}, L. and {Copin}, Y. and {Courbin}, F. and {Courtois}, H.~M. and {Cropper}, M. and {Da Silva}, A. and {Degaudenzi}, H. and {De Lucia}, G. and {Di Giorgio}, A.~M. and {Dolding}, C. and {Dole}, H. and {Dubath}, F. and {Duncan}, C.~A.~J. and {Dupac}, X. and {Dusini}, S. and {Ealet}, A. and {Escoffier}, S. and {Fabricius}, M. and {Farina}, M. and {Farinelli}, R. and {Faustini}, F. and {Ferriol}, S. and {Finelli}, F. and {Fotopoulou}, S. and {Frailis}, M. and {Franceschi}, E. and {Franzetti}, P. and {Galeotta}, S. and {George}, K. and {Gillis}, B. and {Giocoli}, C. and {G{\'o}mez-Alvarez}, P. and {Gracia-Carpio}, J. and {Granett}, B.~R. and {Grazian}, A. and {Grupp}, F. and {Guzzo}, L. and {Gwyn}, S. and {Haugan}, S.~V.~H. and {Hoar}, J. and {Holmes}, W. and {Hook}, I.~M. and {Hormuth}, F. and {Hornstrup}, A. and {Hudelot}, P. and {Jahnke}, K. and {Jhabvala}, M. and {Joachimi}, B. and {Keih{\"a}nen}, E. and {Kermiche}, S. and {Kiessling}, A. and {Kubik}, B. and {K{\"u}mmel}, M. and {Kunz}, M. and {Kurki-Suonio}, H. and {Le Boulc'h}, Q. and {Le Brun}, A.~M.~C. and {Le Mignant}, D. and {Ligori}, S. and {Lilje}, P.~B. and {Lindholm}, V. and {Lloro}, I. and {Mainetti}, G. and {Maino}, D. and {Maiorano}, E. and {Mansutti}, O. and {Marggraf}, O. and {Martinelli}, M. and {Martinet}, N. and {Marulli}, F. and {Massey}, R. and {Masters}, D.~C. and {Maurogordato}, S. and {Medinaceli}, E. and {Mei}, S. and {Melchior}, M. and {Mellier}, Y. and {Meneghetti}, M. and {Merlin}, E. and {Meylan}, G. and {Mora}, A. and {Moresco}, M. and {Moscardini}, L. and {Nakajima}, R. and {Neissner}, C. and {Niemi}, S.-M. and {Nightingale}, J.~W. and {Padilla}, C. and {Paltani}, S. and {Pasian}, F. and {Pedersen}, K. and {Percival}, W.~J. and {Pettorino}, V. and {Pires}, S. and {Polenta}, G. and {Poncet}, M. and {Popa}, L.~A. and {Raison}, F. and {Rebolo}, R. and {Renzi}, A. and {Rhodes}, J. and {Riccio}, G. and {Romelli}, E. and {Roncarelli}, M. and {Rossetti}, E. and {Rusholme}, B. and {Saglia}, R. and {Sakr}, Z. and {S{\'a}nchez}, A.~G. and {Sapone}, D. and {Sartoris}, B. and {Schewtschenko}, J.~A. and {Schirmer}, M. and {Schneider}, P. and {Schrabback}, T. and {Scodeggio}, M. and {Secroun}, A. and {Seidel}, G. and {Serrano}, S. and {Simon}, P. and {Sirignano}, C. and {Sirri}, G. and {Skottfelt}, J. and {Stanco}, L. and {Steinwagner}, J. and {Surace}, C. and {Tallada-Cresp{\'\i}}, P. and {Taylor}, A.~N. and {Teplitz}, H.~I. and {Tereno}, I. and {Toft}, S. and {Toledo-Moreo}, R. and {Torradeflot}, F. and {Tutusaus}, I. and {Valenziano}, L. and {Valiviita}, J. and {Vassallo}, T. and {Verdoes Kleijn}, G.},
        title = "{Euclid Quick Data Release (Q1). A first view of the star-forming main sequence in the Euclid Deep Fields}",
      journal = {arXiv e-prints},
         year = 2025,
        month = mar,
          eid = {arXiv:2503.15314},
        pages = {arXiv:2503.15314},
          doi = {10.48550/arXiv.2503.15314},
archivePrefix = {arXiv},
       eprint = {2503.15314},
 primaryClass = {astro-ph.GA},
       adsurl = {https://ui.adsabs.harvard.edu/abs/2025arXiv250315314E}
}

@ARTICLE{2024ApJ...961..163B,
       author = {{Bluck}, Asa F.~L. and {Conselice}, Christopher J. and {Ormerod}, Katherine and {Piotrowska}, Joanna M. and {Adams}, Nathan and {Austin}, Duncan and {Caruana}, Joseph and {Duncan}, K.~J. and {Ferreira}, Leonardo and {Goubert}, Paul and {Harvey}, Thomas and {Trussler}, James and {Maiolino}, Roberto},
        title = "{Galaxy Quenching at the High Redshift Frontier: A Fundamental Test of Cosmological Models in the Early Universe with JWST-CEERS}",
      journal = {\apj},
         year = 2024,
        month = feb,
       volume = {961},
       number = {2},
          eid = {163},
        pages = {163},
          doi = {10.3847/1538-4357/ad0a98},
archivePrefix = {arXiv},
       eprint = {2311.02526},
 primaryClass = {astro-ph.GA},
       adsurl = {https://ui.adsabs.harvard.edu/abs/2024ApJ...961..163B}
}

@ARTICLE{2024A&A...691A.164K,
       author = {{Koprowski}, M.~P. and {Wijesekera}, J.~V. and {Dunlop}, J.~S. and {McLeod}, D.~J. and {Micha{\l}owski}, M.~J. and {Lisiecki}, K. and {McLure}, R.~J.},
        title = "{Charting the main sequence of star-forming galaxies out to redshifts z {\ensuremath{\lesssim}} 5.7}",
      journal = {\aap},
         year = 2024,
        month = nov,
       volume = {691},
          eid = {A164},
        pages = {A164},
          doi = {10.1051/0004-6361/202449948},
archivePrefix = {arXiv},
       eprint = {2403.06575},
 primaryClass = {astro-ph.GA},
       adsurl = {https://ui.adsabs.harvard.edu/abs/2024A&A...691A.164K}
}

@ARTICLE{2014ApJ...788...28V,
       author = {{van der Wel}, A. and {Franx}, M. and {van Dokkum}, P.~G. and {Skelton}, R.~E. and {Momcheva}, I.~G. and {Whitaker}, K.~E. and {Brammer}, G.~B. and {Bell}, E.~F. and {Rix}, H. -W. and {Wuyts}, S. and {Ferguson}, H.~C. and {Holden}, B.~P. and {Barro}, G. and {Koekemoer}, A.~M. and {Chang}, Yu-Yen and {McGrath}, E.~J. and {H{\"a}ussler}, B. and {Dekel}, A. and {Behroozi}, P. and {Fumagalli}, M. and {Leja}, J. and {Lundgren}, B.~F. and {Maseda}, M.~V. and {Nelson}, E.~J. and {Wake}, D.~A. and {Patel}, S.~G. and {Labb{\'e}}, I. and {Faber}, S.~M. and {Grogin}, N.~A. and {Kocevski}, D.~D.},
        title = "{3D-HST+CANDELS: The Evolution of the Galaxy Size-Mass Distribution since z = 3}",
      journal = {\apj},
         year = 2014,
        month = jun,
       volume = {788},
       number = {1},
          eid = {28},
        pages = {28},
          doi = {10.1088/0004-637X/788/1/28},
archivePrefix = {arXiv},
       eprint = {1404.2844},
 primaryClass = {astro-ph.GA},
       adsurl = {https://ui.adsabs.harvard.edu/abs/2014ApJ...788...28V}
}

@ARTICLE{2023ApJ...950..130S,
       author = {{Song}, Jie and {Fang}, GuanWen and {Gu}, Yizhou and {Lin}, Zesen and {Kong}, Xu},
        title = "{The Effect of Environment on the Properties of the Most Massive Galaxies at 0.5 < z < 2.5 in the COSMOS-DASH Field}",
      journal = {\apj},
         year = 2023,
        month = jun,
       volume = {950},
       number = {2},
          eid = {130},
        pages = {130},
          doi = {10.3847/1538-4357/acd174},
archivePrefix = {arXiv},
       eprint = {2305.10677},
 primaryClass = {astro-ph.GA},
       adsurl = {https://ui.adsabs.harvard.edu/abs/2023ApJ...950..130S}
}

@ARTICLE{2020ApJ...905..170M,
       author = {{Mosleh}, Moein and {Hosseinnejad}, Shiva and {Hosseini-ShahiSavandi}, S. Zahra and {Tacchella}, Sandro},
        title = "{Galaxy Sizes Since z = 2 from the Perspective of Stellar Mass Distribution within Galaxies}",
      journal = {\apj},
         year = 2020,
        month = dec,
       volume = {905},
       number = {2},
          eid = {170},
        pages = {170},
          doi = {10.3847/1538-4357/abc7cc},
archivePrefix = {arXiv},
       eprint = {2011.04656},
 primaryClass = {astro-ph.GA},
       adsurl = {https://ui.adsabs.harvard.edu/abs/2020ApJ...905..170M}
}

@ARTICLE{2019ApJ...877..103S,
       author = {{Suess}, Katherine A. and {Kriek}, Mariska and {Price}, Sedona H. and {Barro}, Guillermo},
        title = "{Half-mass Radii for {\ensuremath{\sim}}7000 Galaxies at 1.0 {\ensuremath{\leq}} z {\ensuremath{\leq}} 2.5: Most of the Evolution in the Mass-Size Relation Is Due to Color Gradients}",
      journal = {\apj},
         year = 2019,
        month = jun,
       volume = {877},
       number = {2},
          eid = {103},
        pages = {103},
          doi = {10.3847/1538-4357/ab1bda},
archivePrefix = {arXiv},
       eprint = {1904.10992},
 primaryClass = {astro-ph.GA},
       adsurl = {https://ui.adsabs.harvard.edu/abs/2019ApJ...877..103S}
}

@ARTICLE{2023ApJ...945..155M,
       author = {{Miller}, Tim B. and {van Dokkum}, Pieter and {Mowla}, Lamiya},
        title = "{Color Gradients and Half-mass Radii of Galaxies Out to z = 2 in the CANDELS/3D-HST Fields: Further Evidence for Important Differences in the Evolution of Mass-weighted and Light-weighted Sizes}",
      journal = {\apj},
         year = 2023,
        month = mar,
       volume = {945},
       number = {2},
          eid = {155},
        pages = {155},
          doi = {10.3847/1538-4357/acbc74},
archivePrefix = {arXiv},
       eprint = {2207.05895},
 primaryClass = {astro-ph.GA},
       adsurl = {https://ui.adsabs.harvard.edu/abs/2023ApJ...945..155M}
}

@ARTICLE{2022ApJ...937L..33S,
       author = {{Suess}, Katherine A. and {Bezanson}, Rachel and {Nelson}, Erica J. and {Setton}, David J. and {Price}, Sedona H. and {van Dokkum}, Pieter and {Brammer}, Gabriel and {Labb{\'e}}, Ivo and {Leja}, Joel and {Miller}, Tim B. and {Robertson}, Brant and {Wel}, Arjen van der and {Weaver}, John R. and {Whitaker}, Katherine E.},
        title = "{Rest-frame Near-infrared Sizes of Galaxies at Cosmic Noon: Objects in JWST's Mirror Are Smaller than They Appeared}",
      journal = {\apjl},
         year = 2022,
        month = oct,
       volume = {937},
       number = {2},
          eid = {L33},
        pages = {L33},
          doi = {10.3847/2041-8213/ac8e06},
archivePrefix = {arXiv},
       eprint = {2207.10655},
 primaryClass = {astro-ph.GA},
       adsurl = {https://ui.adsabs.harvard.edu/abs/2022ApJ...937L..33S}
}

@ARTICLE{2024ApJ...960...53V,
       author = {{van der Wel}, Arjen and {Martorano}, Marco and {H{\"a}u{\ss}ler}, Boris and {Nedkova}, Kalina V. and {Miller}, Tim B. and {Brammer}, Gabriel B. and {van de Ven}, Glenn and {Leja}, Joel and {Bezanson}, Rachel S. and {Muzzin}, Adam and {Marchesini}, Danilo and {de Graaff}, Anna and {Nelson}, Erica J. and {Kriek}, Mariska and {Bell}, Eric F. and {Franx}, Marijn},
        title = "{Stellar Half-mass Radii of 0.5 z < 2.3 Galaxies: Comparison with JWST/NIRCam Half-light Radii}",
      journal = {\apj},
         year = 2024,
        month = jan,
       volume = {960},
       number = {1},
          eid = {53},
        pages = {53},
          doi = {10.3847/1538-4357/ad02ee},
archivePrefix = {arXiv},
       eprint = {2307.03264},
 primaryClass = {astro-ph.GA},
       adsurl = {https://ui.adsabs.harvard.edu/abs/2024ApJ...960...53V}
}

@ARTICLE{2019ApJ...880...57M,
       author = {{Mowla}, Lamiya A. and {van Dokkum}, Pieter and {Brammer}, Gabriel B. and {Momcheva}, Ivelina and {van der Wel}, Arjen and {Whitaker}, Katherine and {Nelson}, Erica and {Bezanson}, Rachel and {Muzzin}, Adam and {Franx}, Marijn and {MacKenty}, John and {Leja}, Joel and {Kriek}, Mariska and {Marchesini}, Danilo},
        title = "{COSMOS-DASH: The Evolution of the Galaxy Size-Mass Relation since z {\ensuremath{\sim}} 3 from New Wide-field WFC3 Imaging Combined with CANDELS/3D-HST}",
      journal = {\apj},
         year = 2019,
        month = jul,
       volume = {880},
       number = {1},
          eid = {57},
        pages = {57},
          doi = {10.3847/1538-4357/ab290a},
archivePrefix = {arXiv},
       eprint = {1808.04379},
 primaryClass = {astro-ph.GA},
       adsurl = {https://ui.adsabs.harvard.edu/abs/2019ApJ...880...57M}
}

@ARTICLE{2021MNRAS.506..928N,
       author = {{Nedkova}, Kalina V. and {H{\"a}u{\ss}ler}, Boris and {Marchesini}, Danilo and {Dimauro}, Paola and {Brammer}, Gabriel and {Eigenthaler}, Paul and {Feinstein}, Adina D. and {Ferguson}, Henry C. and {Huertas-Company}, Marc and {Johnston}, Evelyn J. and {Kado-Fong}, Erin and {Kartaltepe}, Jeyhan S. and {Labb{\'e}}, Ivo and {Lange-Vagle}, Daniel and {Martis}, Nicholas S. and {McGrath}, Elizabeth J. and {Muzzin}, Adam and {Oesch}, Pascal and {Ordenes-Brice{\~n}o}, Yasna and {Puzia}, Thomas and {Shipley}, Heath V. and {Simmons}, Brooke D. and {Skelton}, Rosalind E. and {Stefanon}, Mauro and {van der Wel}, Arjen and {Whitaker}, Katherine E.},
        title = "{Extending the evolution of the stellar mass-size relation at z {\ensuremath{\leq}} 2 to low stellar mass galaxies from HFF and CANDELS}",
      journal = {\mnras},
         year = 2021,
        month = sep,
       volume = {506},
       number = {1},
        pages = {928-956},
          doi = {10.1093/mnras/stab1744},
archivePrefix = {arXiv},
       eprint = {2106.07663},
 primaryClass = {astro-ph.GA},
       adsurl = {https://ui.adsabs.harvard.edu/abs/2021MNRAS.506..928N}
}

@ARTICLE{2024ApJ...962..176W,
       author = {{Ward}, Ethan and {de la Vega}, Alexander and {Mobasher}, Bahram and {McGrath}, Elizabeth J. and {Iyer}, Kartheik G. and {Calabr{\`o}}, Antonello and {Costantin}, Luca and {Dickinson}, Mark and {Holwerda}, Benne W. and {Huertas-Company}, Marc and {Hirschmann}, Michaela and {Lucas}, Ray A. and {Pandya}, Viraj and {Wilkins}, Stephen M. and {Yung}, L.~Y. Aaron and {Arrabal Haro}, Pablo and {Bagley}, Micaela B. and {Finkelstein}, Steven L. and {Kartaltepe}, Jeyhan S. and {Koekemoer}, Anton M. and {Papovich}, Casey and {Pirzkal}, Nor},
        title = "{Evolution of the Size{\textendash}Mass Relation of Star-forming Galaxies Since z = 5.5 Revealed by CEERS}",
      journal = {\apj},
         year = 2024,
        month = feb,
       volume = {962},
       number = {2},
          eid = {176},
        pages = {176},
          doi = {10.3847/1538-4357/ad20ed},
archivePrefix = {arXiv},
       eprint = {2311.02162},
 primaryClass = {astro-ph.GA},
       adsurl = {https://ui.adsabs.harvard.edu/abs/2024ApJ...962..176W}
}

@ARTICLE{2024MNRAS.527.6110O,
       author = {{Ormerod}, K. and {Conselice}, C.~J. and {Adams}, N.~J. and {Harvey}, T. and {Austin}, D. and {Trussler}, J. and {Ferreira}, L. and {Caruana}, J. and {Lucatelli}, G. and {Li}, Q. and {Roper}, W.~J.},
        title = "{EPOCHS VI: the size and shape evolution of galaxies since z   8 with JWST Observations}",
      journal = {\mnras},
         year = 2024,
        month = jan,
       volume = {527},
       number = {3},
        pages = {6110-6125},
          doi = {10.1093/mnras/stad3597},
archivePrefix = {arXiv},
       eprint = {2309.04377},
 primaryClass = {astro-ph.GA},
       adsurl = {https://ui.adsabs.harvard.edu/abs/2024MNRAS.527.6110O}
}

@ARTICLE{2015ApJS..219...15S,
       author = {{Shibuya}, Takatoshi and {Ouchi}, Masami and {Harikane}, Yuichi},
        title = "{Morphologies of {\ensuremath{\sim}}190,000 Galaxies at z = 0-10 Revealed with HST Legacy Data. I. Size Evolution}",
      journal = {\apjs},
         year = 2015,
        month = aug,
       volume = {219},
       number = {2},
          eid = {15},
        pages = {15},
          doi = {10.1088/0067-0049/219/2/15},
archivePrefix = {arXiv},
       eprint = {1503.07481},
 primaryClass = {astro-ph.GA},
       adsurl = {https://ui.adsabs.harvard.edu/abs/2015ApJS..219...15S}
}

@ARTICLE{2024MNRAS.533.3724V,
       author = {{Varadaraj}, R.~G. and {Bowler}, R.~A.~A. and {Jarvis}, M.~J. and {Adams}, N.~J. and {Choustikov}, N. and {Koekemoer}, A.~M. and {Carnall}, A.~C. and {McLeod}, D.~J. and {Dunlop}, J.~S. and {Donnan}, C.~T. and {Grogin}, N.~A.},
        title = "{The sizes of bright Lyman-break galaxies at z ≃ 3-5 with JWST PRIMER}",
      journal = {\mnras},
         year = 2024,
        month = sep,
       volume = {533},
       number = {3},
        pages = {3724-3741},
          doi = {10.1093/mnras/stae2022},
archivePrefix = {arXiv},
       eprint = {2401.15971},
 primaryClass = {astro-ph.GA},
       adsurl = {https://ui.adsabs.harvard.edu/abs/2024MNRAS.533.3724V}
}

@ARTICLE{2025arXiv250407185Y,
       author = {{Yang}, Lilan and {Kartaltepe}, Jeyhan S. and {Franco}, Maximilien and {Ding}, Xuheng and {Achenbach}, Mark J. and {Arango-Toro}, Rafael C. and {Casey}, Caitlin M. and {Drakos}, Nicole E. and {Faisst}, Andreas L. and {Gillman}, Steven and {Gozaliasl}, Ghassem and {Huertas-Company}, Marc and {Jin}, Shuowen and {Liu}, Daizhong and {Magdis}, Georgios and {Massey}, Richard and {Silverman}, John D. and {Tanaka}, Takumi S. and {Yu}, Si-Yue and {Akins}, Hollis B. and {Allen}, Natalie and {Ilbert}, Olivier and {Koekemoer}, Anton M. and {McCracken}, Henry Joy and {Paquereau}, Louise and {Rhodes}, Jason and {Robertson}, Brant E. and {Shuntov}, Marko and {Toft}, Sune},
        title = "{COSMOS-Web: Unraveling the Evolution of Galaxy Size and Related Properties at $2<z<10$}",
      journal = {arXiv e-prints},
         year = 2025,
        month = apr,
          eid = {arXiv:2504.07185},
        pages = {arXiv:2504.07185},
          doi = {10.48550/arXiv.2504.07185},
archivePrefix = {arXiv},
       eprint = {2504.07185},
 primaryClass = {astro-ph.GA},
       adsurl = {https://ui.adsabs.harvard.edu/abs/2025arXiv250407185Y}
}

@ARTICLE{2000MNRAS.319..168C,
       author = {{Cole}, Shaun and {Lacey}, Cedric G. and {Baugh}, Carlton M. and {Frenk}, Carlos S.},
        title = "{Hierarchical galaxy formation}",
      journal = {\mnras},
         year = 2000,
        month = nov,
       volume = {319},
       number = {1},
        pages = {168-204},
          doi = {10.1046/j.1365-8711.2000.03879.x},
archivePrefix = {arXiv},
       eprint = {astro-ph/0007281},
 primaryClass = {astro-ph},
       adsurl = {https://ui.adsabs.harvard.edu/abs/2000MNRAS.319..168C}
}

@ARTICLE{2024ARNPS..74..173P,
       author = {{Primack}, Joel R.},
        title = "{Galaxy Formation in {\ensuremath{\Lambda}}CDM Cosmology}",
      journal = {Annual Review of Nuclear and Particle Science},
         year = 2024,
        month = jun,
       volume = {74},
       number = {1},
        pages = {173-206},
          doi = {10.1146/annurev-nucl-102622-023052},
       adsurl = {https://ui.adsabs.harvard.edu/abs/2024ARNPS..74..173P}
}

@ARTICLE{2018ARA&A..56..435W,
       author = {{Wechsler}, Risa H. and {Tinker}, Jeremy L.},
        title = "{The Connection Between Galaxies and Their Dark Matter Halos}",
      journal = {\araa},
         year = 2018,
        month = sep,
       volume = {56},
        pages = {435-487},
          doi = {10.1146/annurev-astro-081817-051756},
archivePrefix = {arXiv},
       eprint = {1804.03097},
 primaryClass = {astro-ph.GA},
       adsurl = {https://ui.adsabs.harvard.edu/abs/2018ARA&A..56..435W}
}

@ARTICLE{2014A&ARv..22...71S,
       author = {{S{\'a}nchez Almeida}, Jorge and {Elmegreen}, Bruce G. and {Mu{\~n}oz-Tu{\~n}{\'o}n}, Casiana and {Elmegreen}, Debra Meloy},
        title = "{Star formation sustained by gas accretion}",
      journal = {\aapr},
         year = 2014,
        month = jul,
       volume = {22},
          eid = {71},
        pages = {71},
          doi = {10.1007/s00159-014-0071-1},
archivePrefix = {arXiv},
       eprint = {1405.3178},
 primaryClass = {astro-ph.GA},
       adsurl = {https://ui.adsabs.harvard.edu/abs/2014A&ARv..22...71S}
}

@INPROCEEDINGS{2017ASSL..430..145K,
       author = {{Kacprzak}, Glenn G.},
        title = "{Gas Accretion in Star-Forming Galaxies}",
    booktitle = {Gas Accretion onto Galaxies},
         year = 2017,
       editor = {{Fox}, Andrew and {Dav{\'e}}, Romeel},
       series = {Astrophysics and Space Science Library},
       volume = {430},
        month = jan,
        pages = {145},
          doi = {10.1007/978-3-319-52512-9_7},
archivePrefix = {arXiv},
       eprint = {1612.00451},
 primaryClass = {astro-ph.GA},
       adsurl = {https://ui.adsabs.harvard.edu/abs/2017ASSL..430..145K}
}

@ARTICLE{2022MNRAS.517L..92E,
       author = {{Ellison}, Sara L. and {Wilkinson}, Scott and {Woo}, Joanna and {Leung}, Ho-Hin and {Wild}, Vivienne and {Bickley}, Robert W. and {Patton}, David R. and {Quai}, Salvatore and {Gwyn}, Stephen},
        title = "{Galaxy mergers can rapidly shut down star formation}",
      journal = {\mnras},
         year = 2022,
        month = nov,
       volume = {517},
       number = {1},
        pages = {L92-L96},
          doi = {10.1093/mnrasl/slac109},
archivePrefix = {arXiv},
       eprint = {2209.07613},
 primaryClass = {astro-ph.GA},
       adsurl = {https://ui.adsabs.harvard.edu/abs/2022MNRAS.517L..92E}
}

@ARTICLE{2024OJAp....7E.121E,
       author = {{Ellison}, Sara and {Ferreira}, Leonardo and {Wild}, Vivienne and {Wilkinson}, Scott and {Rowlands}, Kate and {Patton}, David R.},
        title = "{Galaxy evolution in the post-merger regime. II {\textendash} Post-merger quenching peaks within 500 Myr of coalescence}",
      journal = {The Open Journal of Astrophysics},
         year = 2024,
        month = dec,
       volume = {7},
          eid = {121},
        pages = {121},
          doi = {10.33232/001c.127779},
archivePrefix = {arXiv},
       eprint = {2410.06357},
 primaryClass = {astro-ph.GA},
       adsurl = {https://ui.adsabs.harvard.edu/abs/2024OJAp....7E.121E}
}

@ARTICLE{2022MNRAS.515.1430D,
       author = {{Davies}, Jonathan J. and {Pontzen}, Andrew and {Crain}, Robert A.},
        title = "{Galaxy mergers can initiate quenching by unlocking an AGN-driven transformation of the baryon cycle}",
      journal = {\mnras},
         year = 2022,
        month = sep,
       volume = {515},
       number = {1},
        pages = {1430-1443},
          doi = {10.1093/mnras/stac1742},
archivePrefix = {arXiv},
       eprint = {2203.08157},
 primaryClass = {astro-ph.GA},
       adsurl = {https://ui.adsabs.harvard.edu/abs/2022MNRAS.515.1430D}
}

@ARTICLE{2014Natur.516...68G,
       author = {{Geach}, J.~E. and {Hickox}, R.~C. and {Diamond-Stanic}, A.~M. and {Krips}, M. and {Rudnick}, G.~H. and {Tremonti}, C.~A. and {Sell}, P.~H. and {Coil}, A.~L. and {Moustakas}, J.},
        title = "{Stellar feedback as the origin of an extended molecular outflow in a starburst galaxy}",
      journal = {\nat},
         year = 2014,
        month = dec,
       volume = {516},
       number = {7529},
        pages = {68-70},
          doi = {10.1038/nature14012},
archivePrefix = {arXiv},
       eprint = {1412.1091},
 primaryClass = {astro-ph.GA},
       adsurl = {https://ui.adsabs.harvard.edu/abs/2014Natur.516...68G}
}

@ARTICLE{2014ApJ...795..104W,
       author = {{Whitaker}, Katherine E. and {Franx}, Marijn and {Leja}, Joel and {van Dokkum}, Pieter G. and {Henry}, Alaina and {Skelton}, Rosalind E. and {Fumagalli}, Mattia and {Momcheva}, Ivelina G. and {Brammer}, Gabriel B. and {Labb{\'e}}, Ivo and {Nelson}, Erica J. and {Rigby}, Jane R.},
        title = "{Constraining the Low-mass Slope of the Star Formation Sequence at 0.5 < z < 2.5}",
      journal = {\apj},
         year = 2014,
        month = nov,
       volume = {795},
       number = {2},
          eid = {104},
        pages = {104},
          doi = {10.1088/0004-637X/795/2/104},
archivePrefix = {arXiv},
       eprint = {1407.1843},
 primaryClass = {astro-ph.GA},
       adsurl = {https://ui.adsabs.harvard.edu/abs/2014ApJ...795..104W}
}

@ARTICLE{2007ApJ...670..156D,
       author = {{Daddi}, E. and {Dickinson}, M. and {Morrison}, G. and {Chary}, R. and {Cimatti}, A. and {Elbaz}, D. and {Frayer}, D. and {Renzini}, A. and {Pope}, A. and {Alexander}, D.~M. and {Bauer}, F.~E. and {Giavalisco}, M. and {Huynh}, M. and {Kurk}, J. and {Mignoli}, M.},
        title = "{Multiwavelength Study of Massive Galaxies at z\raisebox{-0.5ex}\textasciitilde2. I. Star Formation and Galaxy Growth}",
      journal = {\apj},
         year = 2007,
        month = nov,
       volume = {670},
       number = {1},
        pages = {156-172},
          doi = {10.1086/521818},
archivePrefix = {arXiv},
       eprint = {0705.2831},
 primaryClass = {astro-ph},
       adsurl = {https://ui.adsabs.harvard.edu/abs/2007ApJ...670..156D}
}

@ARTICLE{2017ApJ...847...76S,
       author = {{Santini}, Paola and {Fontana}, Adriano and {Castellano}, Marco and {Di Criscienzo}, Marcella and {Merlin}, Emiliano and {Amorin}, Ricardo and {Cullen}, Fergus and {Daddi}, Emanuele and {Dickinson}, Mark and {Dunlop}, James S. and {Grazian}, Andrea and {Lamastra}, Alessandra and {McLure}, Ross J. and {Micha{\l}owski}, Micha{\l}. J. and {Pentericci}, Laura and {Shu}, Xinwen},
        title = "{The Star Formation Main Sequence in the Hubble Space Telescope Frontier Fields}",
      journal = {\apj},
         year = 2017,
        month = sep,
       volume = {847},
       number = {1},
          eid = {76},
        pages = {76},
          doi = {10.3847/1538-4357/aa8874},
archivePrefix = {arXiv},
       eprint = {1706.07059},
 primaryClass = {astro-ph.GA},
       adsurl = {https://ui.adsabs.harvard.edu/abs/2017ApJ...847...76S}
}

@ARTICLE{2025ApJ...981..161R,
       author = {{Rinaldi}, P. and {Navarro-Carrera}, R. and {Caputi}, K.~I. and {Iani}, E. and {{\"O}stlin}, G. and {Colina}, L. and {Alberts}, S. and {{\'A}lvarez-M{\'a}rquez}, J. and {Annunziatella}, M. and {Boogaard}, L. and {Costantin}, L. and {Hjorth}, J. and {Langeroodi}, D. and {Melinder}, J. and {Moutard}, T. and {Walter}, F.},
        title = "{The Emergence of the Star Formation Main Sequence with Redshift Unfolded by JWST}",
      journal = {\apj},
         year = 2025,
        month = mar,
       volume = {981},
       number = {2},
          eid = {161},
        pages = {161},
          doi = {10.3847/1538-4357/adb309},
archivePrefix = {arXiv},
       eprint = {2406.13554},
 primaryClass = {astro-ph.GA},
       adsurl = {https://ui.adsabs.harvard.edu/abs/2025ApJ...981..161R}
}

@ARTICLE{2012ApJ...754L..29W,
       author = {{Whitaker}, Katherine E. and {van Dokkum}, Pieter G. and {Brammer}, Gabriel and {Franx}, Marijn},
        title = "{The Star Formation Mass Sequence Out to z = 2.5}",
      journal = {\apjl},
         year = 2012,
        month = aug,
       volume = {754},
       number = {2},
          eid = {L29},
        pages = {L29},
          doi = {10.1088/2041-8205/754/2/L29},
archivePrefix = {arXiv},
       eprint = {1205.0547},
 primaryClass = {astro-ph.CO},
       adsurl = {https://ui.adsabs.harvard.edu/abs/2012ApJ...754L..29W}
}

@ARTICLE{2020MNRAS.497..698T,
       author = {{Tacchella}, Sandro and {Forbes}, John C. and {Caplar}, Neven},
        title = "{Stochastic modelling of star-formation histories II: star-formation variability from molecular clouds and gas inflow}",
      journal = {\mnras},
         year = 2020,
        month = sep,
       volume = {497},
       number = {1},
        pages = {698-725},
          doi = {10.1093/mnras/staa1838},
archivePrefix = {arXiv},
       eprint = {2006.09382},
 primaryClass = {astro-ph.GA},
       adsurl = {https://ui.adsabs.harvard.edu/abs/2020MNRAS.497..698T}
}

@ARTICLE{2010ApJ...713..686D,
       author = {{Daddi}, E. and {Bournaud}, F. and {Walter}, F. and {Dannerbauer}, H. and {Carilli}, C.~L. and {Dickinson}, M. and {Elbaz}, D. and {Morrison}, G.~E. and {Riechers}, D. and {Onodera}, M. and {Salmi}, F. and {Krips}, M. and {Stern}, D.},
        title = "{Very High Gas Fractions and Extended Gas Reservoirs in z = 1.5 Disk Galaxies}",
      journal = {\apj},
         year = 2010,
        month = apr,
       volume = {713},
       number = {1},
        pages = {686-707},
          doi = {10.1088/0004-637X/713/1/686},
archivePrefix = {arXiv},
       eprint = {0911.2776},
 primaryClass = {astro-ph.CO},
       adsurl = {https://ui.adsabs.harvard.edu/abs/2010ApJ...713..686D}
}

@ARTICLE{2010MNRAS.407.2091G,
       author = {{Genzel}, R. and {Tacconi}, L.~J. and {Gracia-Carpio}, J. and {Sternberg}, A. and {Cooper}, M.~C. and {Shapiro}, K. and {Bolatto}, A. and {Bouch{\'e}}, N. and {Bournaud}, F. and {Burkert}, A. and {Combes}, F. and {Comerford}, J. and {Cox}, P. and {Davis}, M. and {F{\"o}rster Schreiber}, N.~M. and {Garcia-Burillo}, S. and {Lutz}, D. and {Naab}, T. and {Neri}, R. and {Omont}, A. and {Shapley}, A. and {Weiner}, B.},
        title = "{A study of the gas-star formation relation over cosmic time}",
      journal = {\mnras},
         year = 2010,
        month = oct,
       volume = {407},
       number = {4},
        pages = {2091-2108},
          doi = {10.1111/j.1365-2966.2010.16969.x},
archivePrefix = {arXiv},
       eprint = {1003.5180},
 primaryClass = {astro-ph.CO},
       adsurl = {https://ui.adsabs.harvard.edu/abs/2010MNRAS.407.2091G}
}

@ARTICLE{2018MNRAS.479.5083A,
       author = {{Abdurro'uf} and {Akiyama}, Masayuki},
        title = "{Evolution of spatially resolved star formation main sequence and surface density profiles in massive disc galaxies at 0 {\ensuremath{\lesssim}} z {\ensuremath{\lesssim}} 1: inside-out stellar mass buildup and quenching}",
      journal = {\mnras},
         year = 2018,
        month = oct,
       volume = {479},
       number = {4},
        pages = {5083-5100},
          doi = {10.1093/mnras/sty1771},
archivePrefix = {arXiv},
       eprint = {1802.03782},
 primaryClass = {astro-ph.GA},
       adsurl = {https://ui.adsabs.harvard.edu/abs/2018MNRAS.479.5083A}
}

@ARTICLE{2023ApJ...945..117A,
       author = {{Abdurro'uf} and {Coe}, Dan and {Jung}, Intae and {Ferguson}, Henry C. and {Brammer}, Gabriel and {Iyer}, Kartheik G. and {Bradley}, Larry D. and {Dayal}, Pratika and {Windhorst}, Rogier A. and {Zitrin}, Adi and {Meena}, Ashish Kumar and {Oguri}, Masamune and {Diego}, Jose M. and {Kokorev}, Vasily and {Dimauro}, Paola and {Adamo}, Angela and {Conselice}, Christopher J. and {Welch}, Brian and {Vanzella}, Eros and {Hsiao}, Tiger Yu-Yang and {Xu}, Xinfeng and {Roy}, Namrata and {Mulcahey}, Celia R.},
        title = "{Spatially Resolved Stellar Populations of 0.3 < z < 6.0 Galaxies in WHL 0137-08 and MACS 0647+70 Clusters as Revealed by JWST: How Do Galaxies Grow and Quench over Cosmic Time?}",
      journal = {\apj},
         year = 2023,
        month = mar,
       volume = {945},
       number = {2},
          eid = {117},
        pages = {117},
          doi = {10.3847/1538-4357/acba06},
archivePrefix = {arXiv},
       eprint = {2301.02209},
 primaryClass = {astro-ph.GA},
       adsurl = {https://ui.adsabs.harvard.edu/abs/2023ApJ...945..117A}
}

@ARTICLE{2012ApJ...747L..28N,
       author = {{Nelson}, Erica June and {van Dokkum}, Pieter G. and {Brammer}, Gabriel and {F{\"o}rster Schreiber}, Natascha and {Franx}, Marijn and {Fumagalli}, Mattia and {Patel}, Shannon and {Rix}, Hans-Walter and {Skelton}, Rosalind E. and {Bezanson}, Rachel and {Da Cunha}, Elisabete and {Kriek}, Mariska and {Labbe}, Ivo and {Lundgren}, Britt and {Quadri}, Ryan and {Schmidt}, Kasper B.},
        title = "{Spatially Resolved H{\ensuremath{\alpha}} Maps and Sizes of 57 Strongly Star-forming Galaxies at z \raisebox{-0.5ex}\textasciitilde 1 from 3D-HST: Evidence for Rapid Inside-out Assembly of Disk Galaxies}",
      journal = {\apjl},
         year = 2012,
        month = mar,
       volume = {747},
       number = {2},
          eid = {L28},
        pages = {L28},
          doi = {10.1088/2041-8205/747/2/L28},
archivePrefix = {arXiv},
       eprint = {1202.1822},
 primaryClass = {astro-ph.CO},
       adsurl = {https://ui.adsabs.harvard.edu/abs/2012ApJ...747L..28N}
}

@ARTICLE{2016ApJ...828...27N,
       author = {{Nelson}, Erica June and {van Dokkum}, Pieter G. and {F{\"o}rster Schreiber}, Natascha M. and {Franx}, Marijn and {Brammer}, Gabriel B. and {Momcheva}, Ivelina G. and {Wuyts}, Stijn and {Whitaker}, Katherine E. and {Skelton}, Rosalind E. and {Fumagalli}, Mattia and {Hayward}, Christopher C. and {Kriek}, Mariska and {Labb{\'e}}, Ivo and {Leja}, Joel and {Rix}, Hans-Walter and {Tacconi}, Linda J. and {van der Wel}, Arjen and {van den Bosch}, Frank C. and {Oesch}, Pascal A. and {Dickey}, Claire and {Ulf Lange}, Johannes},
        title = "{Where Stars Form: Inside-out Growth and Coherent Star Formation from HST H{\ensuremath{\alpha}} Maps of 3200 Galaxies across the Main Sequence at 0.7 < z < 1.5}",
      journal = {\apj},
         year = 2016,
        month = sep,
       volume = {828},
       number = {1},
          eid = {27},
        pages = {27},
          doi = {10.3847/0004-637X/828/1/27},
archivePrefix = {arXiv},
       eprint = {1507.03999},
 primaryClass = {astro-ph.GA},
       adsurl = {https://ui.adsabs.harvard.edu/abs/2016ApJ...828...27N}
}

@ARTICLE{2020ARA&A..58...99S,
       author = {{S{\'a}nchez}, Sebasti{\'a}n F.},
        title = "{Spatially Resolved Spectroscopic Properties of Low-Redshift Star-Forming Galaxies}",
      journal = {\araa},
         year = 2020,
        month = aug,
       volume = {58},
        pages = {99-155},
          doi = {10.1146/annurev-astro-012120-013326},
archivePrefix = {arXiv},
       eprint = {1911.06925},
 primaryClass = {astro-ph.GA},
       adsurl = {https://ui.adsabs.harvard.edu/abs/2020ARA&A..58...99S}
}

@ARTICLE{2022MNRAS.510.3622B,
       author = {{Baker}, William M. and {Maiolino}, Roberto and {Bluck}, Asa F.~L. and {Lin}, Lihwai and {Ellison}, Sara L. and {Belfiore}, Francesco and {Pan}, Hsi-An and {Thorp}, Mallory},
        title = "{The ALMaQUEST survey IX: the nature of the resolved star forming main sequence}",
      journal = {\mnras},
         year = 2022,
        month = mar,
       volume = {510},
       number = {3},
        pages = {3622-3628},
          doi = {10.1093/mnras/stab3672},
archivePrefix = {arXiv},
       eprint = {2201.03592},
 primaryClass = {astro-ph.GA},
       adsurl = {https://ui.adsabs.harvard.edu/abs/2022MNRAS.510.3622B}
}

@ARTICLE{2015Sci...348..314T,
       author = {{Tacchella}, S. and {Carollo}, C.~M. and {Renzini}, A. and {F{\"o}rster Schreiber}, N.~M. and {Lang}, P. and {Wuyts}, S. and {Cresci}, G. and {Dekel}, A. and {Genzel}, R. and {Lilly}, S.~J. and {Mancini}, C. and {Newman}, S. and {Onodera}, M. and {Shapley}, A. and {Tacconi}, L. and {Woo}, J. and {Zamorani}, G.},
        title = "{Evidence for mature bulges and an inside-out quenching phase 3 billion years after the Big Bang}",
      journal = {Science},
         year = 2015,
        month = apr,
       volume = {348},
       number = {6232},
        pages = {314-317},
          doi = {10.1126/science.1261094},
archivePrefix = {arXiv},
       eprint = {1504.04021},
 primaryClass = {astro-ph.GA},
       adsurl = {https://ui.adsabs.harvard.edu/abs/2015Sci...348..314T}
}

@ARTICLE{2012ApJ...753..114W,
       author = {{Wuyts}, Stijn and {F{\"o}rster Schreiber}, Natascha M. and {Genzel}, Reinhard and {Guo}, Yicheng and {Barro}, Guillermo and {Bell}, Eric F. and {Dekel}, Avishai and {Faber}, Sandra M. and {Ferguson}, Henry C. and {Giavalisco}, Mauro and {Grogin}, Norman A. and {Hathi}, Nimish P. and {Huang}, Kuang-Han and {Kocevski}, Dale D. and {Koekemoer}, Anton M. and {Koo}, David C. and {Lotz}, Jennifer and {Lutz}, Dieter and {McGrath}, Elizabeth and {Newman}, Jeffrey A. and {Rosario}, David and {Saintonge}, Amelie and {Tacconi}, Linda J. and {Weiner}, Benjamin J. and {van der Wel}, Arjen},
        title = "{Smooth(er) Stellar Mass Maps in CANDELS: Constraints on the Longevity of Clumps in High-redshift Star-forming Galaxies}",
      journal = {\apj},
         year = 2012,
        month = jul,
       volume = {753},
       number = {2},
          eid = {114},
        pages = {114},
          doi = {10.1088/0004-637X/753/2/114},
archivePrefix = {arXiv},
       eprint = {1203.2611},
 primaryClass = {astro-ph.CO},
       adsurl = {https://ui.adsabs.harvard.edu/abs/2012ApJ...753..114W}
}

@ARTICLE{2010ApJ...709..644I,
       author = {{Ilbert}, O. and {Salvato}, M. and {Le Floc'h}, E. and {Aussel}, H. and {Capak}, P. and {McCracken}, H.~J. and {Mobasher}, B. and {Kartaltepe}, J. and {Scoville}, N. and {Sanders}, D.~B. and {Arnouts}, S. and {Bundy}, K. and {Cassata}, P. and {Kneib}, J. -P. and {Koekemoer}, A. and {Le F{\`e}vre}, O. and {Lilly}, S. and {Surace}, J. and {Taniguchi}, Y. and {Tasca}, L. and {Thompson}, D. and {Tresse}, L. and {Zamojski}, M. and {Zamorani}, G. and {Zucca}, E.},
        title = "{Galaxy Stellar Mass Assembly Between 0.2 < z < 2 from the S-COSMOS Survey}",
      journal = {\apj},
         year = 2010,
        month = feb,
       volume = {709},
       number = {2},
        pages = {644-663},
          doi = {10.1088/0004-637X/709/2/644},
archivePrefix = {arXiv},
       eprint = {0903.0102},
 primaryClass = {astro-ph.CO},
       adsurl = {https://ui.adsabs.harvard.edu/abs/2010ApJ...709..644I}
}

@ARTICLE{2025arXiv250314591M,
       author = {{Mosleh}, Moein and {Riahi-Zamin}, Mohammad and {Tacchella}, Sandro},
        title = "{Reconstructing Star Formation Histories of High-Redshift Galaxies: A Comparison of Resolved Parametric and Non-Parametric Models}",
      journal = {arXiv e-prints},
         year = 2025,
        month = mar,
          eid = {arXiv:2503.14591},
        pages = {arXiv:2503.14591},
          doi = {10.48550/arXiv.2503.14591},
archivePrefix = {arXiv},
       eprint = {2503.14591},
 primaryClass = {astro-ph.GA},
       adsurl = {https://ui.adsabs.harvard.edu/abs/2025arXiv250314591M}
}

@ARTICLE{2025arXiv250405244H,
       author = {{Harvey}, Thomas and {Conselice}, Christopher J. and {Adams}, Nathan J. and {Austin}, Duncan and {Li}, Qiong and {Rusakov}, Vadim and {Westcott}, Lewi and {Goolsby}, Caio M. and {Lovell}, Christopher C. and {Cochrane}, Rachel K. and {Vijayan}, Aswin P. and {Trussler}, James},
        title = "{Behind the spotlight: a systematic assessment of outshining using NIRCam medium bands in the JADES Origins Field}",
      journal = {\mnras},
         year = 2025,
        month = oct,
       volume = {542},
       number = {4},
        pages = {2998-3027},
          doi = {10.1093/mnras/staf1396},
archivePrefix = {arXiv},
       eprint = {2504.05244},
 primaryClass = {astro-ph.GA},
       adsurl = {https://ui.adsabs.harvard.edu/abs/2025MNRAS.542.2998H}
}

@ARTICLE{2024A&A...686A..63G,
       author = {{Gim{\'e}nez-Arteaga}, C. and {Fujimoto}, S. and {Valentino}, F. and {Brammer}, G.~B. and {Mason}, C.~A. and {Rizzo}, F. and {Rusakov}, V. and {Colina}, L. and {Prieto-Lyon}, G. and {Oesch}, P.~A. and {Espada}, D. and {Heintz}, K.~E. and {Knudsen}, K.~K. and {Dessauges-Zavadsky}, M. and {Laporte}, N. and {Lee}, M. and {Magdis}, G.~E. and {Ono}, Y. and {Ao}, Y. and {Ouchi}, M. and {Kohno}, K. and {Koekemoer}, A.~M.},
        title = "{Outshining in the spatially resolved analysis of a strongly lensed galaxy at z = 6.072 with JWST NIRCam}",
      journal = {\aap},
         year = 2024,
        month = jun,
       volume = {686},
          eid = {A63},
        pages = {A63},
          doi = {10.1051/0004-6361/202349135},
archivePrefix = {arXiv},
       eprint = {2402.17875},
 primaryClass = {astro-ph.GA},
       adsurl = {https://ui.adsabs.harvard.edu/abs/2024A&A...686A..63G}
}

@ARTICLE{2023ApJ...948..126G,
       author = {{Gim{\'e}nez-Arteaga}, Clara and {Oesch}, Pascal A. and {Brammer}, Gabriel B. and {Valentino}, Francesco and {Mason}, Charlotte A. and {Weibel}, Andrea and {Barrufet}, Laia and {Fujimoto}, Seiji and {Heintz}, Kasper E. and {Nelson}, Erica J. and {Strait}, Victoria B. and {Suess}, Katherine A. and {Gibson}, Justus},
        title = "{Spatially Resolved Properties of Galaxies at 5 < z < 9 in the SMACS 0723 JWST ERO Field}",
      journal = {\apj},
         year = 2023,
        month = may,
       volume = {948},
       number = {2},
          eid = {126},
        pages = {126},
          doi = {10.3847/1538-4357/acc5ea},
archivePrefix = {arXiv},
       eprint = {2212.08670},
 primaryClass = {astro-ph.GA},
       adsurl = {https://ui.adsabs.harvard.edu/abs/2023ApJ...948..126G}
}

\appendix
\nolinenumbers

\section{Compare with mock data} \label{sec:7.1}

\begin{figure*}[htb!]
\centering
\includegraphics[width=1.0\textwidth]{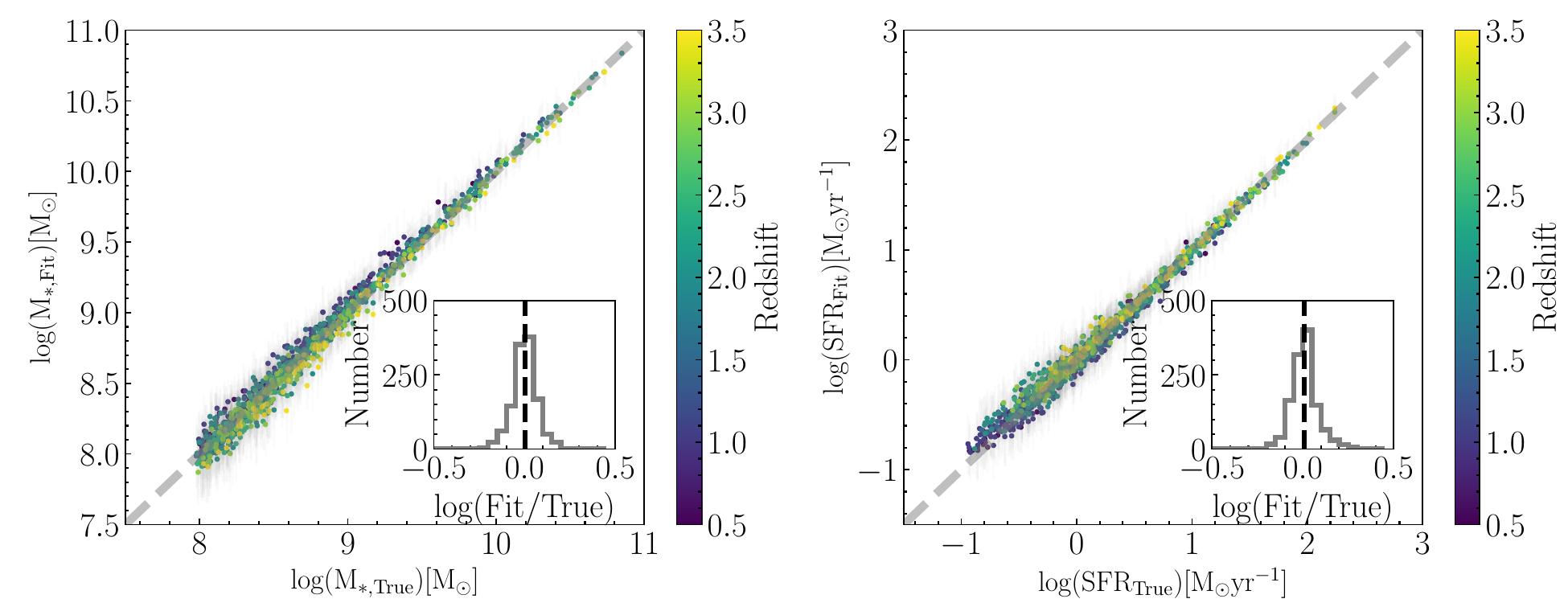}
\caption{Comparison between the best-fit parameters derived from SED-fitting tests using mock SEDs and the corresponding input (ground truth) values. Data points are color-coded by redshift. The light gray error bars represent the uncertainties from the SED fitting process. The left panel shows results for stellar mass, while the right panel displays results for SFR. Insets in each panel present histograms of the logarithmic differences between the recovered and input values, with black dashed lines indicating the median offsets.}
\label{fig12}
\end{figure*} 

Previous studies have emphasized the importance of broad wavelength coverage for accurately recovering galaxy physical properties such as stellar mass and SFR (e.g., \citealt{2010ApJ...709..644I, 2023ApJ...958...82S}). However, recent works by \cite{2023ApJ...945..117A} and \cite{2025ApJ...978L..42C} have demonstrated that combining JWST and HST observations can yield reliable estimates of these properties out to $z\sim 6$. In this section, we assess the robustness of our SED-fitting methodology using mock data. Specifically, we construct a suite of mock SEDs with the CIGALE code, adopting the same set of stellar population templates as described in Section \ref{sec:3.1}. The mock SEDs are generated at seven discrete redshifts, ranging from $z = 0.5$ to $z = 3.5$ in intervals of 0.5, to systematically test the performance of our method across cosmic time. To ensure realistic mock catalogs, we draw galaxy samples that reproduce the stellar mass and SFR distributions observed in real galaxies at each redshift, thereby avoiding unphysical parameter configurations.

We construct the mock SEDs using ten filters consistent with those employed in the JADES-GDS field. To emulate realistic observational conditions, Gaussian noise is added to the model fluxes. Given that the median flux uncertainties vary across filters in real observations, we set the standard deviation of the Gaussian noise in each band to match the corresponding median flux uncertainty measured in the JADES-GDS field. We have verified that adopting noise properties from other fields does not significantly affect the results. Applying the same selection criteria described in Section \ref{sec:2.3}, we randomly select 2,000 mock galaxies that meet the S/N threshold of 3 in at least six JWST bands. These mock SEDs are then processed using the same SED-fitting procedure as outlined in Section \ref{sec:3.1}.

In Figure \ref{fig12}, we compare the recovered physical properties from our SED fitting with the true input values used to generate the mock SEDs. Each data point is color-coded by redshift, while the light gray error bars represent the uncertainties from the SED fitting process. The gray dashed line denotes the one-to-one relation. Insets in each panel display the distribution of residuals, with the vertical black dashed line indicating the median offset.

Overall, our SED-fitting procedure successfully recovers the intrinsic physical parameters. The median offsets between the true and recovered values of stellar mass and SFR are all within 0.05 dex, with corresponding scatters below 0.1 dex. These results confirm that the combination of JWST and HST photometry yields robust constraints on both stellar mass and SFR. Although it is well established that assumptions in stellar population models and star formation histories can introduce systematic uncertainties in derived parameters, our primary objective is to obtain internally consistent measurements of both integrated and spatially resolved galaxy properties. Therefore, the model-dependent systematics is beyond the scope of consideration of this work.

\section{The importance of F090W and F410M data} \label{sec:7.2}

\begin{figure*}[htb!]
\centering
\includegraphics[width=1.0\textwidth]{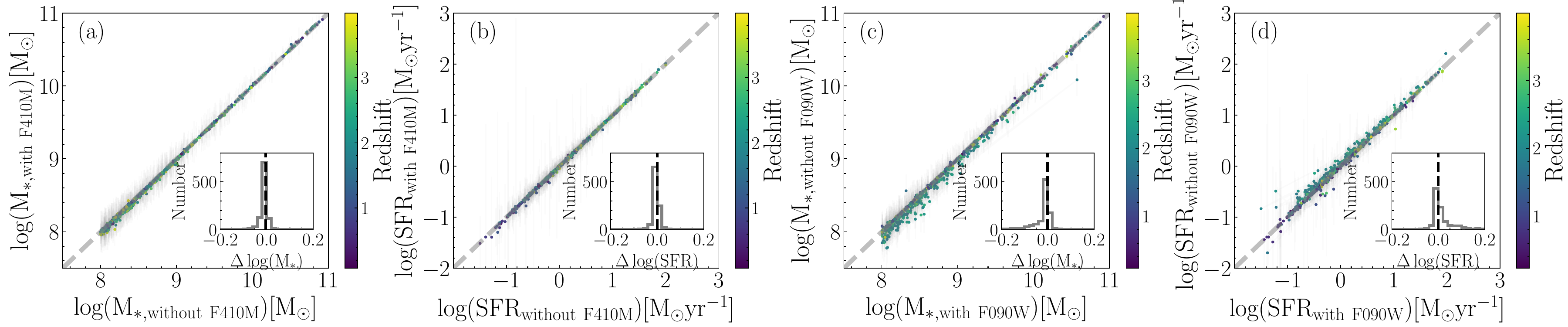}
\caption{Similar to Figure \ref{fig12}, but comparing the best-fit parameters derived from SED fitting using different filter configurations against those obtained with the full 10-filter set from the JADES-GDS field. Panels (a) and (b) show results with the F410M band included, while panels (c) and (d) show results with the F090W band excluded. The addition of F410M has a negligible impact on the derived stellar masses and SFRs. In contrast, the absence of F090W leads to a slight underestimation of stellar mass and a mild overestimation of SFR. Nonetheless, the median offsets remain small, suggesting that robustness of the derived physical parameters.}
\label{fig13}
\end{figure*} 

Although some medium-band filters, such as F410M, are available in our field, we excluded them from the primary analysis due to their relatively shallow depth. However, recent studies have highlighted the importance of medium-band data in accurately constraining galaxy physical properties, particularly at high redshifts where strong emission lines can significantly influence broadband photometry (e.g., \citealt{2024ApJ...974...42D, 2025arXiv250405244H}). To evaluate the potential impact of including medium-band data, we conduct a test using the JADES-GDS dataset, focusing on the effect of incorporating the F410M filter into our SED fitting. We re-fit the physical properties of our selected high-quality sample after adding F410M data using the same methodology outlined in Section \ref{sec:3.1}. The results are presented in Figure \ref{fig13}, panels (a) and (b), in a format consistent with Figure \ref{fig12}. The x-axes correspond to results obtained without the F410M band, while the y-axes show the results with F410M included. 

As illustrated in the figure, the inclusion of F410M has a negligible effect on the derived physical parameters. It is worth noting that the F410M band was included in the photometric redshift measurement step, as medium-band data are known to significantly improve redshift accuracy (e.g., \citealt{2024ApJ...974...42D}). However, our results in Figure \ref{fig13} suggest that, once the redshift is fixed, the inclusion or exclusion of F410M has minimal impact on the estimation of galaxy physical properties.

In addition, we note that the CEERS field lacks imaging in the F090W band, unlike the other fields used in our analysis. Although CEERS data are not included in our primary results, we have confirmed that their inclusion does not substantially affect our conclusions. To specifically evaluate the impact of the missing F090W data on the estimation of galaxy physical properties, we conduct a controlled test using the JADES-GDS dataset, following the same methodology applied to assess the influence of F410M.  We re-fit the physical properties of our selected high-quality sample when missing F090W data. The results are shown in panels (c) and (d) of Figure \ref{fig13}. As demonstrated, the absence of F090W leads to a slight underestimation of stellar mass and a mild overestimation of the SFR. However, the median offsets are close to zero, and the scatter remains below 0.1 dex. These findings indicate that, even without F090W coverage, the derived physical properties remain robust and within acceptable uncertainties.

\section{PSF smearing effect} \label{sec:7.3}

\begin{figure*}[htb!]
\centering
\includegraphics[width=1.0\textwidth]{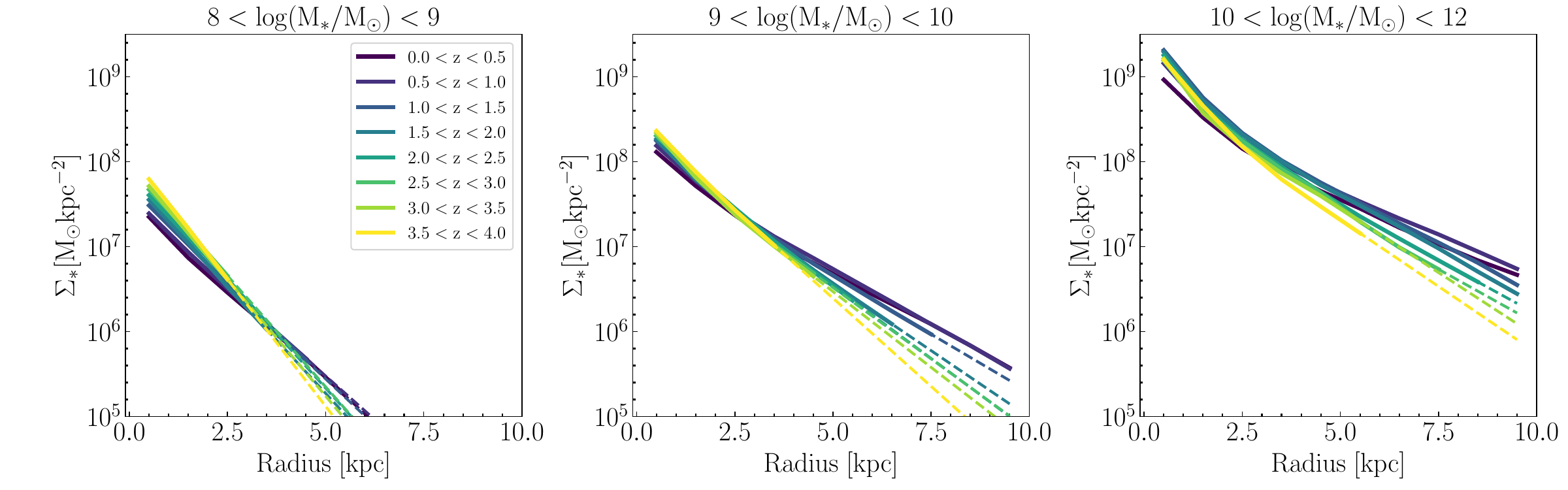}\\
\includegraphics[width=1.0\textwidth]{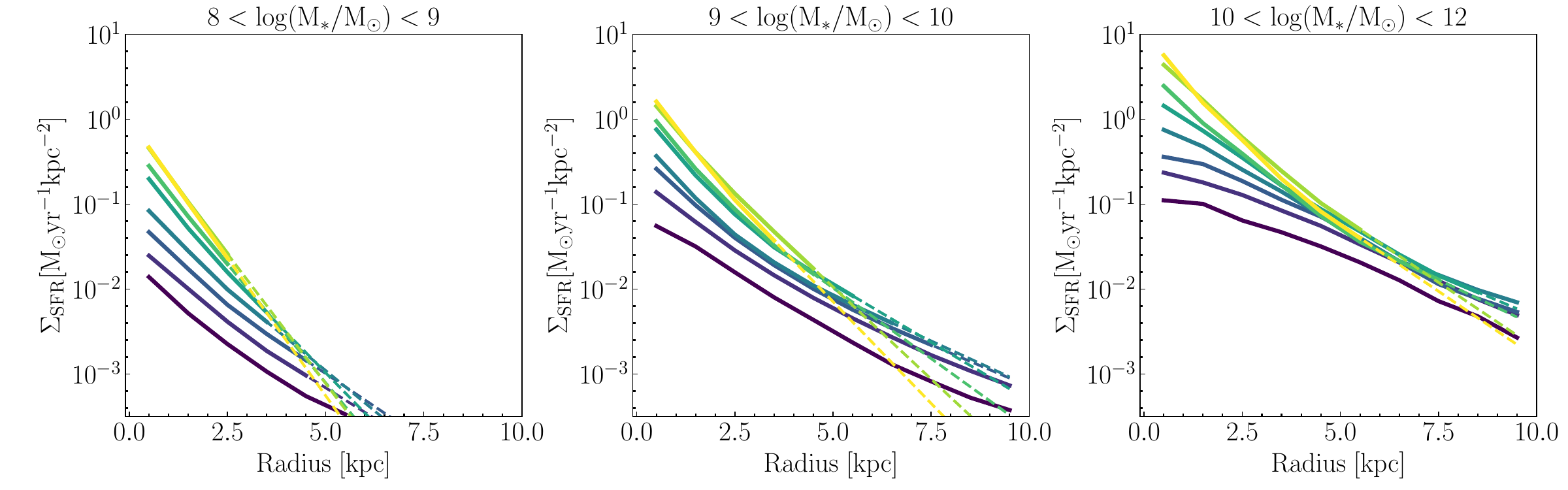}\\
\includegraphics[width=1.0\textwidth]{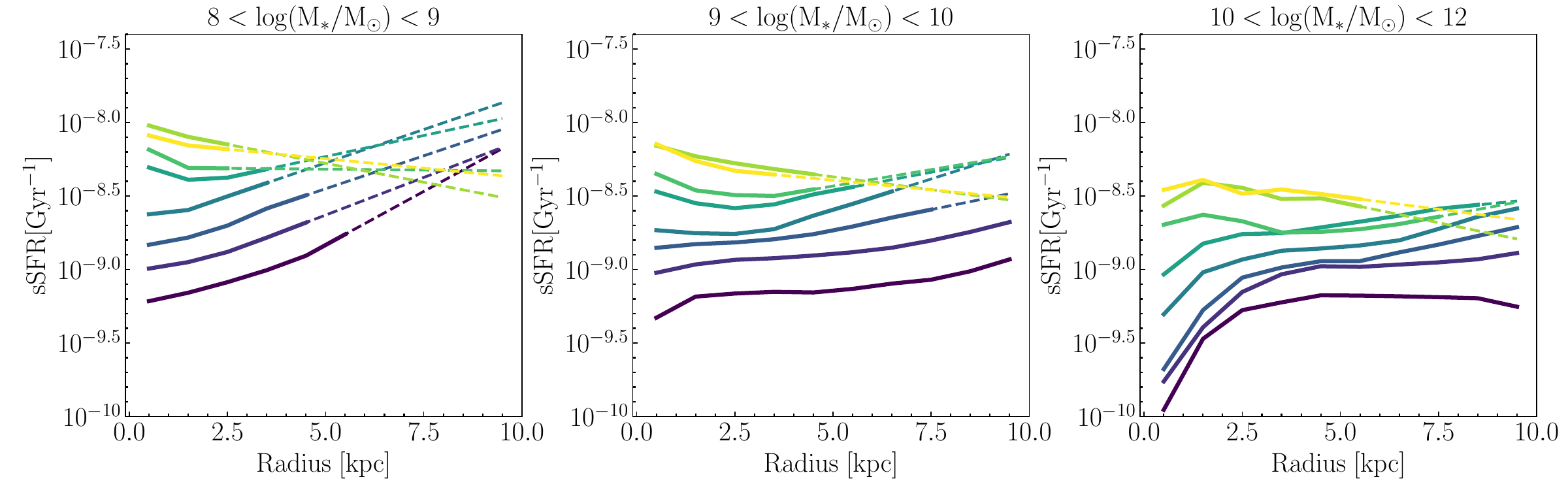}
\caption{Similar to Figure \ref{fig9} - \ref{fig11}, but for radial profiles derived from the deconvolved images.}
\label{fig14}
\end{figure*} 

In this work, we measure galaxy profiles directly from PSF-matched images with a radial step of $0.2R_{\rm e}$. However, PSF smearing can wash out gradient information on scales comparable to the PSF (e.g., \citealt{2018ApJ...860...60L, 2024ApJ...975..252H}). Although, in Section \ref{sec:4.4}, we have measured sSFR gradients using radial separations larger than the PSF FWHM to minimize the impact of PSF smearing on our results, some studies suggest that deconvolved images are more appropriate for mitigating PSF effects (e.g., \citealt{2013ApJ...763...73S, 2019ApJ...877..103S, 2020ApJ...905..170M, 2022ApJ...937L..33S, 2023ApJ...945..155M}).  Additionally, some studies also suggested that the radial steps should be larger than the PSF half-width at half-maximum to account for correlated noise (e.g., \citealt{2025ApJ...980..168D}). In this section, we also present the results obtained from the deconvolved images to test the roboustness of our results\footnote{The corresponding results will also be available at https://github.com/jsong-astro/JWST-CANDELS}.

Following the method described in the previous studies, we obtain the deconvolved images in every band of each galaxy by fitting the corresponding image with a single S\'ersic model. Then a first-order correction is applied to the deconvolved models by adding the residual flux between the observed image and the PSF-convolved model. Based on the rest-frame 1 $\mu$m morphology, we construct a series of concentric elliptical annuli with a radial step of 2 pixels (0.08 arcsec), exceeding the PSF half-width at half-maximum. We then estimate the physical-property profiles following the procedure described in Section \ref{sec:3.4}.

Using the same methodology as in Section \ref{sec:4}, we also derive the median profiles of galaxies across different redshift and stellar-mass bins. The results are presented in Figure \ref{fig14}. It can be seen from this figure, after accounting for PSF smearing, the results are largely consistent with those presented in Figures \ref{fig9}–\ref{fig11}. As in Section \ref{sec:4.4}, we also estimate the corresponding sSFR gradients, with the results presented in Table \ref{tab:1}. For low-mass galaxies, the transition in sSFR gradients becomes even more pronounced. In contrast, for massive galaxies, the central sSFR shows mild suppression, and the transition in sSFR gradients between 0.5 kpc and 2.5 kpc is less evident. However, when a larger radial range is considered to derive the sSFR gradient, negative gradients can also be observed at high redshift.

\end{document}